\documentclass[twocolumn]{aastex62}

\usepackage{amssymb,amsmath}
\usepackage{natbib}
\newcommand{\Teff}{T_{\mathrm{eff}}}
\newcommand{\logg}{\log{g}}
\newcommand{\logRHK}{\log{R'_{\mathrm{HK}}}}

\begin{document}

\title{Retired A Stars and Their Companions VIII: 15 New Planetary Signals Around Subgiants and Transit Parameters for California Planet Search Planets with Subgiant Hosts}

\author[0000-0002-4927-9925]{Jacob K. Luhn}
\altaffiliation{NSF Graduate Research Fellow}
\affiliation{Department of Astronomy, The Pennsylvania State University,  525 Davey Lab, University Park, PA 16802, USA}

\author{Fabienne A. Bastien}
\affiliation{Department of Astronomy, The Pennsylvania State University,  525 Davey Lab, University Park, PA 16802, USA}

\author{Jason T. Wright}
\affiliation{Department of Astronomy, The Pennsylvania State University,  525 Davey Lab, University Park, PA 16802, USA}

\author{John A. Johnson}
\affiliation{Institute for Theory and Computation, Harvard-Smithsonian Center for Astrophysics, 60 Garden Street, Cambridge, MA 02138, USA}

\author{Andrew W. Howard}
\affiliation{Department of Astronomy, California Institute of Technology, Pasadena, CA, USA}

\author{Howard Isaacson}
\affiliation{Astronomy Department, University of California, Berkeley, CA, USA}


\correspondingauthor{Jacob Luhn}
\email{jluhn@psu.edu}

\begin{abstract}
We present the discovery of 7 new planets and 8 planet candidates around subgiant stars, as additions to the known sample of planets around ``retired  A stars" \citep{Johnson2006}.  Among these are the possible first 3-planet systems around subgiant stars, HD 163607 and HD 4917. Additionally, we present calculations of possible transit times, durations, depths, and probabilities for all known planets around subgiant ($3 < \log{g} < 4$) stars, focused on possible transits during the \emph{TESS} mission. While most have transit probabilities of 1-2\%, we find that there are 3 planets with transit probabilities $> 9\%$.
\end{abstract}

\section{Introduction}
The occurrence rate of Jupiter-mass planets has been observed to increase with both metallicity and mass of the host star \citep{Ida2004,Fischer2005,Johnson2010, Bowler2010}. Despite the increased occurrence rates, more massive \emph{stars} show a paucity in the number of massive \emph{planets} at short periods, both ``hot Jupiters" on very close orbits and even those out to separations of $\sim$1 au \citep{Johnson2007, Sato2008}. As these planets are the easiest to detect via radial velocity (RV) and transit photometry methods, this does not represent an observational bias, indicating that stellar mass plays a large role in shaping the formation and orbital evolution of planets. 

\citet{Johnson2006} targeted intermediate-mass evolved stars in an effort to observe and study the properties of planets around stars more massive than the sun. The stars selected for the survey come from just below a section of the main sequence known as the Hertzprung gap (HG), which lies between the main sequence and the red giant branch. The sample was selected to include mostly intermediate-mass stars ($M_{\star} \gtrsim1.3~M_{\odot}$), often referred to as ``retired A-type" stars since they had A spectral types when they were on the main sequence. On the main sequence, these stars are difficult targets for precise radial velocity measurements for two reasons: first, because they rotate rapidly, any absorption features they have are significantly Doppler-broadened and second, because of their high effective temperatures, they lack strong absorption lines observed in cooler stars. As a result, typical RV surveys avoid main-sequence stars with intermediate to high masses. However, stars of this mass which have left the main sequence become suitable for RV measurements due to their cooler atmospheres and slower rotational velocities, which lead to narrower absorption features in their spectra. The Retired A-star survey has been responsible for the discovery of more than 40 exoplanets around subgiant stars to date\footnote{\url{http://exoplanets.org}}. 

The planets discovered by the Retired A-star survey exhibit two key differences from planets discovered around lower mass main sequence stars: 1) an increased abundance of giant planets and 2) a decrease in planets with shorter periods. These differences between the planets around more massive stars and those on the main sequence has sparked debate about the true masses of the stars themselves \citep{Lloyd2011,Johnson2012,Lloyd2013,Johnson2013}. The essence of the debate revolves around the statistical likelihood of finding a large population of relatively massive stars in the region of the HR diagram selected for the Retired A-star survey. Different galactic models and using volume-limited versus magnitude-limited samples produce conflicting results for the expected number of massive stars. Further, \citet{Schlaufman2013} investigated the kinematic of these stars, concluding that the velocity dispersions were too high for them to in fact be massive stars. Using asteroseismology, \citet{Stello2017} showed that there was indeed an overestimate in mass for stars above 1.6~M$_{\odot}$ among their sample of 8 stars. More recently, \citet{Ghezzi2018} reanalyzied a subset of the Retired A Star sample and determined atmospheric, rotational, evolutionary and kinematic parameters, finding errors much lower than the 50\% overestimate suggested by \citet{Lloyd2011,Lloyd2013}. By accounting for reddening, they find that the velocity dispersions are consistent with those of more massive main sequence stars with an offset of 0.04~M$_{\odot}$, suggesting that these are in fact massive stars. Putting aside the mass argument, since the Retired A-star sample is composed of entirely \emph{evolved} stars\footnote{selected based on their position near the Hertzsprung gap.}, the ``desert" of short-period planets appears regardless of the true mass of these stars, as there is a noticeable lack of short-period planets around post-main sequence stars. In fact, there appears to be a pileup of planets around post-main sequence stars at 1-2 au, all with masses $\sim$ 1-5 M$_{\mathrm{Jup}}$, possibly indicating that these types of planets are those most likely to remain after post-main sequence stellar evolution while most planets interior to this region are lost due to tidal capture \citep{Villaver2009}. 

Given the lack of short period planets around subgiants, there are few planets around subgiants that are easily detectable by transit surveys. Despite this, a handful of short-period planets with subgiant hosts have been discovered by \emph{Kepler} \citep{Borucki2010}, namely \emph{Kepler} 435~b \citep{Borucki2010}; \emph{Kepler} 56~b,c \citep{Borucki2011}; \emph{Kepler} 108~b,c, 278~b,c, 391~b,c \citep{Rowe2014}; \emph{Kepler} 432~b \citep{Ciceri2015}; and \emph{Kepler} 637~b, 815~b, 1270~b, 774~b, 1004~b, 1394~b, and 643~b \citep{Morton2016} as well one planet in the KELT survey, KELT-11~b \citep{Pepper2017}\footnote{KELT-6~b \citep{Collins2014} also orbits what appears to be a very slightly evolved subgiant.}. Unfortunately, the \emph{Kepler} subgiants are generally too faint and were lower priority than dwarfs for RV followup and so we do not have measured masses for most of these planets\footnote{Those with measured masses are \emph{Kepler} 56~b,c \citep{Otor2016}, \emph{Kepler} 432~c \citep{Quinn2015}}. However, given the large sample of existing \emph{radial velocity} subgiants with known planets, we have the opportunity to examine the transit probabilities for these lower priority transit search targets, in the context of future transit surveys, particularly on the upcoming \emph{TESS} mission \citep{Ricker2014}. Observing transits of existing RV planets has a number of advantages, the biggest being removing the $\sin{i}$ degeneracy for the masses of these planets. Additionally, transits provide a model-independent measure of the stellar density, which could be useful for confirming the masses of the host stars. Another benefit is the tendency of RV surveys to target bright stars, which enables easier ground-based follow-up and provides targets for transmission spectroscopy, for which we have very few studies of temperate, long period Jupiter-sized planets \citep{Kane2009}. Despite large transit surveys (KELT, HAT, \emph{Kepler}, \emph{K2}) in this time, long period Jupiters with hosts bright enough for transmission spectroscopy remain scarce. With these ideas in mind, we present transit parameters for subgiant stars already observed by the California Planet Search \citep{Howard2010} radial velocity survey.

In \autoref{sec:observations}, we describe our sample of subgiant stars, RV measurements, and stellar properties. We then present the discovery of 15 new planetary signals around subgiant stars and several stellar companions, spanning a spectrum of secure detections. \autoref{sec:new_companions} contains those which we determine to be secure planet detections. \autoref{sec:planet_candidates} list those that are ``planet candidates" to better illustrate the varying degrees of security. We include all probable planet candidates in the interest of listing the possible transit times for planets around subgiant hosts. \autoref{sec:binaries} contains a list of the stellar companions. The new planetary signals in this work increase the sample of radial velocity planets around subgiants by more than 25\%. In \autoref{sec:transit_params} we present the transit parameters for these stars. Despite the typically small transit probability of 1-2\%, it is likely, given the sample size of 85 planets, that several do indeed transit. Finally, we present a summary and conclusions in \autoref{sec:conclusions}.

\section{Observations}\label{sec:observations}
\subsection{Sample Selection and Stellar Properties}
\begin{figure}
\includegraphics[width=\columnwidth]{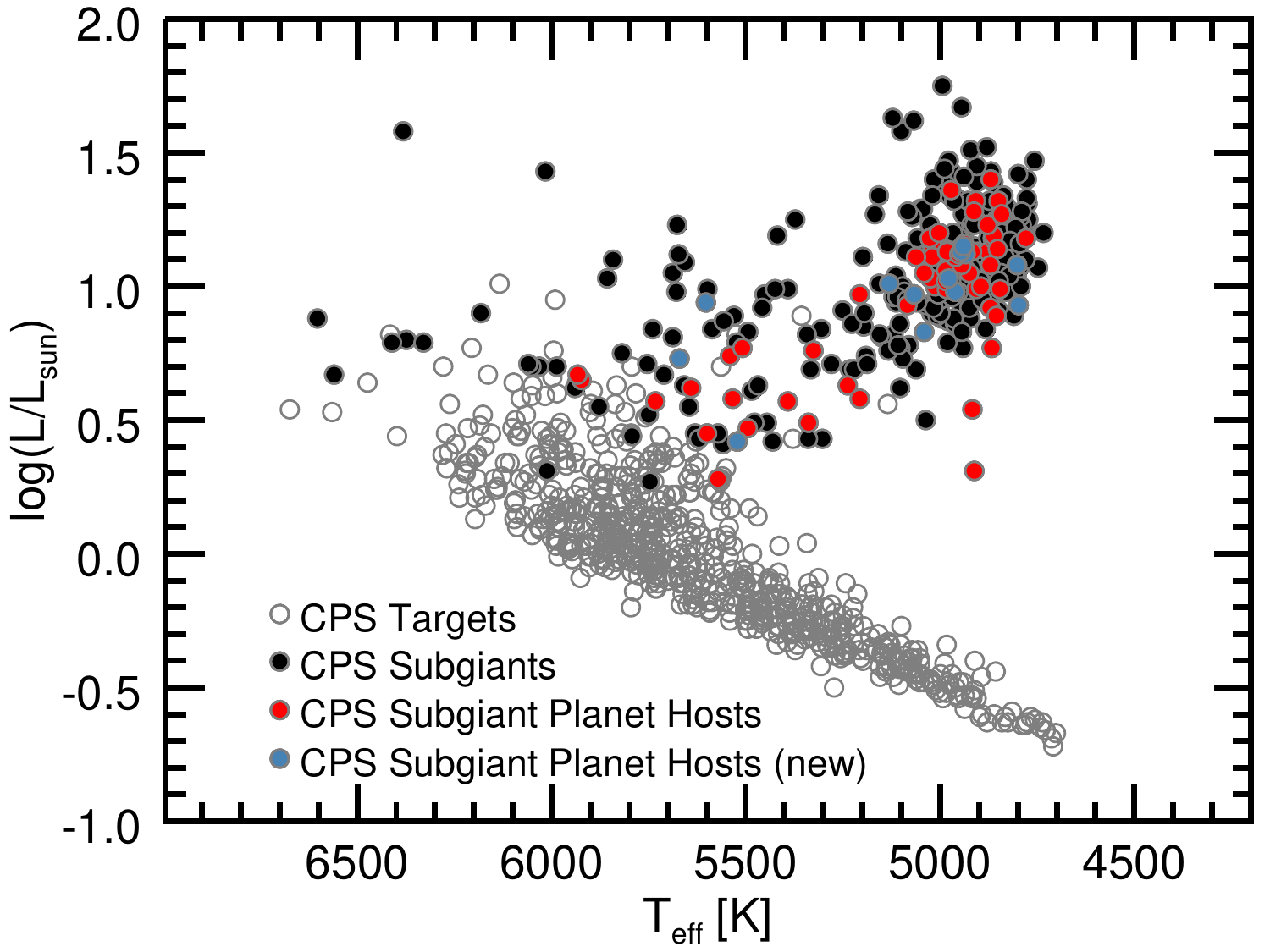}
\caption{HR diagram showing the sample of subgiants among all CPS stars and the subset of those that are planet-host stars. In gray, all CPS target stars with $\Teff$ and $\log(L/L_{\odot})$ values from \citet{Brewer2016} are shown. The black points show the subsample of CPS subgiant stars which are here chosen to be stars with $ 3.0 < \logg < 4.0$. In red are shown the planet host stars (known and new in this work).}
\label{fig:HR_sample}
\end{figure}

Our sample is composed of stars observed as part of the California Planet Search (CPS) with $3.0 < \logg < 4.0$.  From this sample of over 400 stars, we have identified those that are known to host planets, as identified in either \url{www.exoplanets.org} or \url{www.exoplanet.eu}. This list comprises 42 stars in our sample. \autoref{fig:HR_sample} shows an HR diagram of the entire California Planet Search sample, the subset of CPS subgiant stars, and the planet-host stars in our sample. 

All stellar properties come from \citet{Brewer2016} (erratum \citep{Brewer2017} cited as B17 hereafter), who used 1 dimensional LTE model spectra to fit to a star's observed spectrum to determine effective temperatures, metallicities, surface gravities, and elemental abundances. In conjunction with \emph{Hipparcos} parallaxes and V-band magnitudes, these spectral measurements were used to derive masses, radii, and luminosities. Their iterative fitting technique and improved line list correct for systematic discrepancies in $\logg$ between spectroscopy and asteroseismology \citep{Huber2013,Bastien2014b} and their spectroscopic methods are now consistent with the values of $\logg$ obtained from asteroseismology \citep{Brewer2015}. The stellar properties for the stars in this sample are given in \autoref{tbl:stellar_params}. We note that while \citet{Ghezzi2018} determined masses and radii with smaller uncertainties than \citet{Brewer2016} for a subset of the Retired A sample, more than $25\%$ of the stars in this work were not re-analyzed by \citet{Ghezzi2018}. Since the stellar parameters are largely consistent within errors between \citet{Ghezzi2018} and \citet{Brewer2016}, we use stellar properties from \citet{Brewer2016} for consistency \footnote{The radius measurement of HD 193342 is the only parameter for which the two samples are not consistent within the errors, with radius $6~R_{\odot}$ from \citet{Brewer2016} and $8.5~R_{\odot}$ from \citet{Ghezzi2018}. Despite this large discrepancy, the radius difference will only affect the predicted transit parameters for HD 193342~b, which is unlikely to transit (0.27\% transit probability). Using the radius from \citet{Ghezzi2018} would result in a slightly increased transit probability ($\sim 0.4\%$) as well as a smaller transit depth and longer transit durations if it does indeed transit.}.

\subsection{Spectra and Radial Velocity Measurements}\label{sec:other_telescopes}
Observations were taken at Keck Observatory using the High Resolution Echelle Spectrometer (HIRES) with a resolution of $R \approx 55,000$. For a V= 8 magnitude star, the exposure required is 90s to reach a signal-to-noise ratio of 190 at 5800 \r{A}. Radial velocities are calculated using the iodine-cell calibration technique and the forward-modeling procedure described in \citet{Butler1996} and later \citet{Howard2011}. For several stars that were known planet hosts, we included the non-Keck RV measurements as published with the planet discovery. These stars and the instruments used are briefly described below.

\paragraph{HD 1502}
In addition to Keck/HIRES observations HD 1502 was also observed using the Tull Coude Spectrograph \citep{Tull1995} on the Harlan J. Smith Telescope as well as the High Resolution Spectrograph \citep{Tull1998} on the Hobby Eberly Telescope (HET) \citep{Ramsey1998}, both of which have $R=60,000$. Differential RVs were computed using the \emph{Austral} I$_2$-data modeling algorithm \citep{Endl2000}. The velocities are given in Table 2 of \citet{Johnson2011}.

\paragraph{HD 159868}
HD 159868 was observed mainly with the Anglo-Australian Telescope (AAT) \citep{Jones2002} using the UCLES echelle spectrograph. Radial velocities are given in Table 2 of \citet{Wittenmyer2012}.

\paragraph{HD 192699}
HD 192699 was also observed with Lick Observatory's Shane 3~m and 0.6~m Coude Auxiliary Telescopes, which feed into the Hamilton spectrometer \citep{Vogt1987}. The velocities for this star are given in Table 3 of \citet{Johnson2007b}.

\paragraph{HD 114613}
HD 114613 is another star that was observed as part of the Anglo-Australian Planet Search and has RV measurements from the UCLES echelle spectrograph\citep{Diego1990}. The radial velocities are given in Table 1 of \citet{Wittenmyer2014}.

\paragraph{HD 38801}
Additional RV observations of HD 38801 come from the high dispersion spectrograph on the 8.2~m Subaru Telescope \citep{Noguchi2002}. The RV measurements can be found in Table 2 of \citet{Harakawa2010}.

\paragraph{HD 181342}
HD 181342 (also called HIP 95124) has a number of observations from various instruments, including two telescopes at the Cerro Tololo Inter-American Observatory: the 1.5~m telescope using the CHIRON spectrograph \citep{Tokovinin2013} and the 2.2~m telescope using the FEROS spectrograph \citep{Kaufer1999}. In addition, it was observed as part of the Pan Pacific Planet Search \citep{Wittenmyer2011} using the UCLES spectrograph \citep{Diego1990}. All velocities from these instruments are listed in Table A.4 of \citet{Jones2016}.

\paragraph{HD 5608}
HD 5608 was observed as part of the Okayama Planet Search Program \citep{Sato2005}, which uses the HIgh Dispersion Echelle Spectrograph (HIDES) on the1.88~m telescope at Okayama Astrophysical Observatory (OAO). The velocities are given in Table 2 of \citet{Sato2012}.

\paragraph{HD 10697}
HD 10697 was also observed using the two telescopes at McDonald observatory: HET and the 2.7~m Harlan J. Smith telescope. The velocities are given in the electronic version of Table 10 in \citet{Wittenmyer2009}.

\paragraph{HD 210702}
HD 210702 has observations from both Lick Observatory and OAO. The velocities from Lick are given in \citet{Johnson2007}. The velocities from OAO are given in Table 8 of \citet{Sato2012}

\paragraph{HD 214823}
HD 214823 was observed using ELODIE and SOPHIE/SOPHIE+ instruments on the 1.93~m telescope at Observatoire de Haute-Provence. Given ELODIE's large instrumental uncertainty (15-30 m/s) and that there are only 5 measurements from ELODIE we use only the radial velocities from SOPHIE and upgraded SOPHIE+ instruments. This accounts for an additional 24 observations for this star. These velocities are given in \citet{Diaz2016}.

\begin{deluxetable*}{c c c c c}
\tablecaption{Summary of Additional Non-Keck HIRES velocities}
\tabletypesize{\scriptsize}
\tablehead{
	 \colhead{Star} 	&  \colhead{Telescope}    & \colhead{Instrument}  	 & \colhead{N$_{\mathrm{obs}}$} & \colhead{Ref} }
	  \startdata
	 HD 1502 & Harlan J. Smith Telescope & TCS & 25 & \citet{Johnson2011} \\
	 HD 1502 & Hobby-Eberly Telescope & HRS & 20 & \citet{Johnson2011} \\
	 HD 159868 & Anglo-Australian Telescope & UCLES & 47 & \citet{Wittenmyer2012} \\
	 HD 192699 & Lick Observatory & Hamilton spectrometer & 34 & \citet{Johnson2007b} \\
	 HD 114613 & Anglo-Australian Telescope & UCLES & 223 & \citet{Wittenmyer2014} \\
	 HD 38801 & Subaru & High Dispersion Spectrograph & 11 & \citet{Harakawa2010} \\
	 HD 181342 & CTIAO 1.5~m & CHIRON & 11 & \citet{Jones2016} \\
	 HD 181342 & CTIAO 2.2~m & FEROS & 20 & \citet{Jones2016} \\
	 HD 181342 & Anglo-Australian Telescope & UCLES & 5 & \citet{Wittenmyer2011} \\
	 HD 5608 & OAO 1.88~m & HIDES & 43 & \citet{Sato2012} \\
	 HD 10697 & Harlan J. Smith Telescop & TCS & 32 & \citet{Wittenmyer2009} \\
	 HD 10697 & Hobby-Eberly Telescope & HRS & 40 & \citet{Wittenmyer2009} \\
	 HD 210702 & Lick Observatory & Hamilton spectrometer & 29 & \citet{Johnson2007} \\
	 HD 210702 & OAO 1.88~m & HIDES & 36 & \citet{Sato2012} \\
	 HD 214823 & Observatoire de Haute-Provence 1.93~m & SOPHIE & 13 & \citet{Diaz2016} \\
	 HD 214823 & Observatoire de Haute-Provence 1.93~m & SOPHIE+ & 11 & \citet{Diaz2016} \\
	  \enddata
\end{deluxetable*}

\section{7 New Planets around Subgiants}\label{sec:new_companions}
Here we present the new planets discovered around subgiant stars. All radial velocity time series for every star used in this work are given in \autoref{tbl:all_rvs}. We note that our quoted values of $\chi^2$ from the fitting procedure RVLIN throughout this paper are significantly larger than 1. The reason is because we have use only the reported internal measurement errors in our fits. Previous works have inflated individual measurement errors by including a ``jitter'' term in quadrature to account for increased RV variations due to intrinsic stellar variability \citep{Wright2005}. However, given the large known uncertainty in jitter estimates in \citet{Wright2005}, we do not follow this approach. Finally, all stellar parameters quoted here come from B17 and were used to obtain planet parameters where necessary (planet mass, transit probabilities, etc.).

The majority of planets described below have periods $> 300$ days. At these periods, the greatest concern for the planet validity is stellar activity cycles that operate on similar timescales. To avoid misidentifying a stellar activity cycle as a planet, we have examined the simultaneous s-index time series for each of these planets. Except for HD 180053, none of the new planet hosts in this work (including planet candidates in \autoref{sec:planet_candidates}) show significant activity cycles that correlate with the radial velocities. Furthermore, none of the stars show evidence of \emph{non-cycling} activity that correlates with the radial velocity measurements. Thus, we are confident that we have identified planet signals rather than activity-induced radial velocities.

Another timescale of importance is the stellar rotation timescale, which can be hundreds of days for giant stars. For the subgiants in this work we expect that the rotation periods are all much less than the planet periods given that most planet periods here are $>300$ days. While it is possible that some of the planet periods are on timescales similar to the rotation timescales, the coherence and amplitudes of these signals make it unlikely that they are due to rotationally modulated inhomogeneities.

Lastly, we have examined the spectral window functions of the radial velocity time series as described in \citet{Dawson2010} to be certain that none of the planets in this work are a result of sparsely sampled data that could introduce spurious periodicity that lead to false peaks in the periodogram. We find that none of the planets in this work correspond to significant peaks in the spectral window function.

We have ordered the new planets in this paper in descending order of detection security, with the most secure detections first. In general we based our threshold for planet status as meeting three main criteria. First, we need to have observed a full period such that N$_{\mathrm{p}} > 1$, where N$_{\mathrm{p}}$ is the number of periods observed, found by simply dividing the baseline of the time series by the best-fit period. We generally wish N$_{\mathrm{p}} \gg 1$ to be certain about a planet but our threshold of 1 makes it necessary to have observed two instances where the velocities have turned over. Second, we examine the false alarm probability (FAP) and choose a 1\% FAP threshold. Third, and most important, is the ratio of the semi-amplitude to the RMS, $K/\sigma$. This ratio alone tells the ability to detect the planet in a single measurement. We expect then that our ability to detect a planet scales with the number of observations and define a detection threshold $D$ 
\begin{equation}
D \equiv \frac{K}{\sigma} \sqrt{N-M},
\end{equation}
where N is the number of observations and M is the number of free parameters in the planet fit (6 for a one planet fit plus 5 for each additional planet and 1 additional free parameter if the fit includes a linear trend). For a ``10$\sigma$ detection'', $D$ must be greater than 10\footnote{We find that these thresholds combine such that a single threshold of $D>17$ is able to distinguish planet vs. candidate. Although arbitrary, this singular threshold is able to match our intuition.}

\subsection{A 1.7 $M_{\mathrm{Jup}}$ planet around HD 72490}
HD 72490 is a G5 subgiant with V$=7.83$, B-V$=0.95$ and a parallax-based distance of 124.22 pc \citep{SIMBAD}. It has a radius of $\sim$5 R$_{\odot}$, an effective temperature $\Teff = 4934$~K, and surface gravity $\logg = 3.210$. A summary of its stellar parameters can be found in \autoref{tbl:stellar_params}.

\begin{figure*}
\centering
\includegraphics[width=\columnwidth]{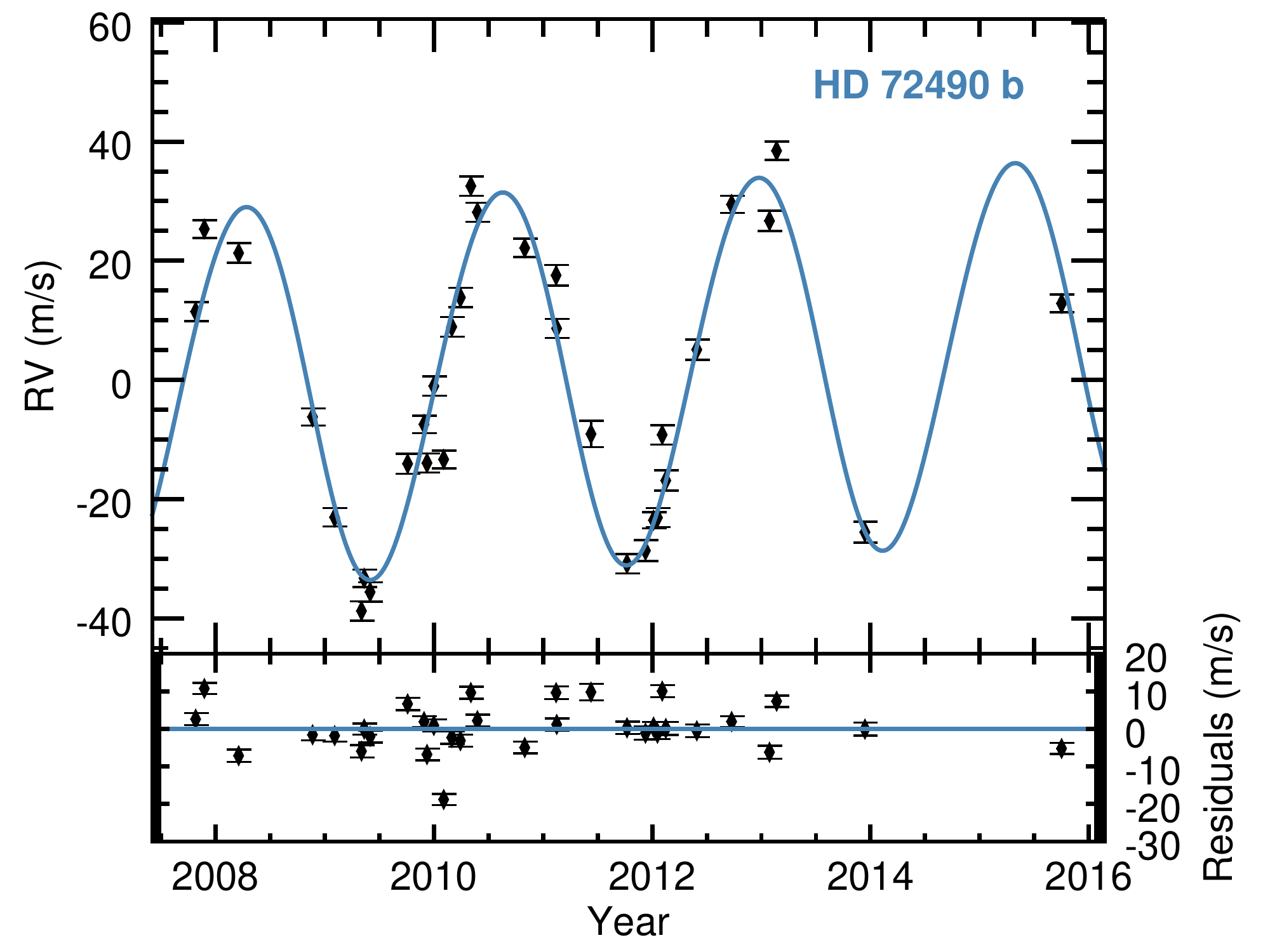}
\includegraphics[width=\columnwidth]{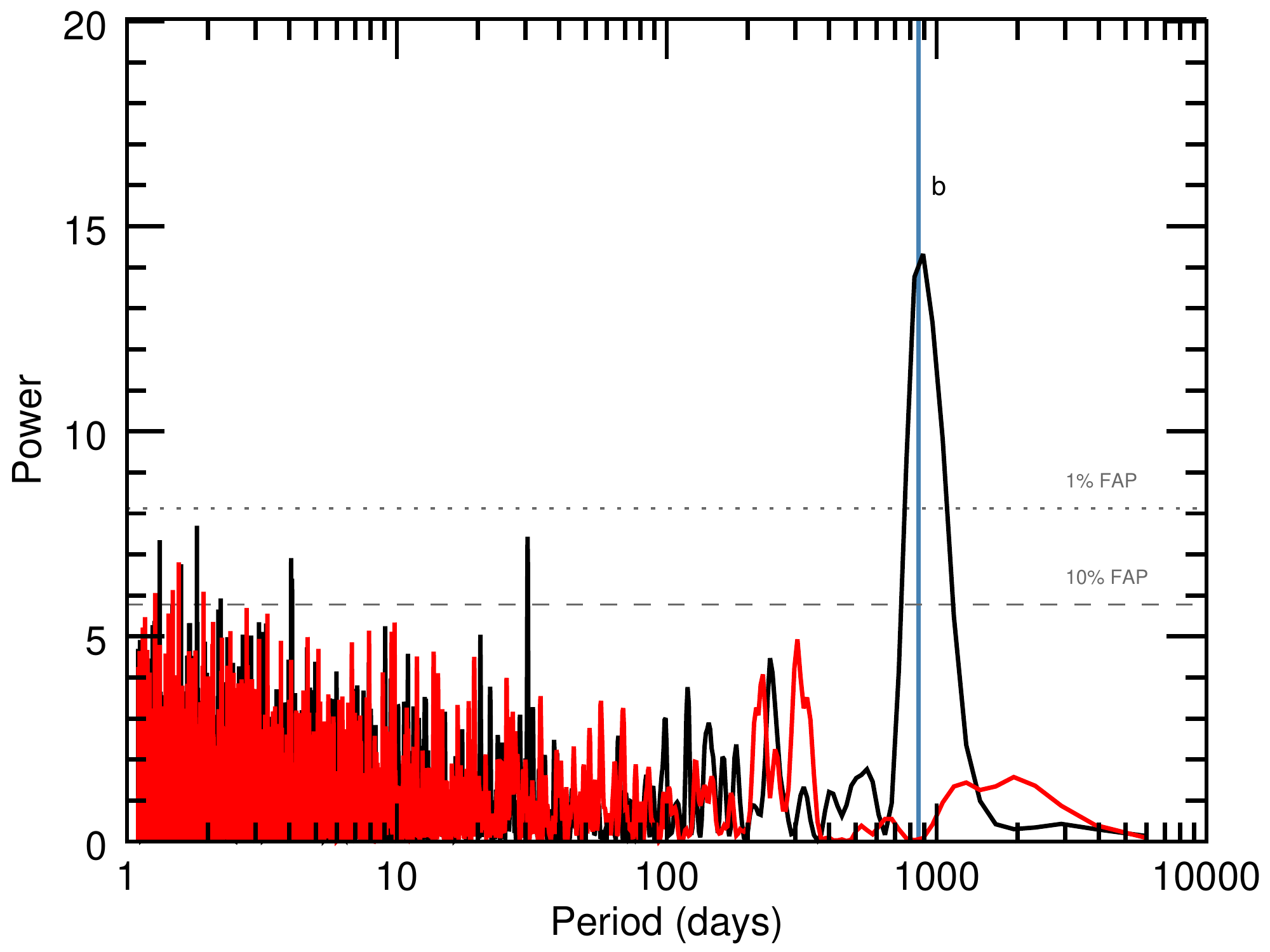}
\caption{(\emph{Left}) Time series and best-fit orbital solution for HD 72490~b, which has period P~$= 858.67$ days, eccentricity  $e=0.0569$, and minimum mass m$_{p}\sin{i} =1.709$~M$_{\mathrm{Jup}}$. The residuals are shown in the bottom panel, which have an RV RMS~$ = 6.43$~m$\,$s$^{-1}$. The remaining best-fit parameters can be found in \autoref{tbl:orbital_params}. (\emph{Right}) Periodogram of HD 72490 RV data before (black) and after (red) subtracting the best-fit planet parameters for HD 72490~b. The vertical line indicates the best-fit period of HD 72490~b.}
\label{fig:72490_time_series}
\end{figure*}

The initial observations of this star began in late 2007 and finished in 2014, with only 2 observations since 2014.  It was identified in \citet{Butler2017} as having a planet candidate. The best-fit Keplerian orbital solution yields an orbital period of $P= 858\pm12$ days (2.35 years), velocity semiamplitude $K=33.5\pm1.5 $~m$\,$s$^{-1}$, and eccentricity $e=0.124\pm0.046$. From the stellar mass M$_{\star} = 1.21$~M$_{\mathrm{\odot}}$ we derive the minimum mass of the planet m$_{p}\sin{i} =1.768\pm0.080$~M$_{\mathrm{Jup}}$ and semi-major axis $a=1.88 \pm 0.17$~AU. The full set of orbital parameters and corresponding uncertainties are given in \autoref{tbl:orbital_params}. The time series showing the signal of HD 72490~b is shown in the left panel of \autoref{fig:72490_time_series}. A search for a possible second planet yielded nothing of significance, which is supported by the periodogram of the raw RV data along with the periodogram of the data with the best-fit model of HD 72490~b subtracted out (shown in the right panel of \autoref{fig:72490_time_series}).  For HD 72490~b, the preliminary next predicted transit is BJD 2459074.350~$\pm~41.296$ (08/12/2020). Its most recent predicted transit was early April 2018 and before that late November 2015, unfortunately missing its \emph{K2} Campaign 5 observations by a mere 4 months.

\subsection{A Jupiter-mass planet around HD 94834}
HD 94834 is a K1 subgiant with V$=7.61$, B-V$=0.99$ \citep{SIMBAD}. It has effective temperature $\Teff = 4798$~K, and surface gravity $\logg = 3.22$. A summary of its stellar parameters can be found in \autoref{tbl:stellar_params}.

\begin{figure*}
\centering
\includegraphics[width=\columnwidth]{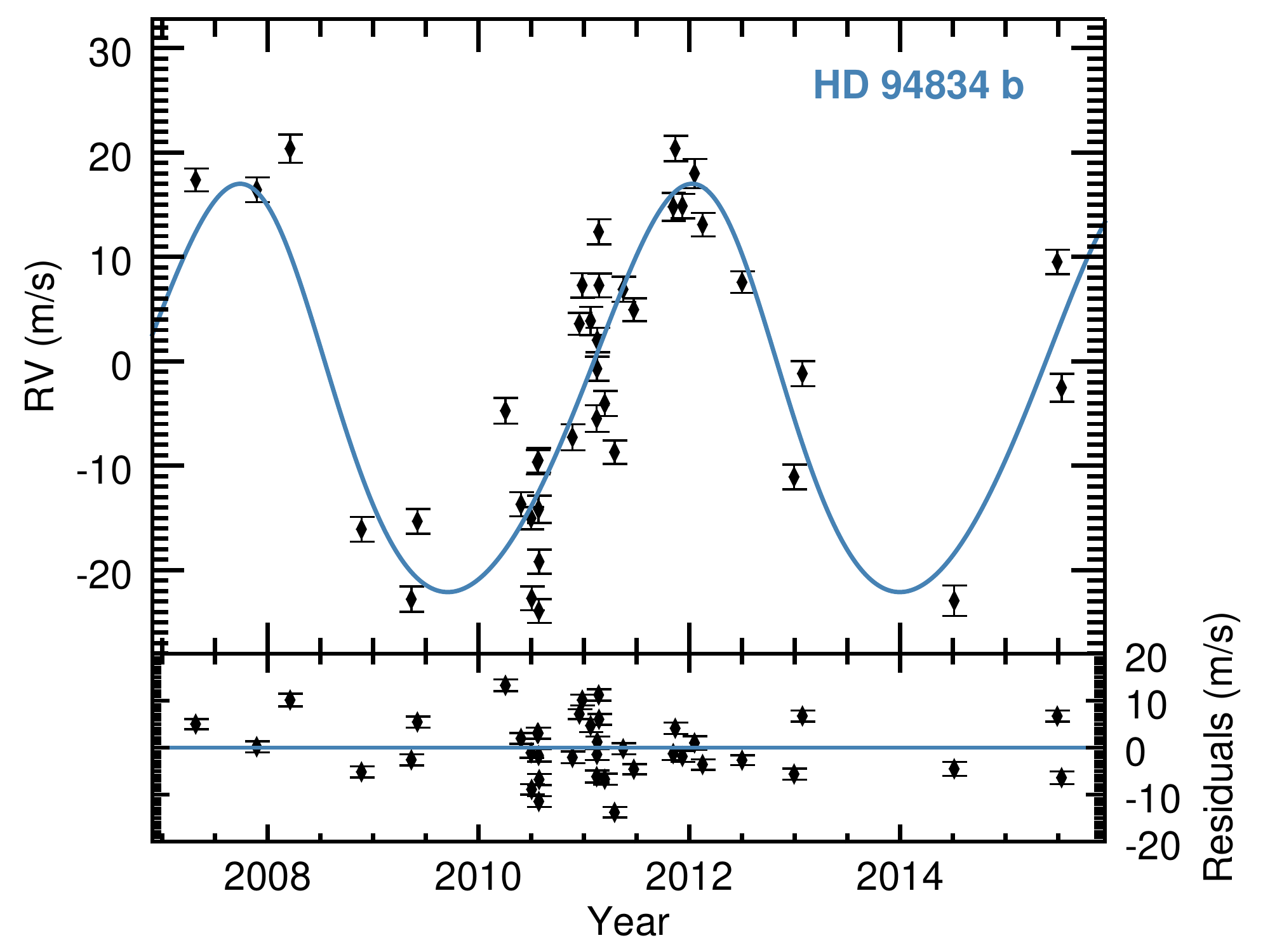}
\includegraphics[width=\columnwidth]{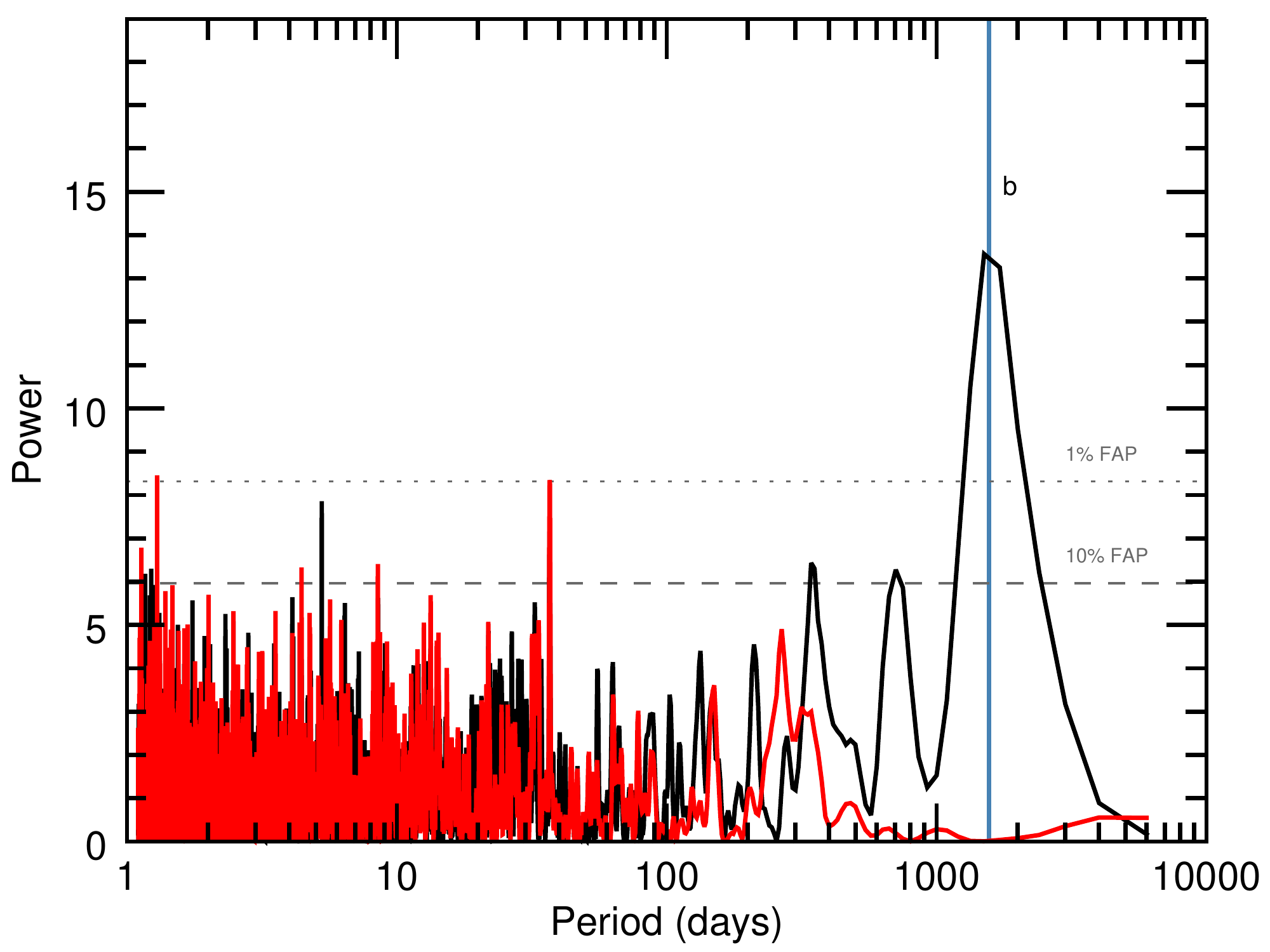}
\caption{(\emph{Left}) Time series and best-fit orbital solution for HD 94834~b, which has period P~$= 1576$ days, eccentricity  $e=0.107$, and minimum mass m$_{p}\sin{i} =1.2$~M$_{\mathrm{Jup}}$. The residuals are shown in the bottom panel, which have an RV RMS $ = 6.32$~m$\,$s$^{-1}$. The remaining best-fit parameters and uncertainties can be found in \autoref{tbl:orbital_params} (\emph{Right}) Periodogram of HD 94834 RV data before (black) and after (red) subtracting the best-fit planet parameters for HD 94834~b. The vertical line indicates the best-fit period of HD 94834~b.}
\label{fig:94834_time_series}
\end{figure*}

Observations of this star span roughly 8.5 years with the majority of observations between 2010 and 2012. The final 5 observations demonstrate the periodicity of the signal. HD 94834 was identified in \citet{Butler2017} as having a planet candidate. Our best-fit orbital solution shows a period of $1576\pm76$ days ($\sim 4.3$ years), velocity semi-amplitude $K=20.7\pm2.9$~m$\,$s$^{-1}$, and eccentricity $e=0.14\pm0.10$. The minimum mass of this planet is $m_{p}\sin{i} =1.192\pm0.017~M_{\mathrm{Jup}}$ with a semi-major axis $a=2.74 \pm0.19$~AU. The full set of orbital parameters are given in \autoref{tbl:orbital_params}. \autoref{fig:94834_time_series} shows the time series of HD 94834~b, with the initial and final periodogram for this system.

\subsection{A Jupiter orbiting HD 14787}
\begin{figure*}
\centering
\includegraphics[width=\columnwidth]{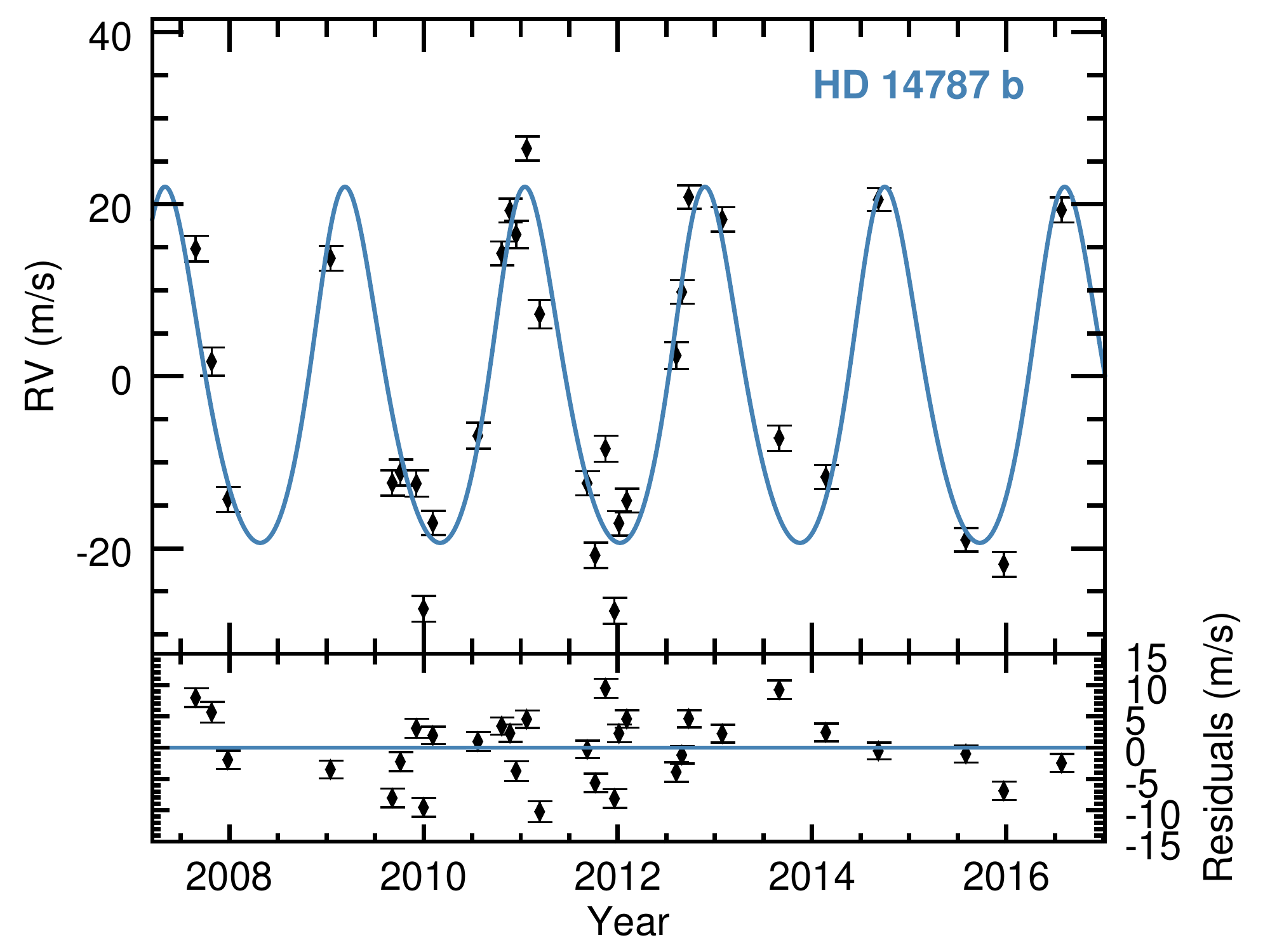}
\includegraphics[width=\columnwidth]{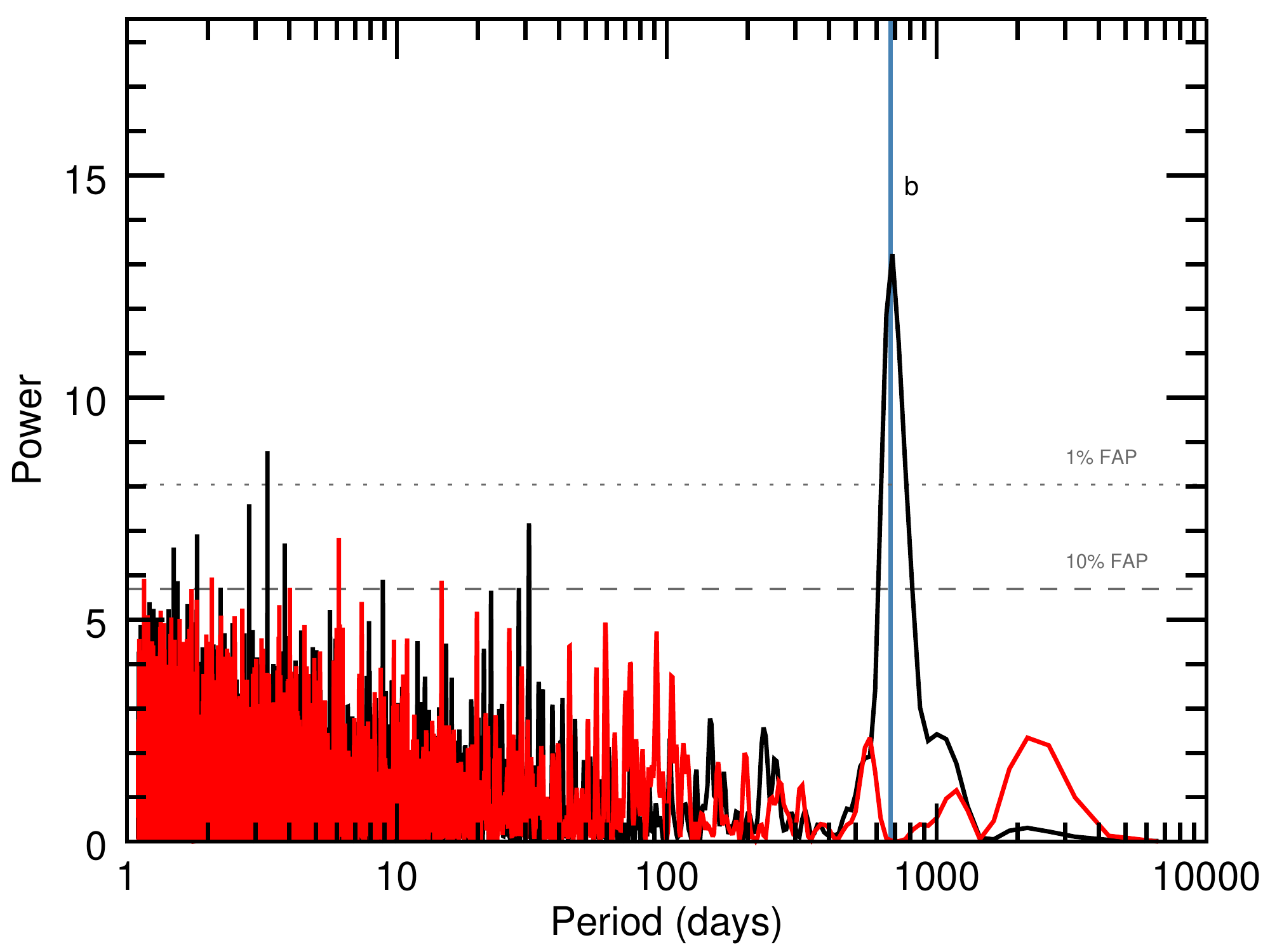}
\caption{(\emph{Left}) Time series and best-fit orbital solution for HD 14787~b, which has which has period P~$= 676$ days, eccentricity  $e=0.155$, and minimum mass m$_{p}\sin{i} =1.12$~M$_{\mathrm{Jup}}$. The residuals are shown in the bottom panel, which have an RV RMS $ =5.3$~m$\,$s$^{-1}$. The remaining best-fit parameters can be found in \autoref{tbl:orbital_params}. (\emph{Right}) Periodogram of HD 14787 RV data before (black) and after (red) subtracting the best-fit planet parameters for HD 14787~b. The vertical line indicates the best-fit period of HD 14787~b.}
\label{fig:14787_time_series}
\end{figure*}

HD 14787 is a G5 subgiant star with V=7.63 and B-V=0.93 \citep{SIMBAD}. It has a mass of M$_{\star} = 1.43$~M$_{\mathrm{\odot}}$, surface gravity $\logg = 3.23$, and $\Teff = 4946$~K. A summary of its stellar parameters can be found in \autoref{tbl:stellar_params}. Observations of this star span about 9 years starting in 2007. The best-fit Keplerian solution yields an orbital period of $676.6\pm8.1$ days, velocity semi-amplitude $20.7\pm1.3$ m/s, and eccentricity 0.155.  It is only slightly more massive than Jupiter-mass with a minimum mass of $1.121\pm0.069$ M$_{\mathrm{Jup}}$. The full set of orbital parameters can be found in \autoref{tbl:orbital_params}. We show in \autoref{fig:14787_time_series} the time series for HD 14787~b as well as the periodogram before and after subtracting out HD 14787~b.

\subsection{A Jupiter orbiting HD 13167}
\begin{figure*}
\centering
\includegraphics[width=\columnwidth]{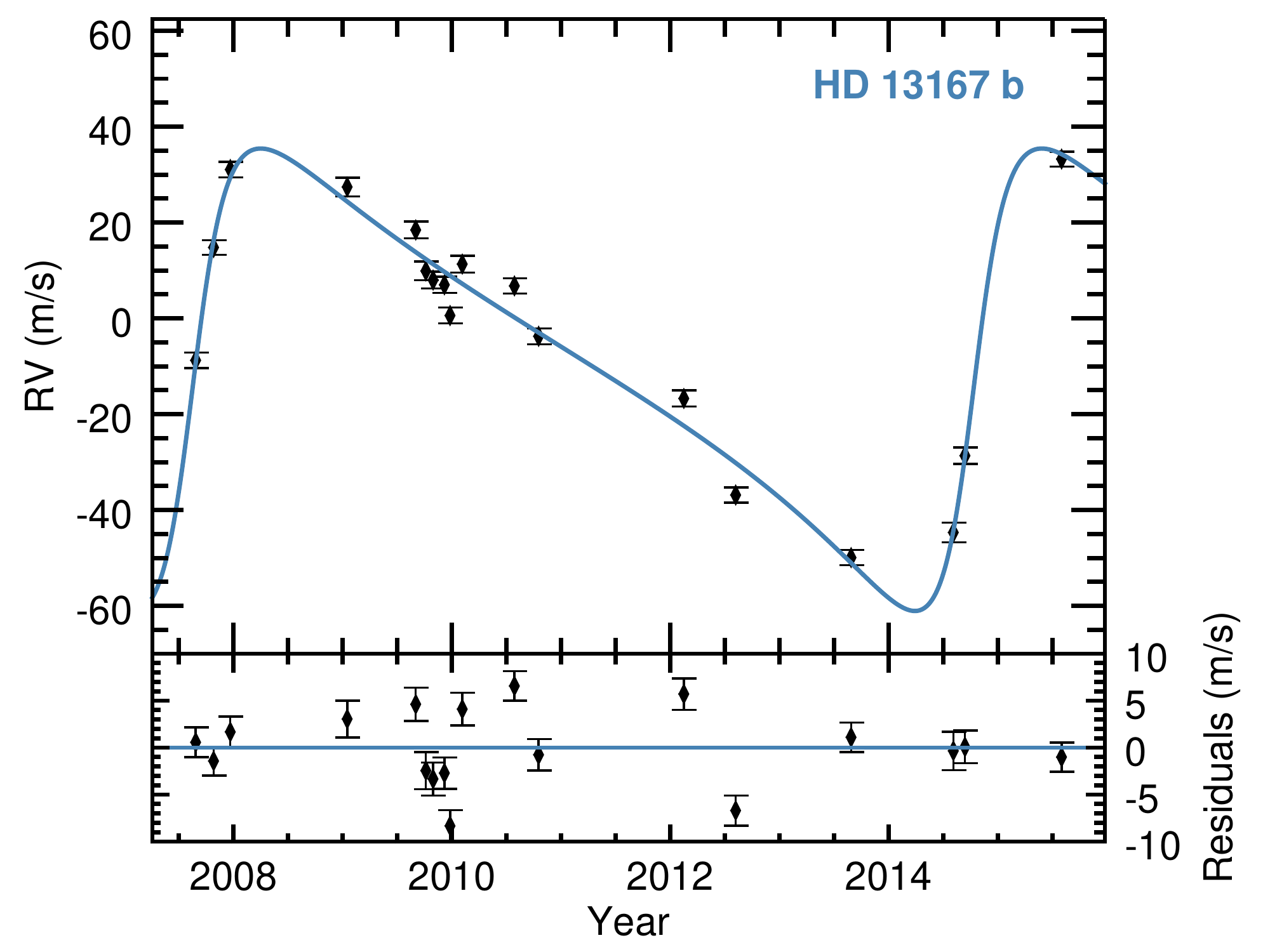}
\includegraphics[width=\columnwidth]{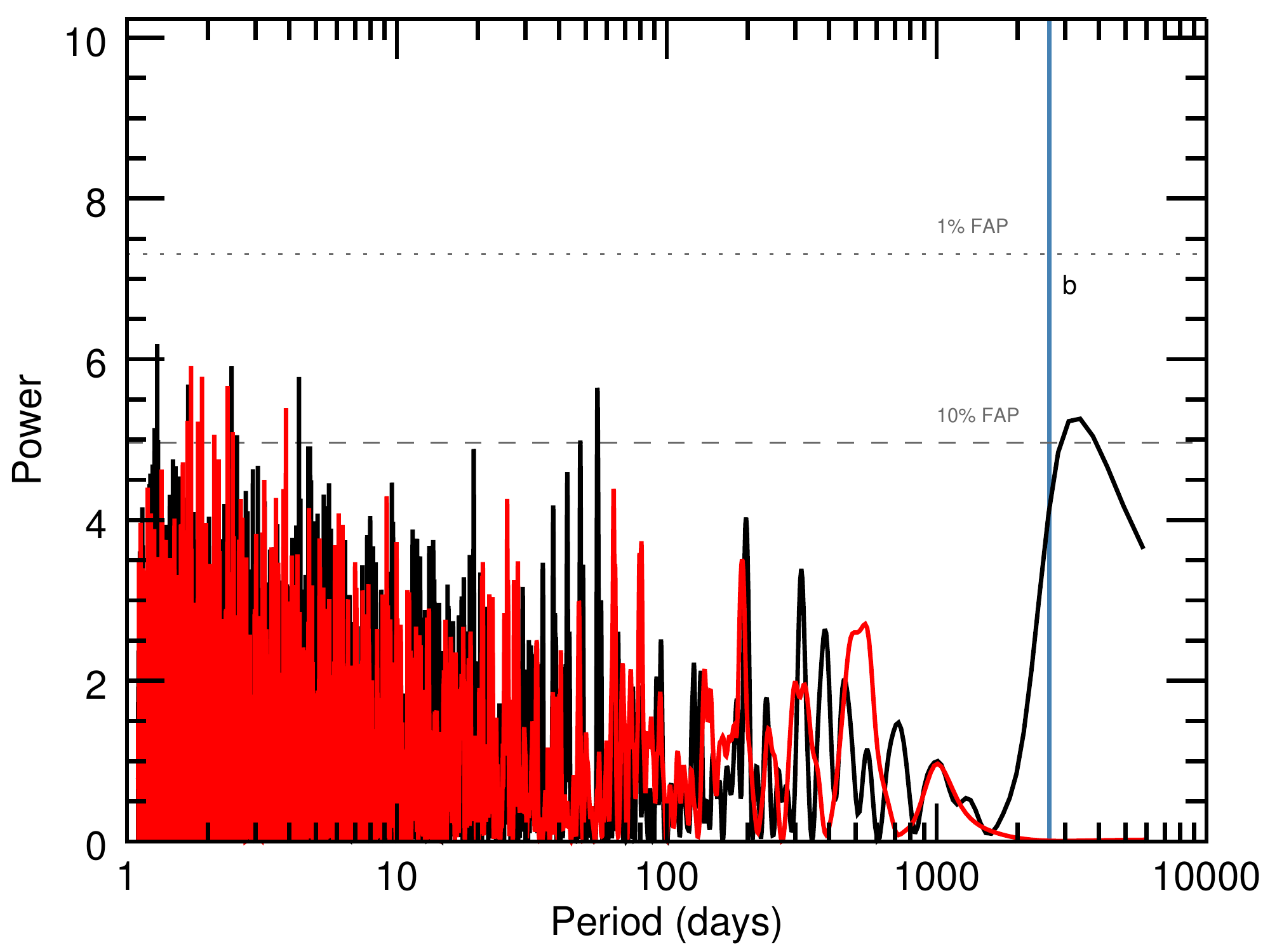}
\caption{(\emph{Left}) Time series and best-fit orbital solution for HD 13167~b, with period P~$= 2613$ days, eccentricity  $e=0.563$, and minimum mass m$_{p}\sin{i} =3.31$~M$_{\mathrm{Jup}}$. The residuals are shown in the bottom panel, which have an RV RMS $ =4.0$~m$\,$s$^{-1}$. The remaining best-fit parameters can be found in \autoref{tbl:orbital_params}. (\emph{Right}) Periodogram of HD 13167 RV data before (black) and after (red) subtracting the best-fit planet parameters for HD 13167~b. The vertical line indicates the best-fit period of HD 13167~b. The peak in the periodogram is only barely visible because the time baseline of observations is only slightly longer than 1 period.}
\label{fig:13167_time_series}
\end{figure*}

HD 13167 is a G3 subgiant star with V=8.34 and B-V=0.65 \citep{SIMBAD}. It has a mass of M$_{\star} = 1.35$~M$_{\mathrm{\odot}}$, surface gravity $\logg = 3.72$, and $\Teff = 5671$~K. A summary of its stellar parameters can be found in \autoref{tbl:stellar_params}. Observations of this star span about 8 years starting in late 2007. The best-fit Keplerian solution yields an orbital period of $2613\pm17$ days, and velocity semi-amplitude $48.2\pm2.7$ m/s on a fairly eccentric orbit (e$=0.563\pm0.033$). It has a minimum mass of $3.31\pm0.16$ M$_{\mathrm{Jup}}$. The full set of orbital parameters can be found in \autoref{tbl:orbital_params}. We show in \autoref{fig:13167_time_series} the time series for HD 13167~b as well as the periodogram before and after subtracting out HD 13167~b. The final observation in the time series (\autoref{fig:13167_time_series}) is what secures this detection, as we have now observed it reaching a second maximum. 

\subsection{A Jupiter orbiting HD 18015}
\begin{figure*}
\includegraphics[width=\columnwidth]{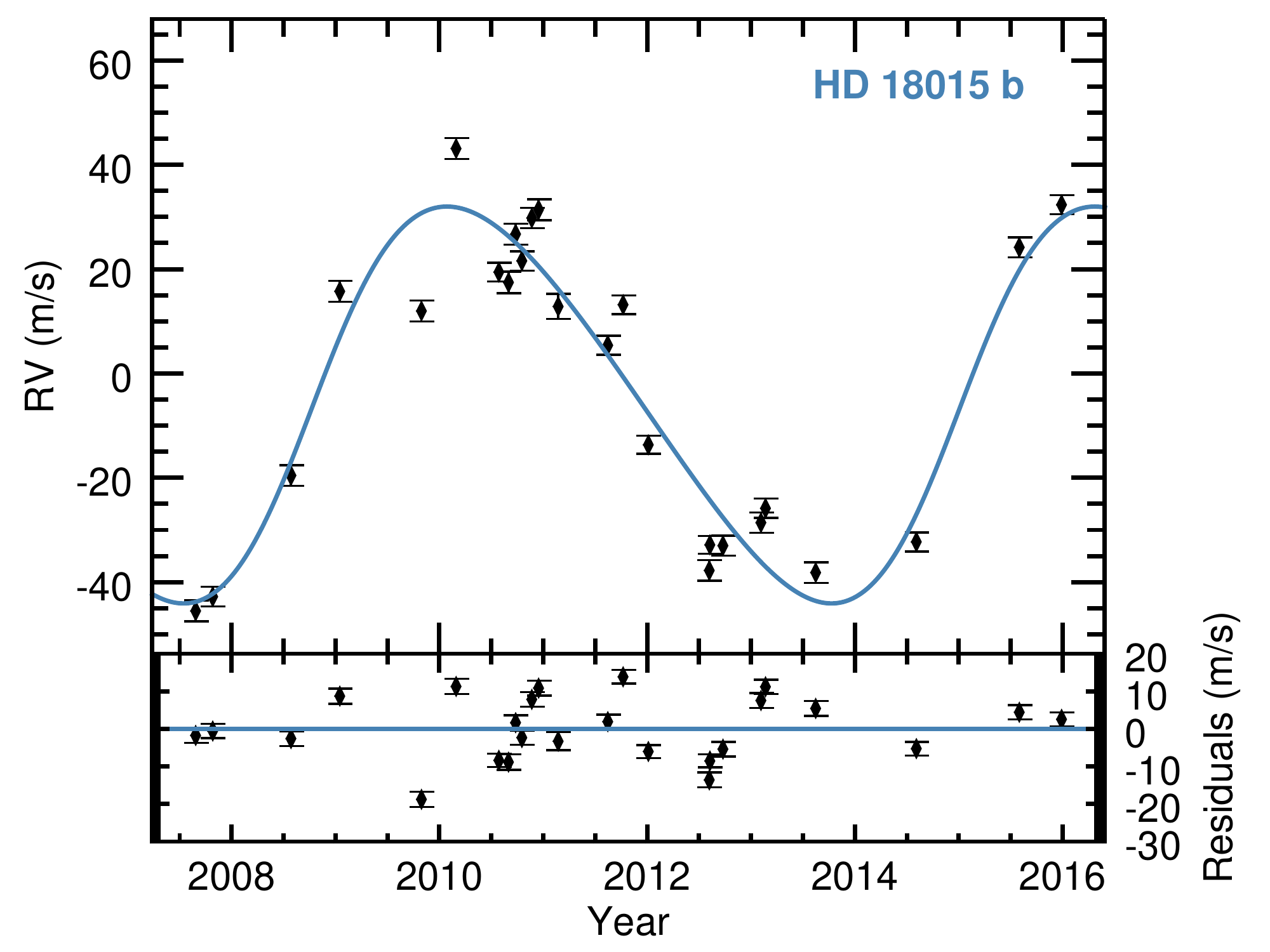}
\includegraphics[width=\columnwidth]{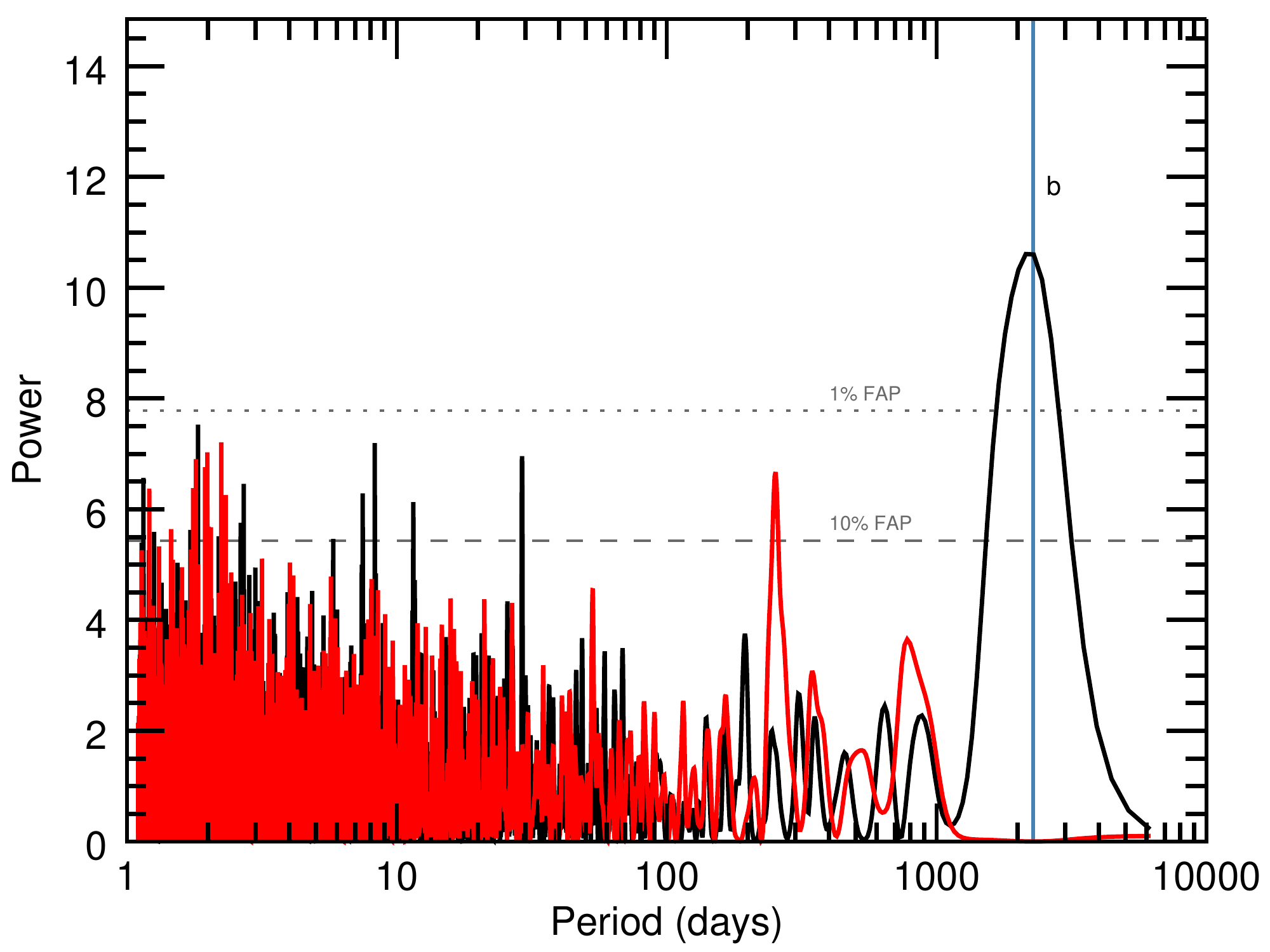}
\caption{(\emph{Left}) Time series and best-fit orbital solution for HD 18015~b, which has period P~$= 2278$ days, eccentricity  $e=0.148$, and minimum mass m$_{p}\sin{i} =3.18$~M$_{\mathrm{Jup}}$. The residuals are shown in the bottom panel, which have an RV RMS $ =8.4$~m$\,$s$^{-1}$. The remaining best-fit parameters can be found in \autoref{tbl:orbital_params}. (\emph{Right}) Periodogram of HD 18015 RV data before (black) and after (red) subtracting the best-fit planet parameters for HD 18015~b. The vertical line indicates the best-fit period of HD 18015~b.}
\label{fig:18015_time_series}
\end{figure*}

HD 18015 is a G6 subgiant star with V=7.89 and B-V=0.68 \citep{SIMBAD}. It has a mass of M$_{\star} = 1.49$~M$_{\mathrm{\odot}}$, surface gravity $\logg = 3.64$, and $\Teff = 5603$~K. A summary of its stellar parameters can be found in \autoref{tbl:stellar_params}. Observations of this star span about 8 years starting in late 2007. The best-fit Keplerian solution yields an orbital period of $2278\pm71$ days, and velocity semi-amplitude $38.0\pm2.7$ m/s, and  eccentricity $0.148\pm0.061$. It has a minimum mass of $3.18\pm0.23$~M$_{\mathrm{Jup}}$. The full set of orbital parameters can be found in \autoref{tbl:orbital_params}. We show in \autoref{fig:18015_time_series} the time series of HD 18015~b as well as the periodogram before and after subtracting out HD 18015~b.

\subsection{A Jupiter orbiting HD 180053}
\begin{figure*}
\centering
\includegraphics[width=\columnwidth]{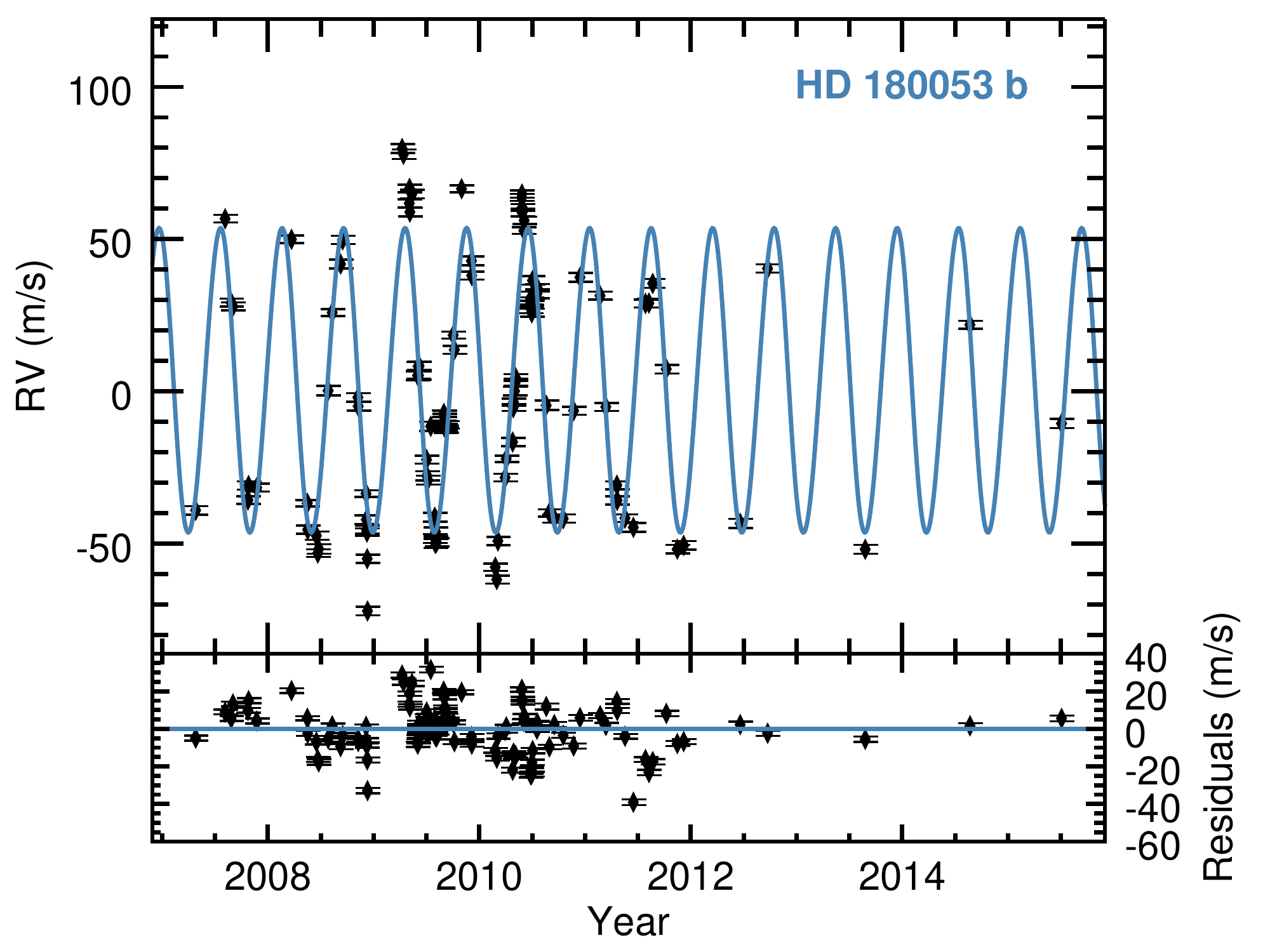}
\includegraphics[width=\columnwidth]{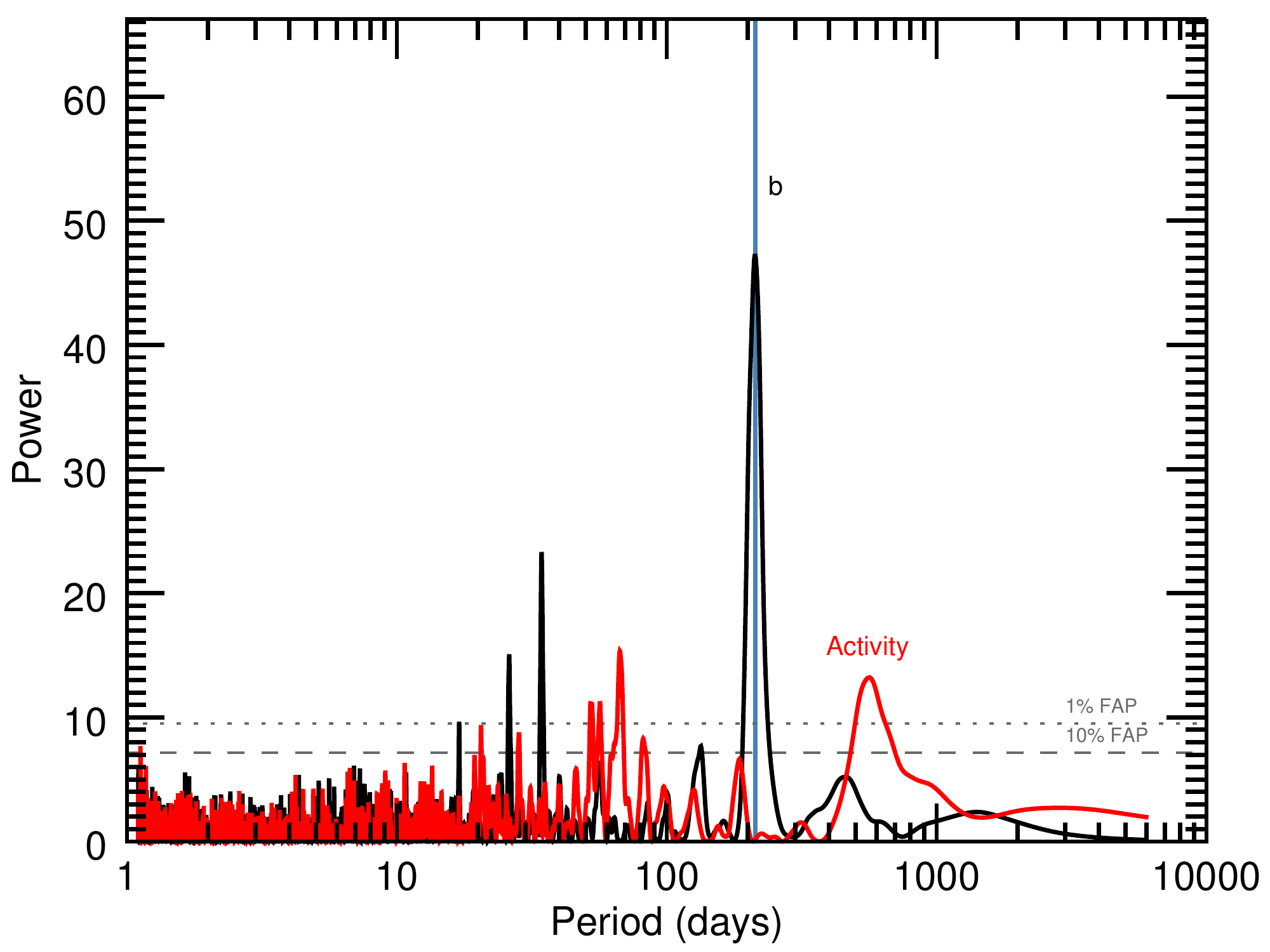}
\caption{(\emph{Left}) Time series of the best-fit orbital solution for HD 180053~b, which has period P~$= 213.72$ days, eccentricity  $e=0.081$, and minimum mass m$_{p}\sin{i} =2.194$~M$_{\mathrm{Jup}}$. The residuals are shown in the bottom panel, which have an RV RMS $ =13.8$~m$\,$s$^{-1}$, likely due to intrinsic variability induced by stellar activity. The remaining best-fit parameters can be found in \autoref{tbl:orbital_params}. (\emph{Right})  Periodogram of HD 180053 RV data before (black) and after (red) subtracting the best-fit planet parameters for HD 180053~b. The vertical line indicates the best-fit period of HD 180053~b. The remaining peaks in the periodogram are likely due to RV variations induced by stellar activity.}
\label{fig:180053_time_series}
\end{figure*}

\begin{figure}
\includegraphics[width=\columnwidth]{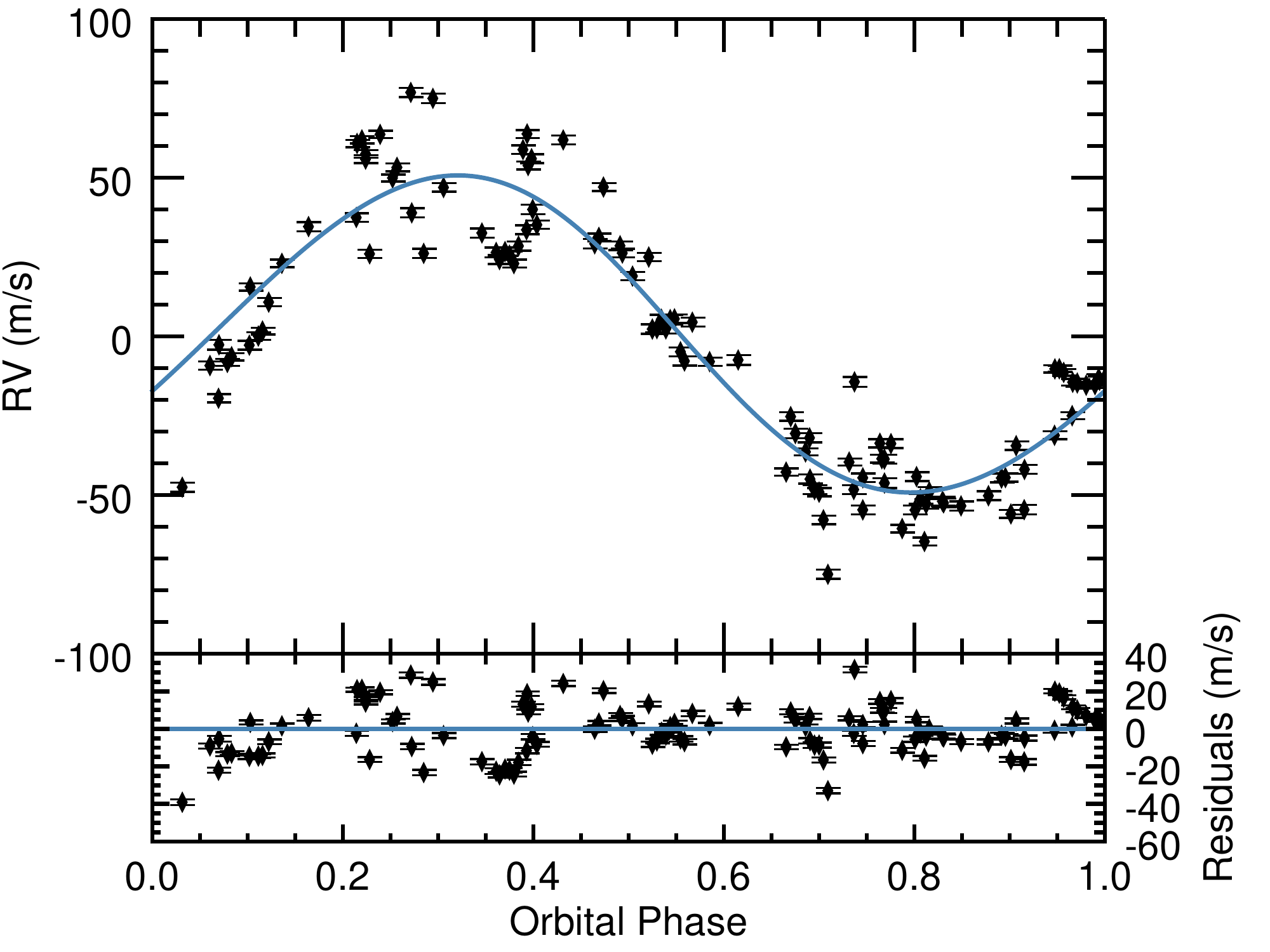}
\caption{Phase-folded velocities of HD 180053~b.}
\label{fig:180053_phase_curve}
\end{figure}

HD 180053 is a K0 subgiant star with V=7.93 and B-V=0.92 \citep{SIMBAD}. It has a mass of M$_{\star} = 1.75$~M$_{\mathrm{\odot}}$, surface gravity $\logg = 3.54$, and $\Teff = 5131$~K. A summary of its stellar parameters can be found in \autoref{tbl:stellar_params}. Observations of this star span about 8 years starting in 2007.  HD 180053 was identified in \citet{Butler2017} as having a planet candidate. The best-fit Keplerian solution yields an orbital period of $213.72\pm0.47$ days, and velocity semi-amplitude $51.5\pm1.4$ m/s, and  eccentricity $0.081\pm0.029$. It has a minimum mass of $2.194\pm0.063$ M$_{\mathrm{Jup}}$. The full set of orbital parameters can be found in \autoref{tbl:orbital_params}. We show in \autoref{fig:180053_time_series} the time series for HD 180053~b as well as the periodogram before and after subtracting out HD 180053~b. We also include in \autoref{fig:180053_phase_curve} the phase-folded RV curve to help show the planet signal more clearly. We note that both the phase curve and final periodogram show evidence of additional RV variations beyond simply 1 planet. Once we remove the best single-planet fit, we find a correlation between the radial velocities and the chromospheric activity as measured by the \ion{Ca}{2} H \&K lines using the Mount Wilson s-index, S$_{\mathrm{HK}}$, measured following the same procedure as in \citet{Isaacson2010}. Given that we obtain poor two- and three-planet fits, we expect that the additional RV variations are simply activity-induced, with timescales near 70 and 600 days, reasonable timescales for modulation from stellar rotation and stellar activity cycles.

\subsection{A Jupiter orbiting HD 4917 }
\begin{figure*}
\centering
\includegraphics[width=\columnwidth]{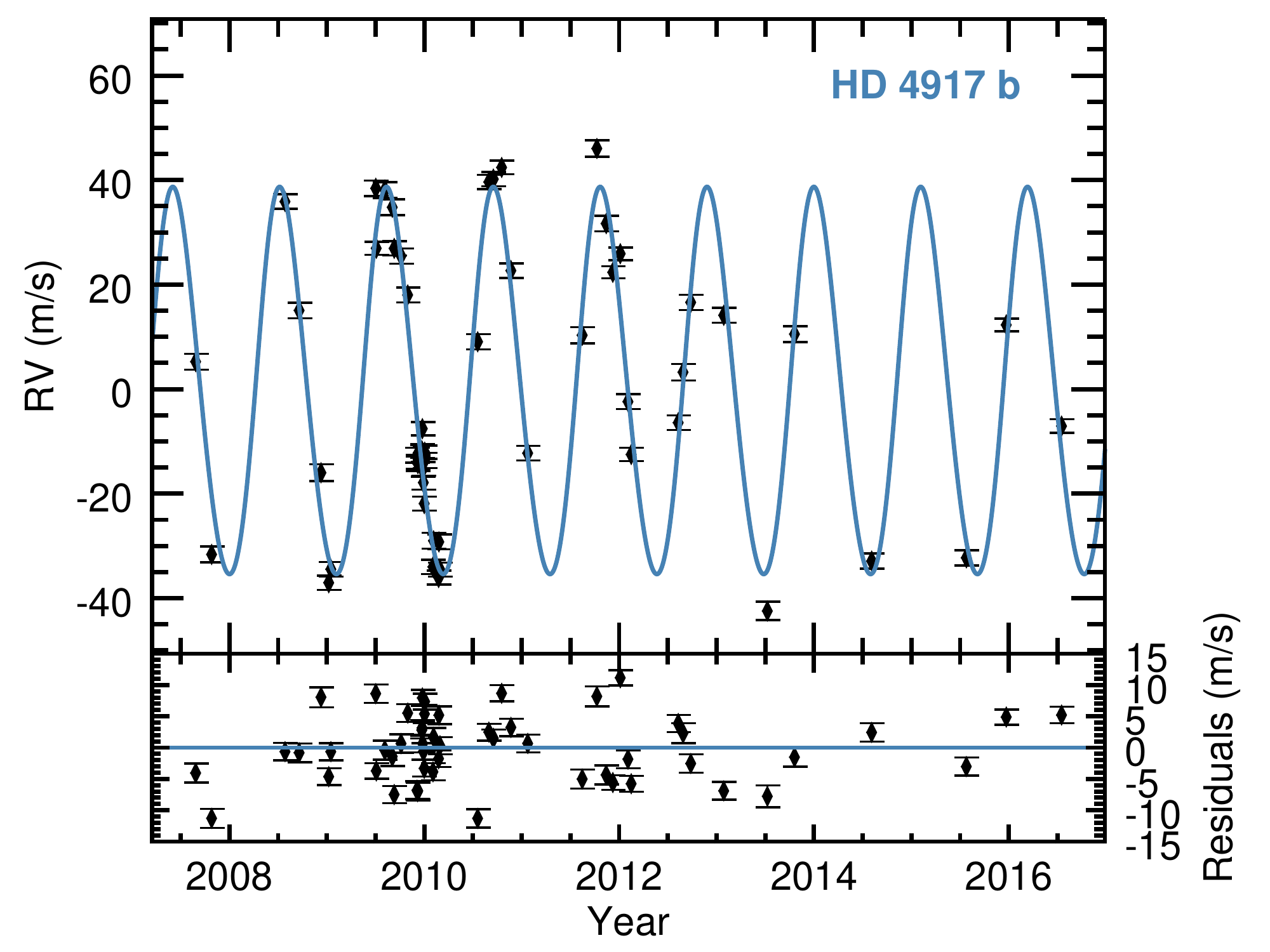}
\includegraphics[width=\columnwidth]{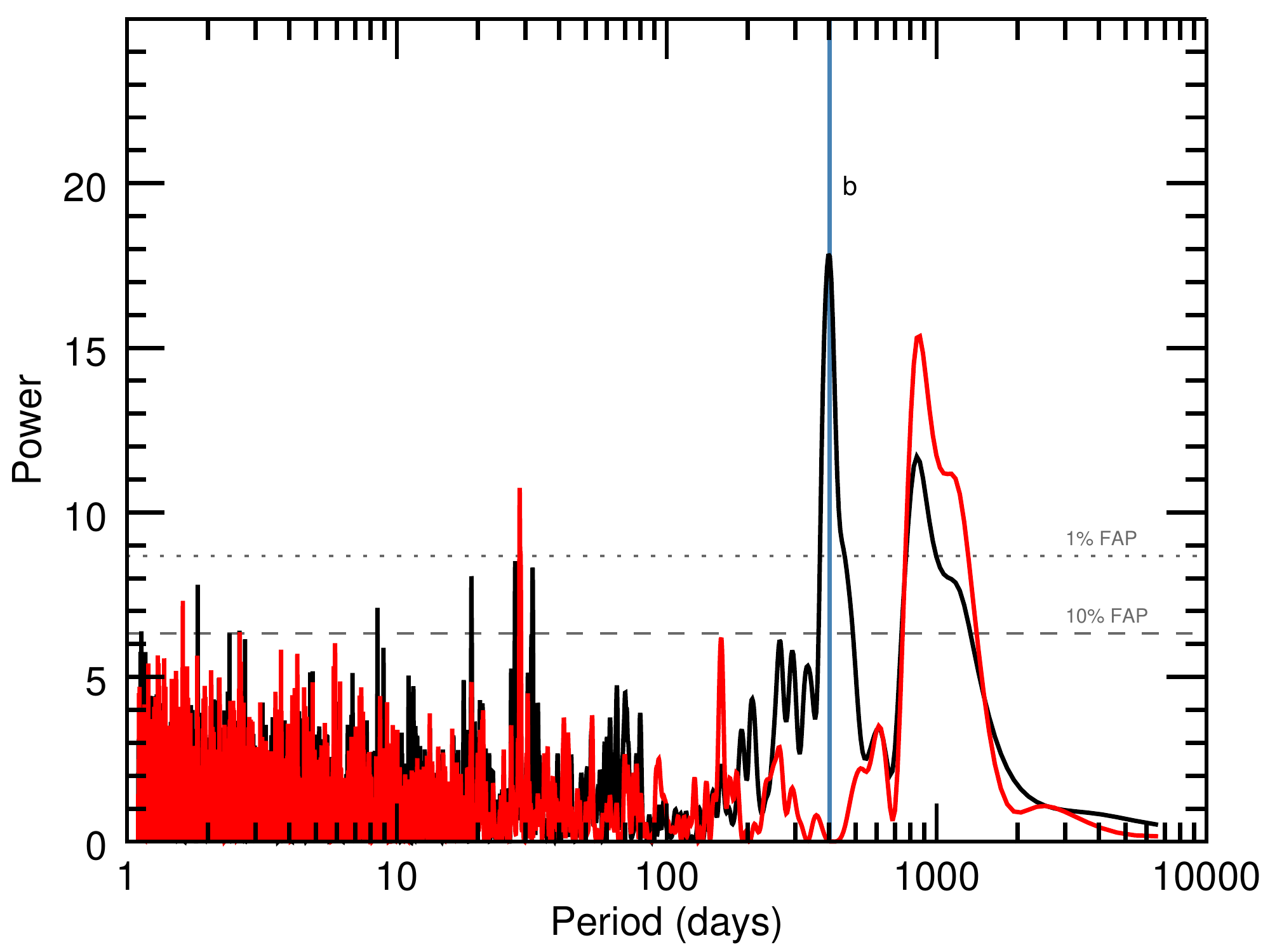}
\caption{ (\emph{Left}) The isolated time series of HD 4917~b (period $P= 400.5\pm1.7$ days, eccentricity $e= 0.066\pm0.041$, and minimum mass m$_{p}\sin{i} =1.615\pm0.093$~M$_{\mathrm{Jup}}$). In this panel the signal from planet candidates c (period $P= 821\pm13$ days, eccentricity $e= 0.467\pm0.088$, and minimum mass m$_{p}\sin{i} =1.37\pm0.13$~M$_{\mathrm{Jup}}$)\& d (period $P= 1093\pm37$ days, eccentricity $e= 0.28\pm0.14$, and minimum mass m$_{p}\sin{i} =0.89\pm0.1$~M$_{\mathrm{Jup}}$) have been subtracted out. (\emph{Right}) Initial (black) and final (red) periodogram of HD 4917 after subtracting out the signal from planet b.  The final periodogram shows the two peaks corresponding to planet candidates c \& d.}
\label{fig:4917_time_series}
\end{figure*}

HD 4917 is a K0 subgiant \citep{SIMBAD}. It has a temperature of $\Teff = 4802$~K and mass 1.32~M$_{\odot}$. Additional stellar properties can be found in \autoref{tbl:stellar_params}. There are nearly 50 observations of this star that span close to 10 years. Here we present the discovery of a m$_{p}\sin{i} =1.615\pm0.093$~M$_{\mathrm{Jup}}$ planet on a $400.5\pm1.7$ day orbit and its phase curve is shown in the left panel of \autoref{fig:4917_time_series}. The right panel shows the periodogram before and after subtracting out the best fit planet.


We note that the final periodogram of the residual shows two intriguing signals: one near 800 days and one near 1000 days. Our best 3-planet fit provided the best overall fit, however, we decline to call the outer two real planets for now because we have not done the requisite dynamical analysis to show that the orbits we derive for them are stable. Instead, we consider the outer two signals as planet candidates, which are discussed in \autoref{sec:planet_candidates}.

\section{8 Candidate Planetary Signals Around Subgiant Stars}\label{sec:planet_candidates}
The following signals are likely due to planetary companions. They are intriguing and likely correct, however they do not rise to the level of the others by not meeting one or more of our detection thresholds, as described in \autoref{sec:new_companions}. We include them here for the purposes of listing possible transits parameters and as long period planets, as it is uncertain whether more data is forthcoming. We encourage continued observations of these systems to fully confirm these planets. The reason for each planet candidate's status as a candidate is listed individually.

\subsection{Two possible additional planets around HD 4917}
\begin{figure*}
\centering
\includegraphics[width=0.275\paperwidth]{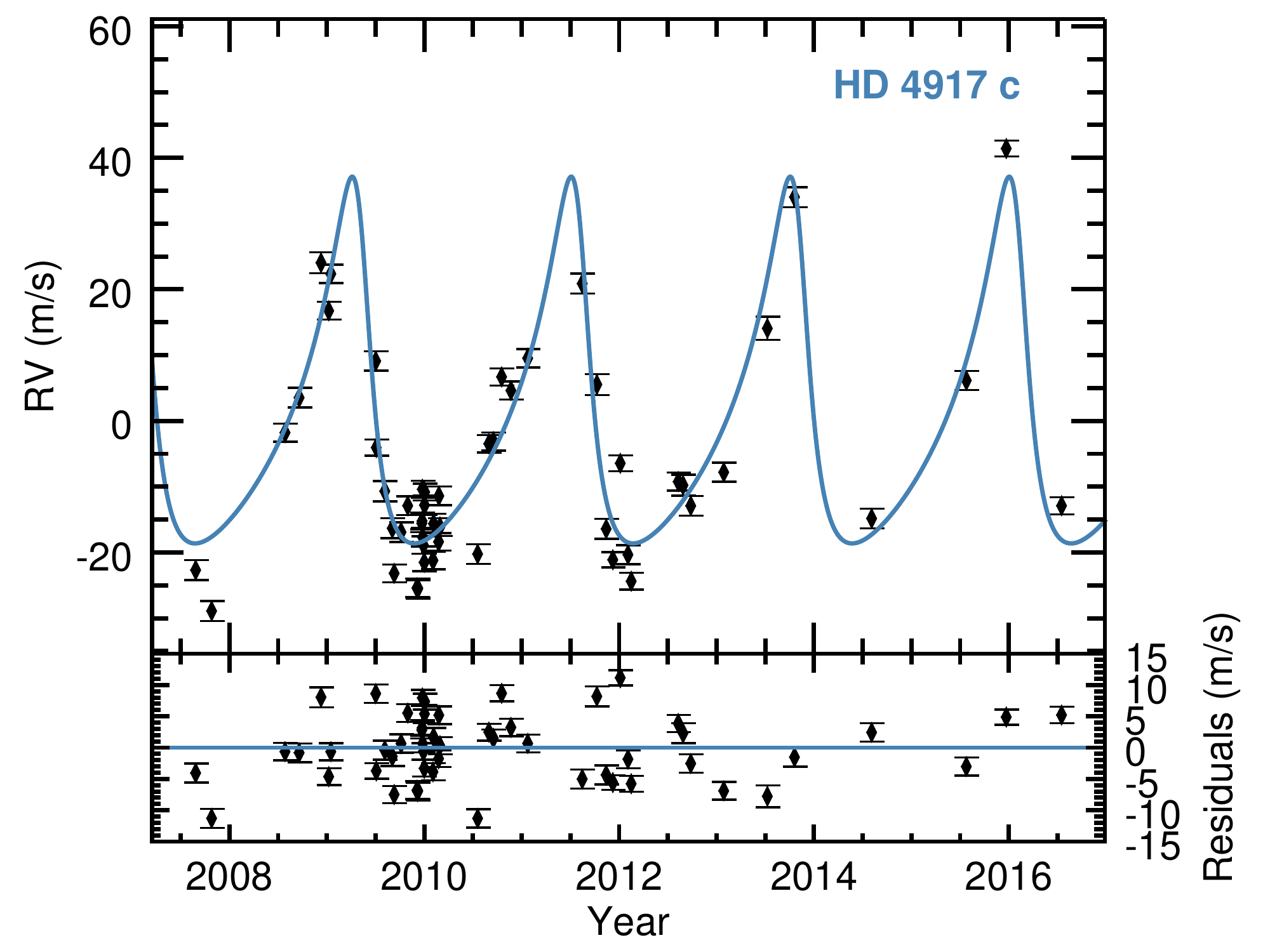}
\includegraphics[width=0.275\paperwidth]{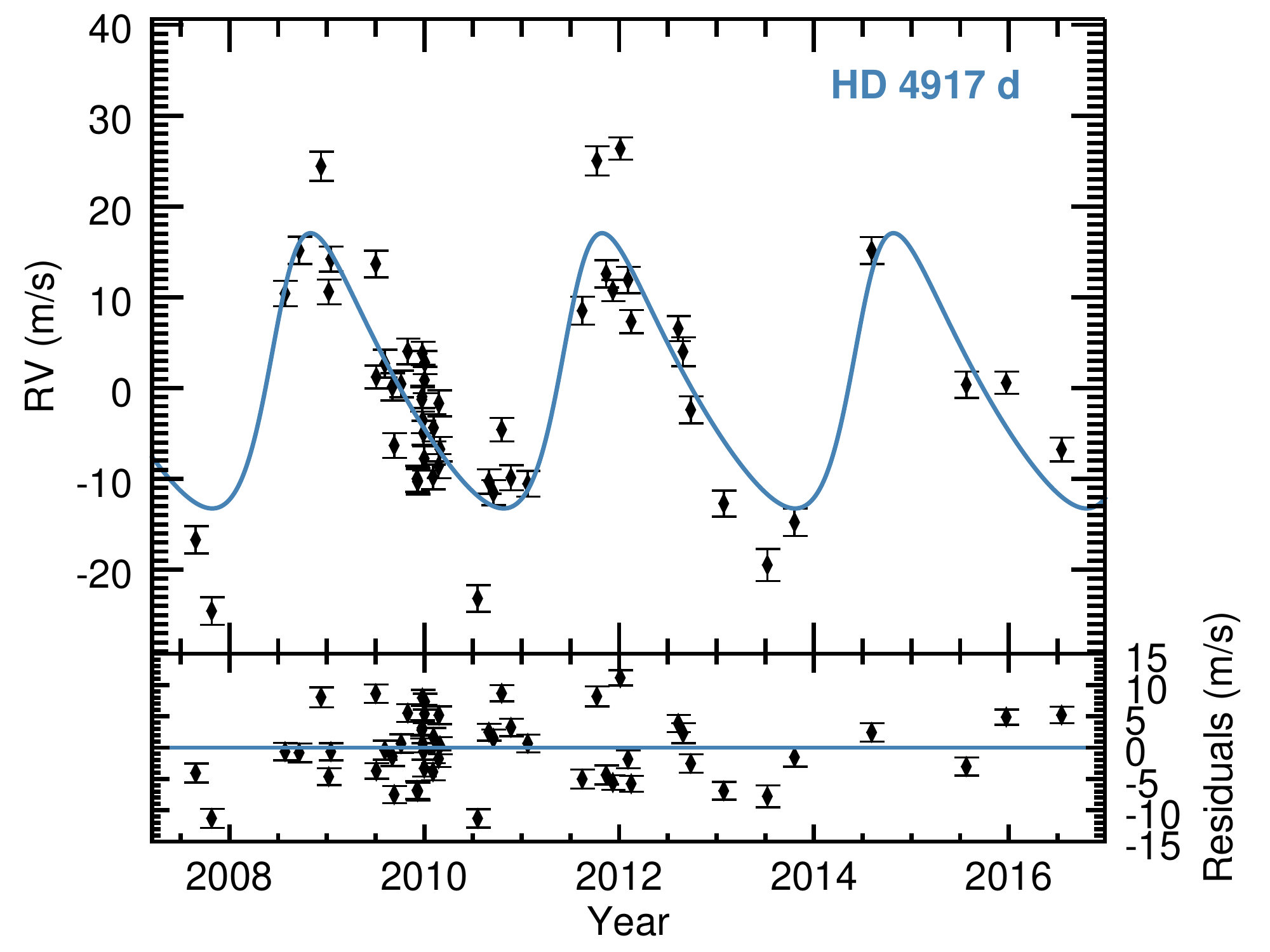}
\includegraphics[width=0.275\paperwidth]{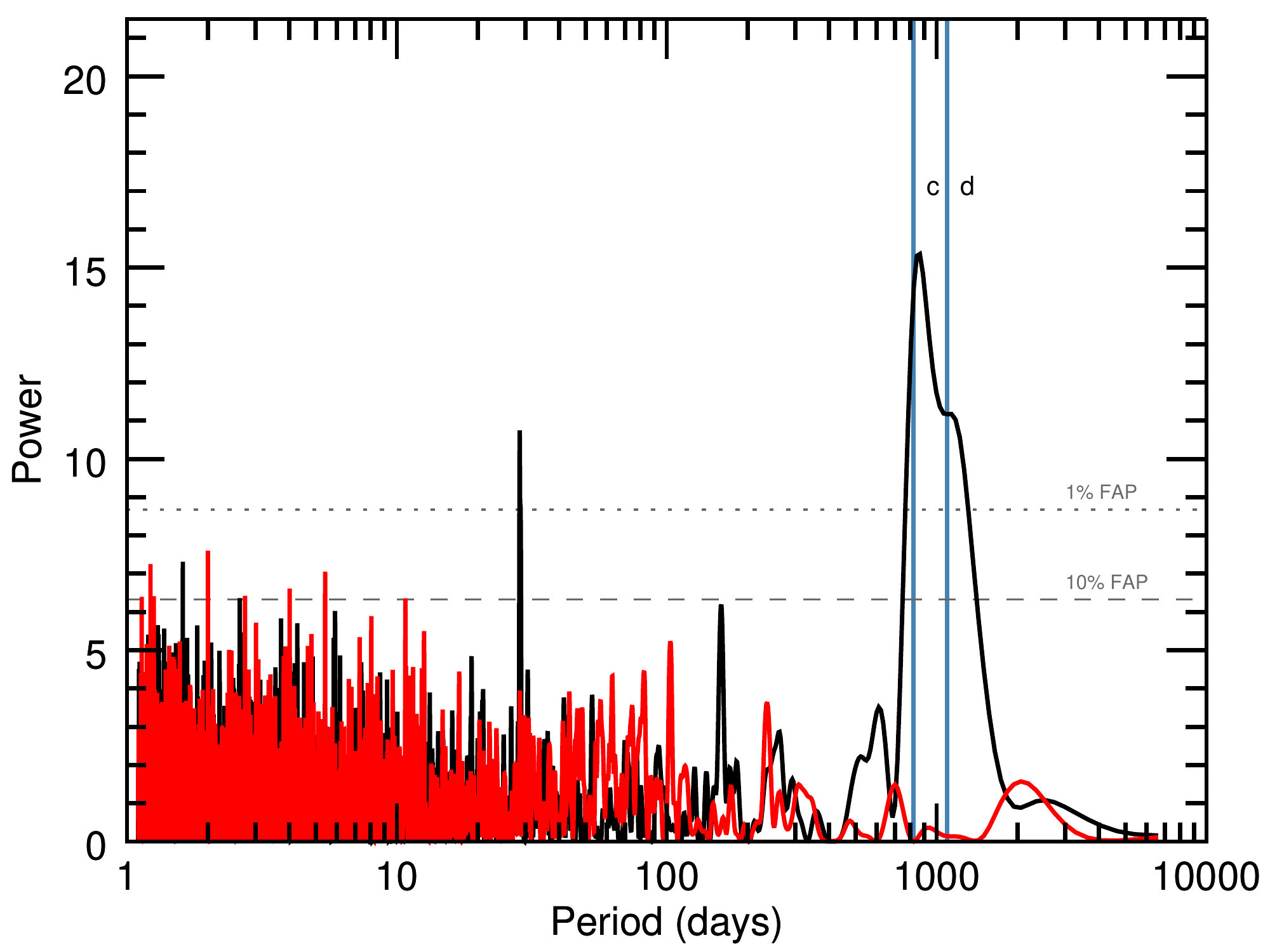}
\caption{Two additional planet candidates around HD 4917. (\emph{Left}) The isolated time series of HD 4917~c (period $P= 821\pm13$ days, eccentricity $e= 0.467\pm0.088$, and minimum mass m$_{p}\sin{i} =1.37\pm0.13$~M$_{\mathrm{Jup}}$). In this panel, the signals from planet b (period $P= 400.5\pm1.7$ days, eccentricity $e= 0.066\pm0.041$, and minimum mass m$_{p}\sin{i} =1.615\pm0.093$~M$_{\mathrm{Jup}}$) and planet candidate d (period $P= 1093\pm37$ days, eccentricity $e= 0.28\pm0.14$, and minimum mass m$_{p}\sin{i} =0.89\pm0.1$~M$_{\mathrm{Jup}}$) have been subtracted out. (\emph{Middle}) The isolated time series of HD 4917~d with signals from planet b and planet candidate c subtracted out. (\emph{Right}) Periodogram before (black) and after (red) subtracting out the two additional planet candidates HD 4917~c\&d. In this case, the initial periodogram is after subtracting out planet b, and is identical to the red periodogram in \autoref{fig:4917_time_series}.}
\label{fig:HD4917c_time_series}
\end{figure*}

The two additional signals around HD 4917, planets c \& d, described above are both convincing signals. HD 4917~c has an orbital period $P= 821\pm13$ days, eccentricity $e= 0.467\pm0.088$, and minimum mass m$_{p}\sin{i} =1.37\pm0.13$~M$_{\mathrm{Jup}}$ and HD 4917~d has orbital period $P= 1093\pm37$ days, eccentricity $e= 0.28\pm0.14$, and minimum mass m$_{p}\sin{i} =0.89\pm0.1$~M$_{\mathrm{Jup}}$. The time series for each of these planets can be found in \autoref{fig:HD4917c_time_series} as well as a periodogram before and after subtracting out these two additional signals (HD 4917~b was discussed in \autoref{sec:new_companions}). However, given the proximity of these two additional signals, we feel that this system warrants additional dynamical investigations to ascertain the stability of this planetary system. That detailed investigation is beyond the scope of this work and we so we consider these two signals as planetary candidates, despite passing our planet thresholds.

\begin{figure}
\includegraphics[width=\columnwidth]{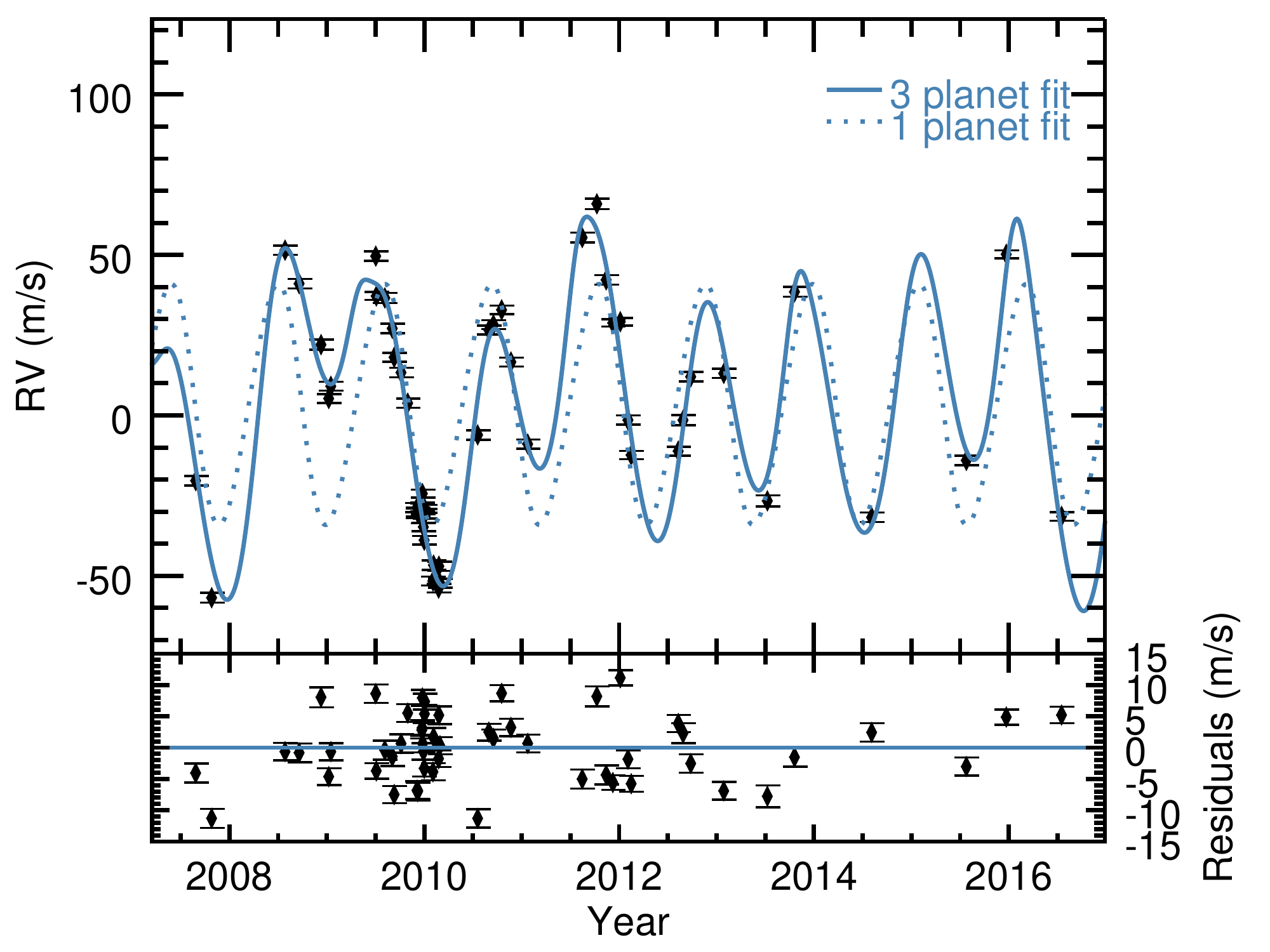}
\caption{Full time series for HD 4917 and best 3-planet fit compared to the best 1-planet fit. A single planet does not fit well, especially evidenced by the observations between mid-2008 and mid-2010. The bottom panel shows the residuals to the 3-planet fit.}
\label{fig:4917_full_time_series}
\end{figure}

\subsection{A possible second planet around HD 18742}
\begin{figure*}
\includegraphics[width=\columnwidth]{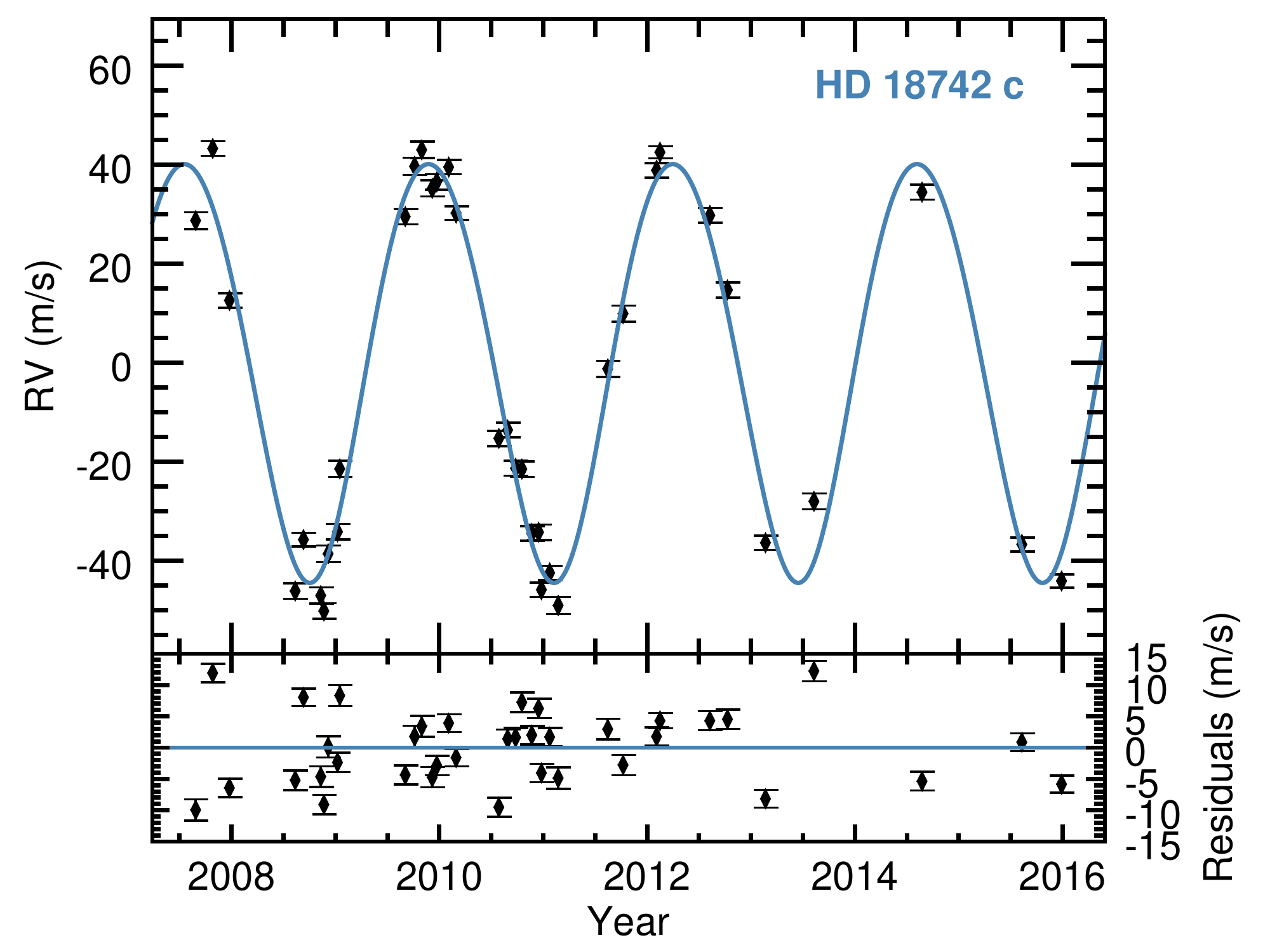}
\includegraphics[width=\columnwidth]{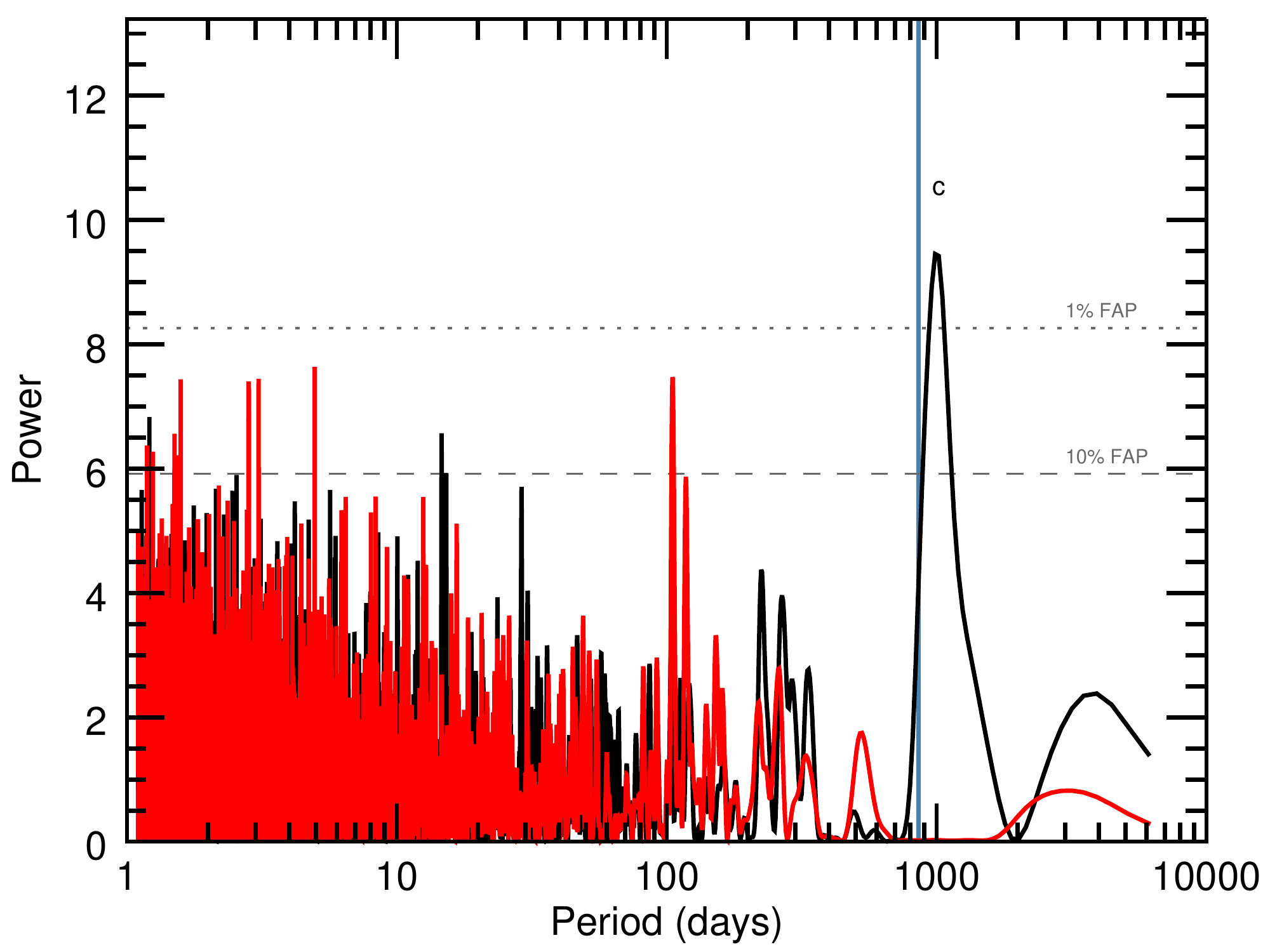}
\caption{Residuals after subtracting out the previously discovered planet HD 18742~b (period P~$=766$ days, eccentricity $e=0.040$, and minimum mass m$_{p}\sin{i} =3.4~$M$_{\mathrm{Jup}}$), showing evidence of a second planet.  (\emph{Left}) Time series for HD 18742~c with period $859$ days, eccentricity $e=0.056$, and minimum mass m$_{p}\sin{i}=2.4~$M$_{\mathrm{Jup}}$. (\emph{Right}) Periodogram after subtracting out the best fit orbital parameters for HD 18742~b (black) and after subtracting out the best 2-planet fit (black). Note the peak in the black curve at about 900 days, the starting guess used in the 2-planet fit. The vertical line of the best-fit planet period does not quite match the periodogram peak. This is likely due to fact that the vertical line comes from the best-fit period of a two-planet joint fit, rather than simply a fit to the residuals of the first planet which is shown in the periodogram. We note that the best 2-planet fit results in an unphysical 9:10 resonance, which is why we consider HD 18742~c a planet candidate.}
\label{fig:18742_time_series}
\end{figure*}

\begin{figure}
\includegraphics[width=\columnwidth]{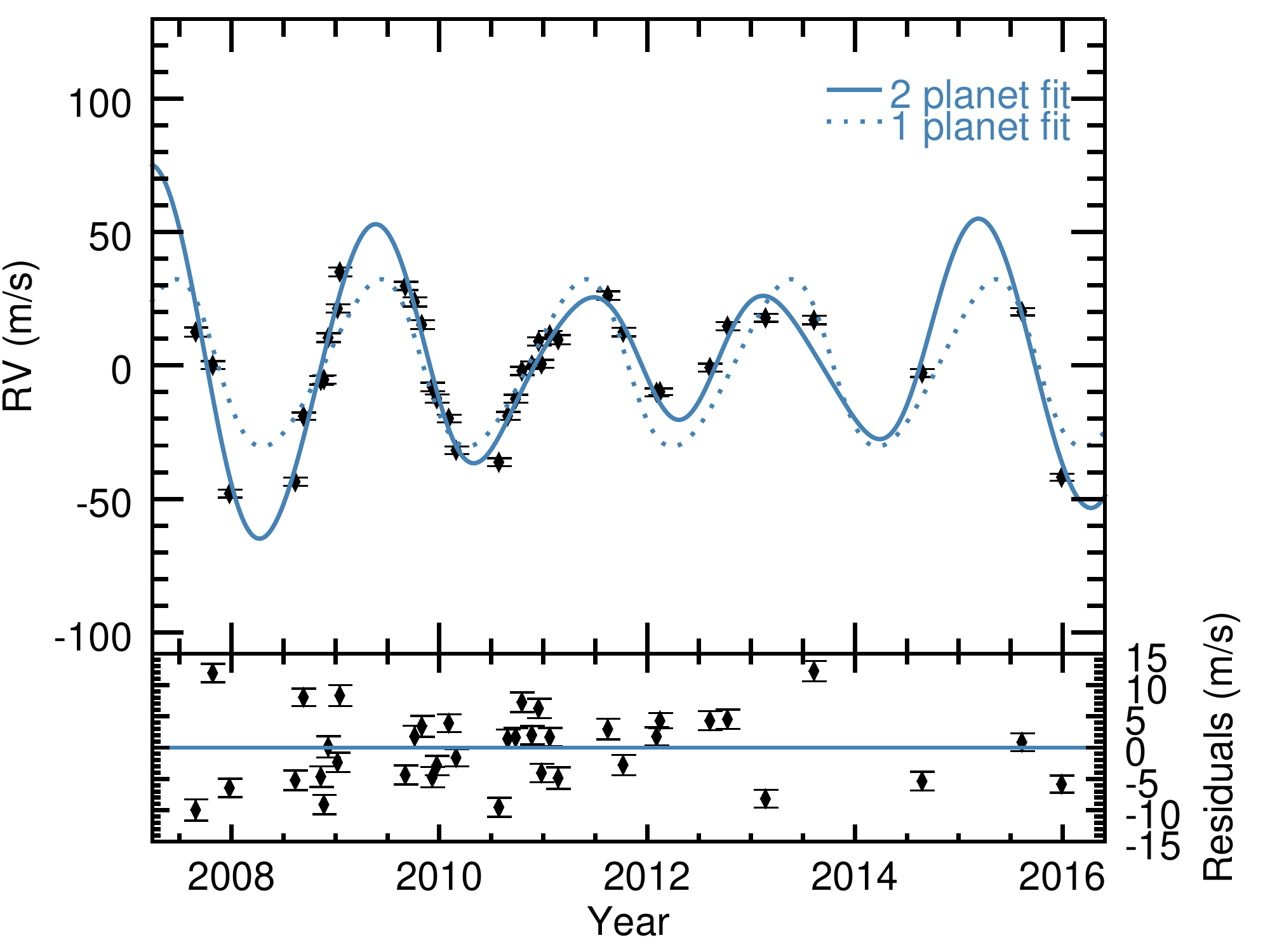}
\caption{Full time series for HD 18742 and best 2-planet fit. The dotted line shows for reference the best one-planet fit. The bottom panel shows the residuals to the 2-planet fit. We note that the best 2-planet fit results in an unphysical 9:10 resonance, which is why we consider HD 18742~c a planet candidate.}
\label{fig:18742_full_time_series}
\end{figure}

HD 18742 is a G8/K0 IV star \citep{SIMBAD} with 1 previously known planet \citep{Johnson2011}. It has a V magnitude V$ = 7.81$, effective temperature $\Teff = 4940$~K, and $\logg=3.09$. Its mass is 1.36~M$_{\odot}$ and it has a radius of 5.13~R$_{\odot}$. Additional stellar properties and uncertainties can be found in \autoref{tbl:stellar_params}. There are 37 RV observations for this star, all from Keck, which span more than 8 years (2007 to late 2015). Here we make use of the additional 11 points that span the last 4.5 years of the 8 years of observations. The original fit for this planet in \citet{Johnson2011} included a linear trend. Indeed, after refitting this planet and subtracting out the new best fit to HD 18742~b (period P~$=766\pm25$ days, eccentricity $e=0.040\pm0.035$, and minimum mass m$_{p}\sin{i} =3.4\pm1.2~$M$_{\mathrm{Jup}}$), there was a substantial peak in the periodogram near 900 days, shown in \autoref{fig:18742_time_series}. Refitting the RVs with a two-planet fit starting with a second planet near 900 days resulted in an improvement in the reduced $\chi^2$ from 78 to 20 and finds HD 18742~c to have a period of $859\pm41$ days, eccentricity $e=0.056\pm52$, and minimum mass m$_{p}\sin{i}=2.4\pm1.2~$M$_{\mathrm{Jup}}$. See \autoref{tbl:orbital_params} for more orbital and planet parameters and uncertainties. The time series for the second planet (HD 18742~c) is shown in the left panel of  \autoref{fig:18742_time_series} with the full time series and 2-planet fit shown in \autoref{fig:18742_full_time_series}. 

Despite providing a much better fit and meeting all of our planet detection thresholds, we have listed it as a candidate signal because the resulting planet would be in a 9:10 resonance with the inner planet, which is non-physical and would be the first set of planets in such a resonance. As a result, further scrutiny is needed in order to confirm or reject this planet. Given the location of the periodogram peak is closer to 1000 days than 860 days (see \autoref{fig:18742_time_series}), it could simply be that we have found a shallow minimum in the two-planet fit that results in the unphysical resonance. Additional observations should uncover the true nature of this signal. Particularly, the one-planet and two-planet models reach their largest divergence in mid 2018 (differ by 60 m/s) and mid 2019 (differ by nearly 90 m/s). Past observations mainly lie in regions where the two solutions differ by 10-20 m/s with a few points out to 30-40 m/s (all points more closely tracing the two-planet solution).

\subsection{A possible third planet around HD 163607}
\begin{figure*}
\includegraphics[width=\columnwidth]{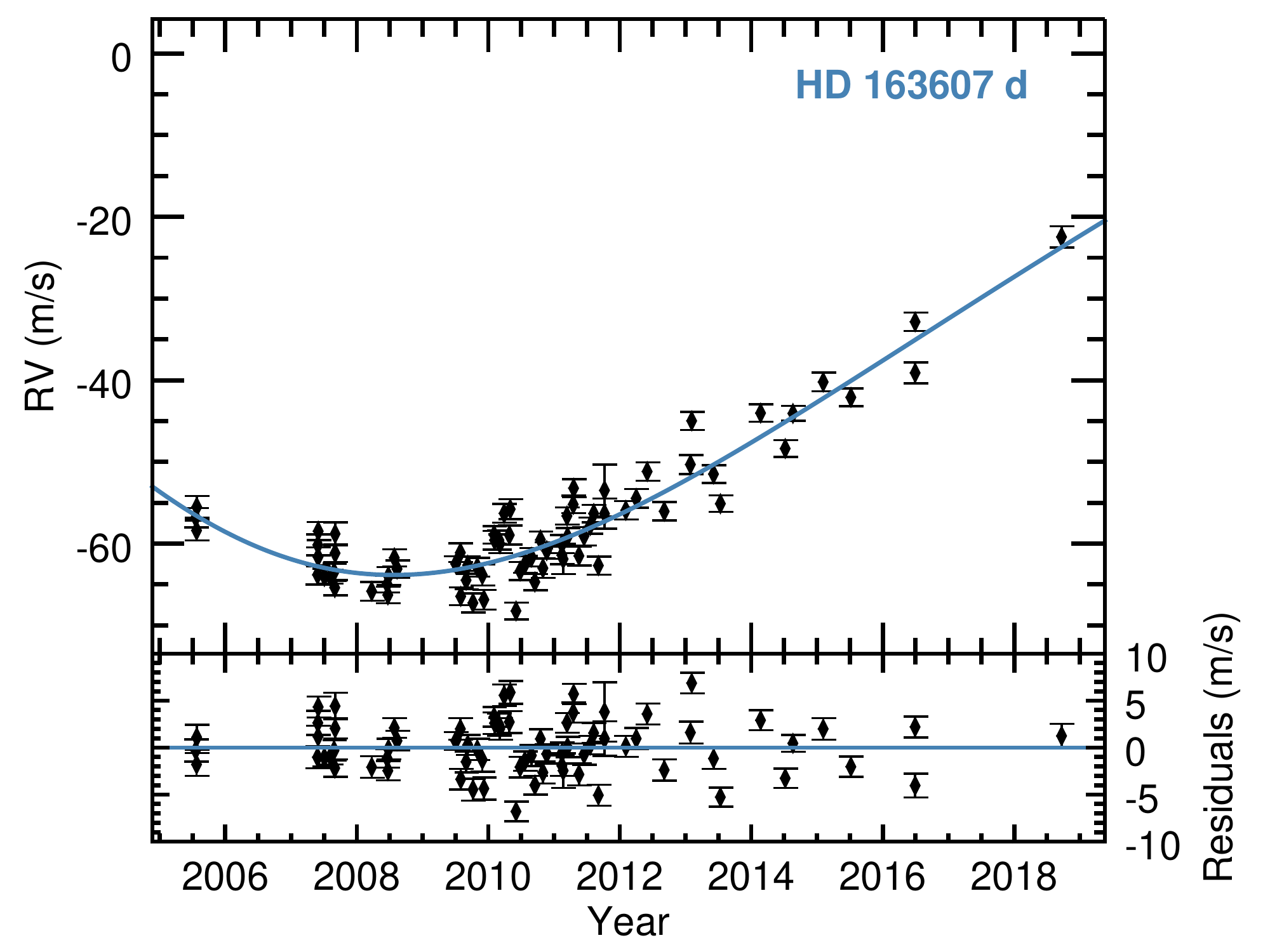}
\includegraphics[width=\columnwidth]{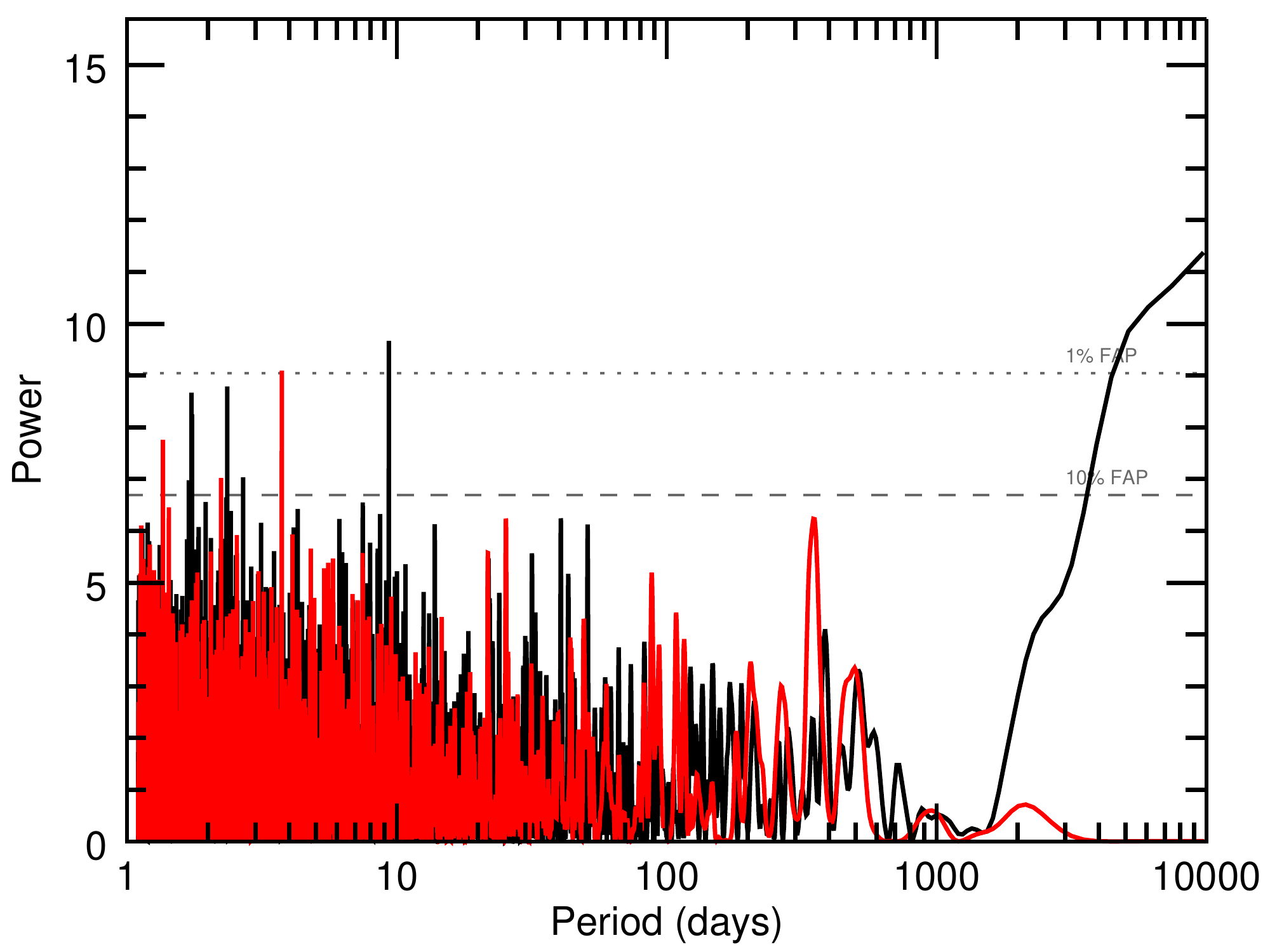}
\caption{Residuals after subtracting out the two previously discovered planets HD 163607~b (period P~$=75.199$ days, eccentricity $e=0.7502$, and minimum mass m$_{p}\sin{i} =0.787~$M$_{\mathrm{Jup}}$) and HD 163607~c (period P~$=1266.3$ days, eccentricity $e=0.075$, and minimum mass m$_{p}\sin{i} =2.193~$M$_{\mathrm{Jup}}$), showing evidence of a third planet.(\emph{Left}) Time series for HD 163607~d with period $22000\pm15000$ days, eccentricity $0.25\pm0.25$ and minimum mass 7.6 M$_{\mathrm{Jup}}$. Incomplete phase coverage means that the uncertainty in the period and eccentricity is relatively high. (\emph{Right}) Periodogram after subtracting out the best fit orbital parameters for HD 163607~b and c (black) and the final periodogram after subtracting out planet d in the best 3-planet fit (red). Planet d is seen as a wide peak at long periods, extending beyond the plot.}
\label{fig:163607_time_series}
\end{figure*}

\begin{figure}
\includegraphics[width=\columnwidth]{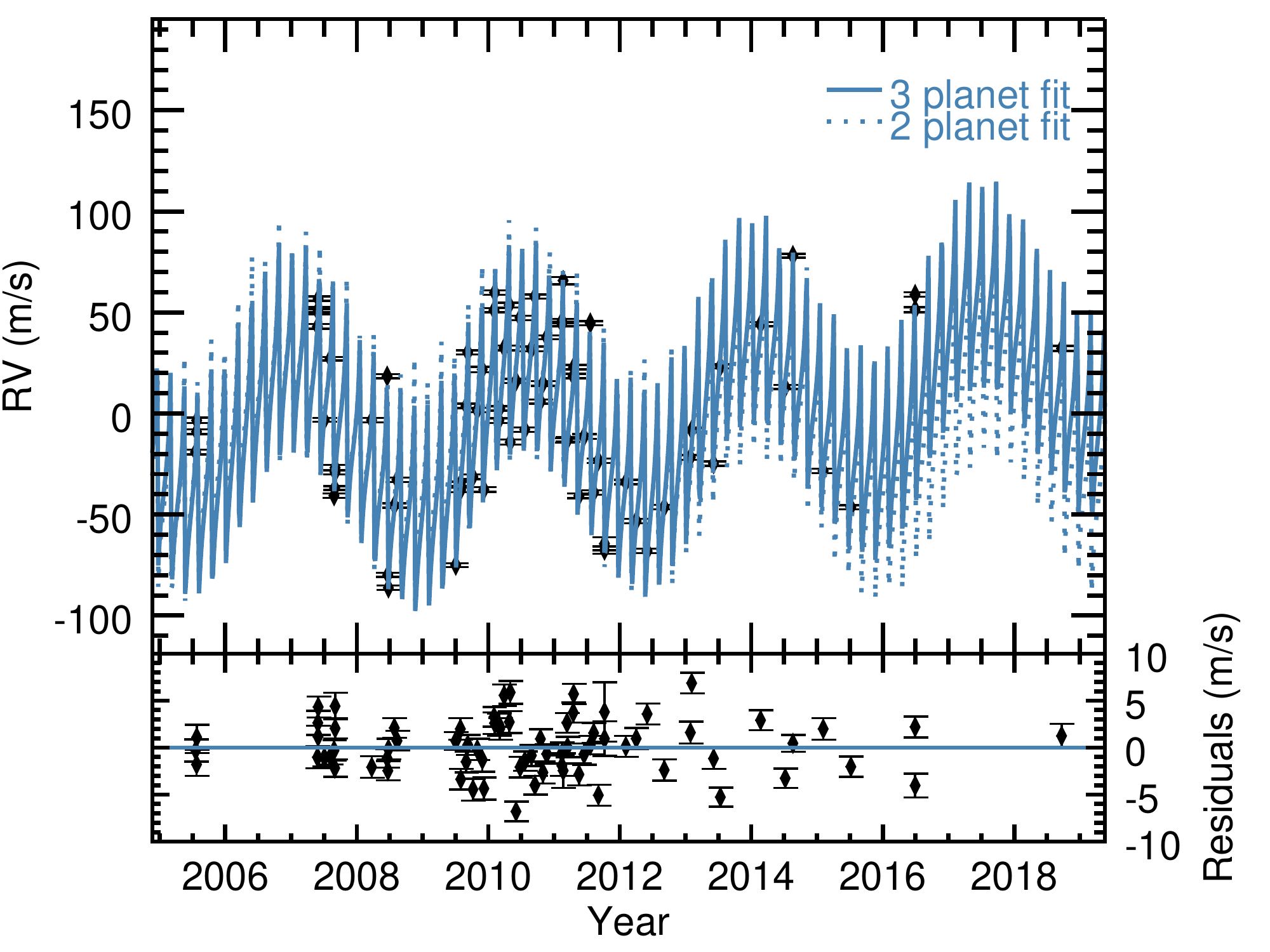}
\caption{Full time series for HD 163607 and best 3-planet fit. The new planet, HD 163607~d is the long period trend that reaches minimum around 2009 and maximum in 2016, evident by the differing minima of the ~1000 day planet). The difference between the two- and three-planet fit is not as easily seen in the time series. The bottom panel shows the residuals to the 3-planet fit.}
\label{fig:163607_full_time_series}
\end{figure}

HD 163607 is a G5 IV star \citep{SIMBAD} with 2 previously known planets \citep{Giguere2012}. It has a V magnitude V$ = 7.979$, effective temperature $\Teff = 5522$~K, and $\logg  = 3.97$. It is a 1.12~M$_{\odot}$ star with radius 1.76~R$_{\odot}$. For additional stellar parameters and uncertainties, see \autoref{tbl:stellar_params}. There are 73 RV observations of this star, the final 20 of which are additional points since the original discovery of HD 163607~b and c. All 73 RV observations come from Keck. After refitting the two planets and subtracting out the new best fits, there was a strong long period signal in the periodogram, shown in the right panel of \autoref{fig:163607_time_series}. Refitting the time series with the inclusion of a third, long-period planet resulted in an improvement in the reduced $\chi^2$ from 37 to 8. The time series for HD 163607~d is shown in \autoref{fig:163607_time_series} with the full time series for all 3 planets shown in \autoref{fig:163607_full_time_series}. As is obvious from the time series, we have not covered a full orbit for this planet, which is why it is a planet candidate. We note that due to the long period of the planet and incomplete phase coverage, the derived orbital parameters for planet d are fairly uncertain, which will be resolved with future observations. In particular, the addition of the most recent observation (September 2018) has shown that the outer planet is indeed very long period and massive. Future observations will continue to constrain the outer planet, which in the current best fit has a mass $7.6\pm4.3$~M$_{\mathrm{Jup}}$, period 22000 $\pm$ 15000 days, and eccentricity 0.25 $\pm$ 0.25. Additional orbital parameters can be found in \autoref{tbl:orbital_params}. If confirmed, this (or similarly HD 4917 above) will be the first 3-planet system around a subgiant, which could be useful in determining evolutionary properties of planets around subgiant stars. The novelty of the 3 planet subgiant system and probability of transit warrants its inclusion here despite lacking a fully constrained orbit.

\subsection{A possible short period planet orbiting HD 180902}\label{sec:hd180902}
\begin{figure*}
\centering
\includegraphics[width=\columnwidth]{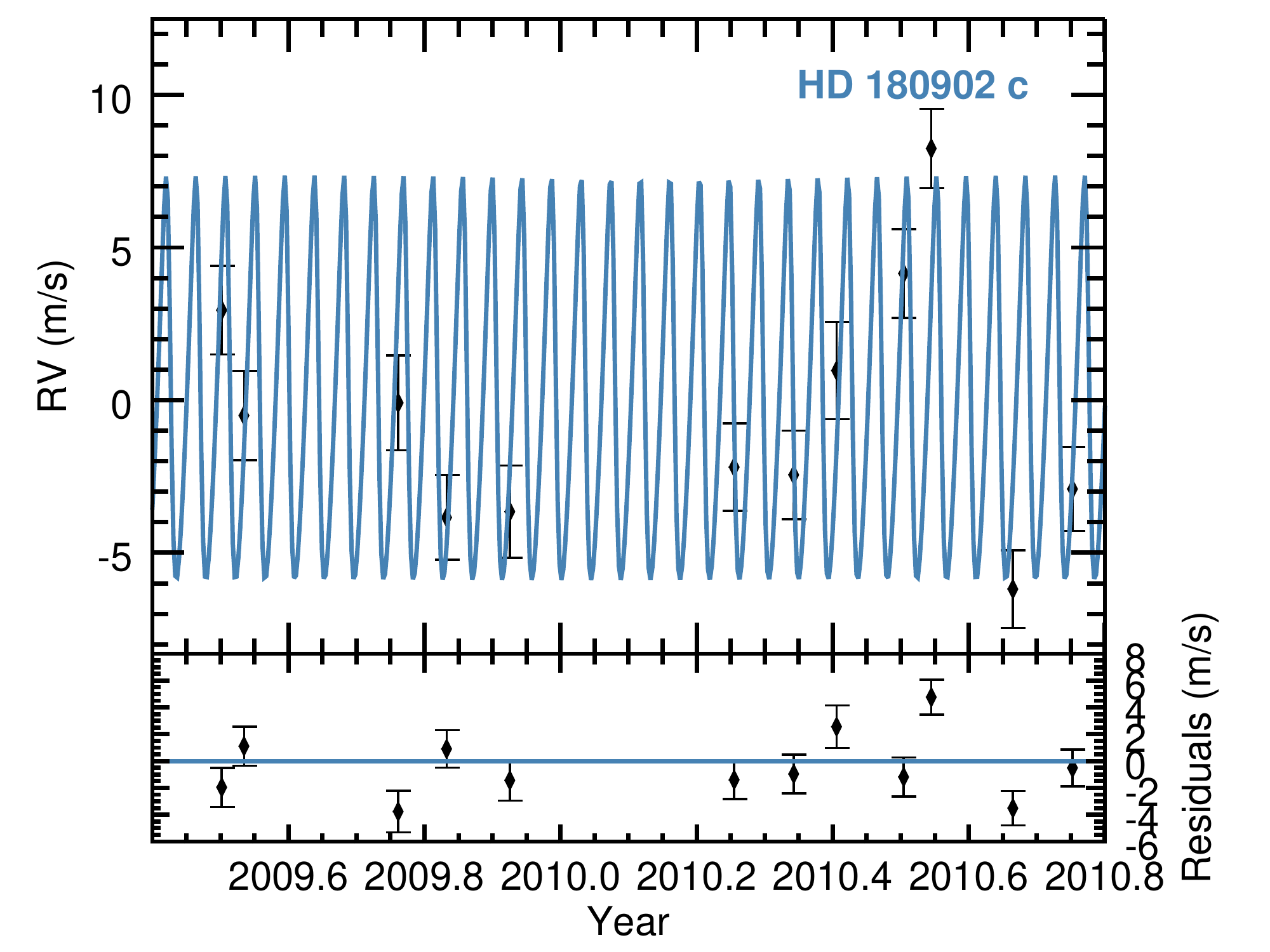}
\includegraphics[width=\columnwidth]{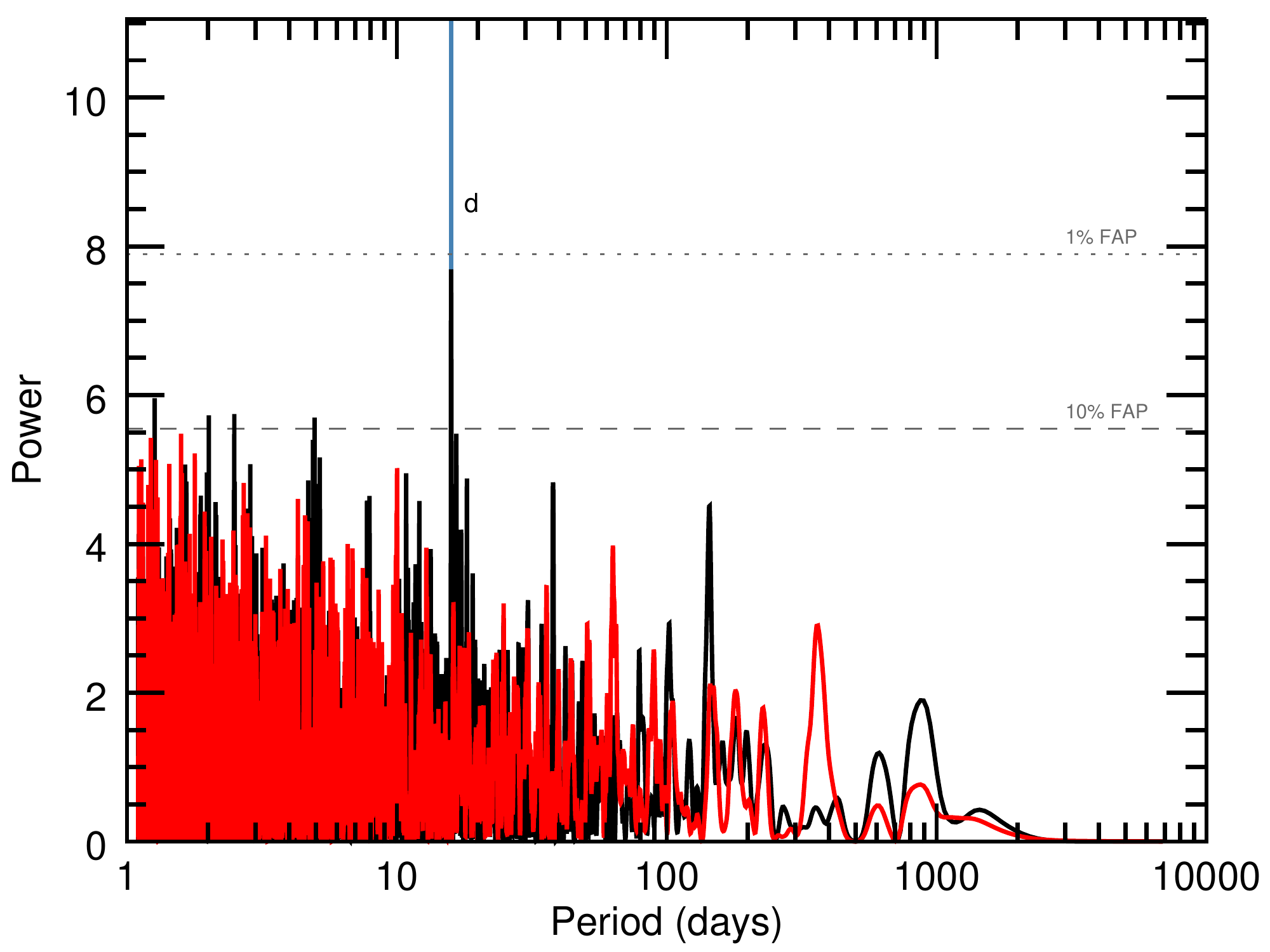}
\caption{Residuals after subtracting out the previously discovered planet HD 180902~b (period P~$=510.9$ days, eccentricity $e=0.107$, and minimum mass m$_{p}\sin{i} =1.685~$M$_{\mathrm{Jup}}$) and the best fit Keplerian for the stellar binary HD 180902~B (period P~$=5880$ days, eccentricity $e=0.107$, and minimum mass m$_{p}\sin{i} =98.7~$M$_{\mathrm{Jup}}$), showing evidence of a second planet. (\emph{Left}) Time series for HD 180902~c with period $15.9058$ days, eccentricity $e=0.28$, and minimum mass m$_{p}\sin{i} =0.099~$M$_{\mathrm{Jup}}$. Due to the short period, the time series on the left has been zoomed in on a high-density region of observations, but \autoref{fig:180902_B_time_series} shows the full set of observations for this star with the longer period stellar binary companion. The phase curve for this planet is shown in \autoref{fig:180902_c_phase_curve}. (\emph{Right}) Periodogram before (black) and after (red) subtracting the 15 day signal for HD 180902~c.}
\label{fig:180902_time_series}
\end{figure*}

\begin{figure}
\includegraphics[width=\columnwidth]{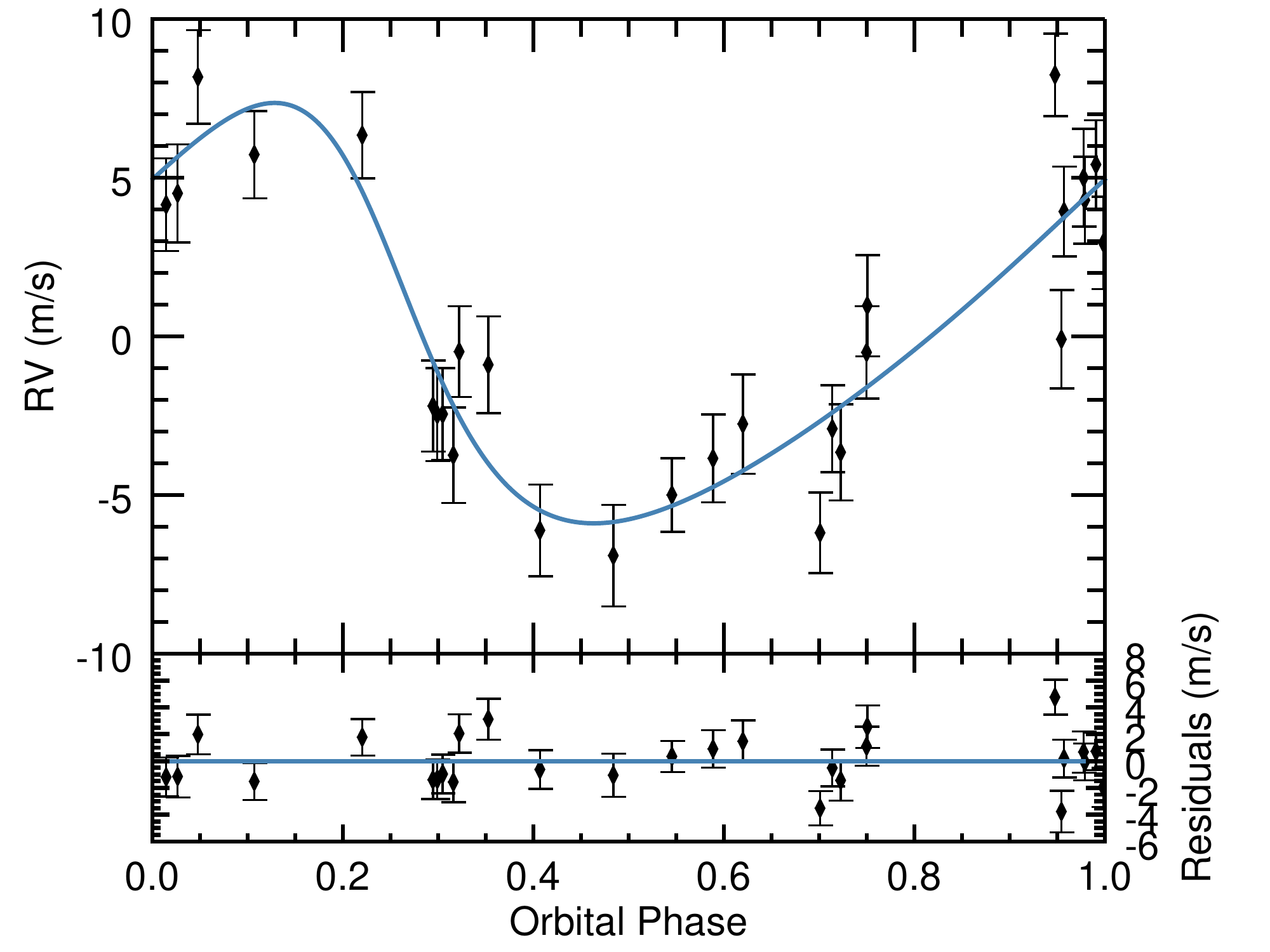}
\caption{The phase curve for HD 180902~c with period 15 days.}
\label{fig:180902_c_phase_curve}
\end{figure}

\begin{figure}
\includegraphics[width=\columnwidth]{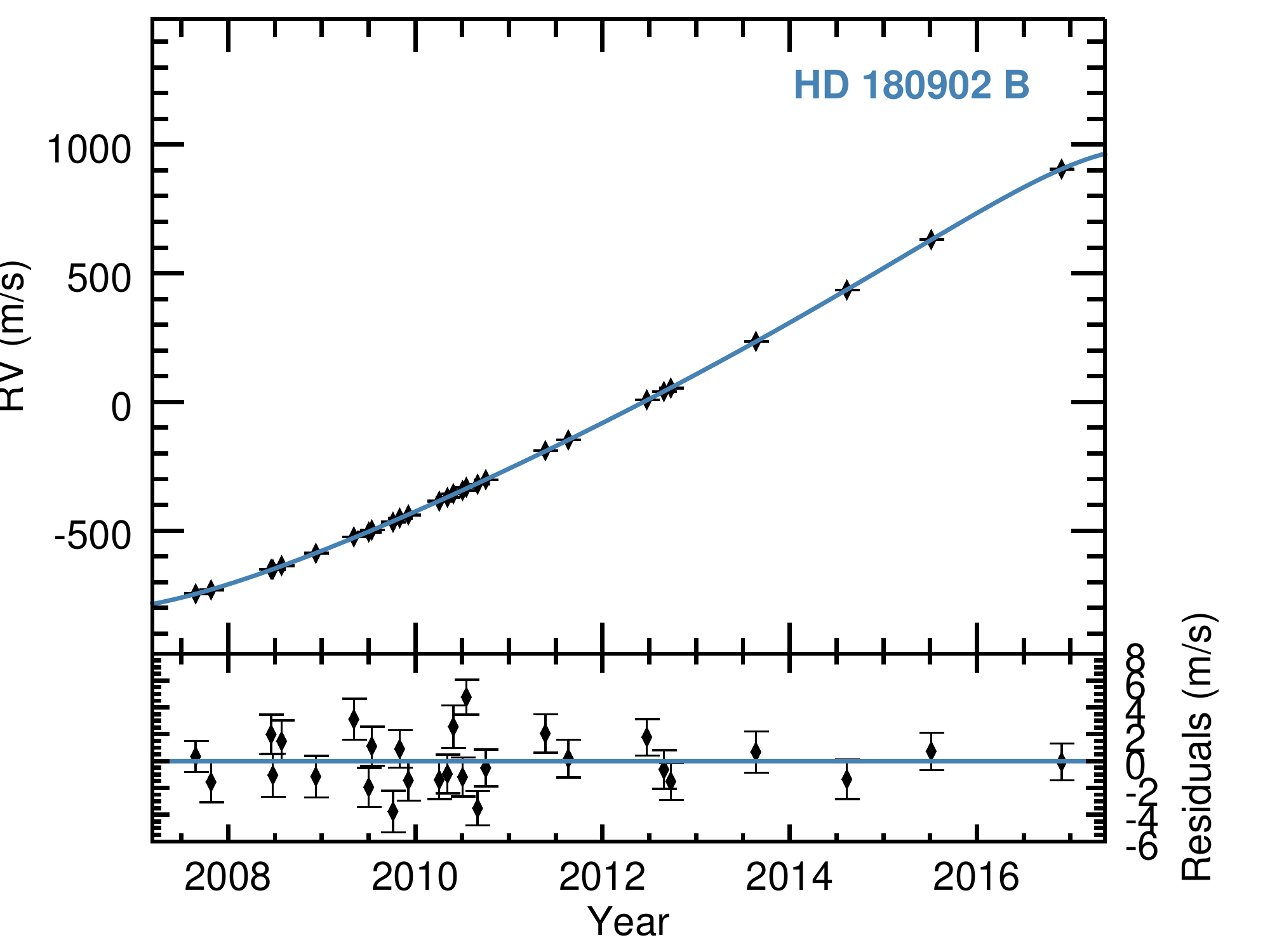}
\caption{Time series for HD 180902~A, showing the signal due to the stellar companion HD 180902~B, where the signals from HD 180902~b and c have been subtracted out. Here we show that the previously identified linear trend has some curvature, and can be fit by a Keplerian with period 5880 days, however phase coverage is incomplete, so errors for estimated parameters are large. For now, it appears to be a low mass star or brown dwarf.}
\label{fig:180902_B_time_series}
\end{figure}

HD 180902 is a K0 star \citep{SIMBAD} with 1 previously known planet \citep{Johnson2010b}. Its V-band magnitude is V$ = 7.78$, its effective temperature is $\Teff = 4961$~K, and it has a surface gravity $\logg = 3.36$. It has a mass of 1.41~M$_{\odot}$ and radius 4.16~R$_{\odot}$. The full list of stellar parameters and uncertainties is given in \autoref{tbl:stellar_params}. There are 28 total RV observations for HD 180902, the final 17 of which are new observations that span over 6 years since the original discovery of HD 180902 \citep{Johnson2010b}. It was noted in \citet{Johnson2010b} and later \citet{Bryan2016} that there was a long-term linear trend that was subtracted out, which was speculated to be a long-period companion. With the additional observations available since its discovery, we are now able to obtain a Keplerian signal to this trend. The trend is indeed due to a companion, which we estimate to be a low mass star with minimum mass $98.7\pm7.6$ M$_{\mathrm{Jup}}$. However, after performing this fit, an additional 15 day signal appeared in the periodogram. Fitting the system with 3 companions resulted in the best fit (bringing the reduced $\chi^2$ from 12 in the two companion fit to 4 in the 3 companion fit).  With this fit, the system is therefore composed of 2 planets (HD 180902~A~b,c), with a low mass stellar companion (HD 180902~B). Like the long-period planet HD 163607~c, the phase coverage for stellar companion HD 180902 B is incomplete and so period and mass uncertainties are quite high. We expect this to be resolved as more observations are obtained. The periodogram showing the peak at 15 days for HD 180902~c is shown in the right panel of \autoref{fig:180902_time_series}, with the time series shown in the left panel. The phase curve for HD 180902~c is shown in \autoref{fig:180902_c_phase_curve}. HD 180902~c has a minimum mass roughly twice the mass of Neptune, at $0.099\pm0.014$ M$_{\mathrm{Jup}}$. Finally, the best fit Keplerian signal for HD 180902~B is given in \autoref{fig:180902_B_time_series}, which shows the signal from HD 180902~B with the RV signals from HD 180902~b and c subtracted out. More phase coverage is needed to place tighter constraints on the orbit of this companion. Orbital parameters for both stellar companion HD 180902~B and planet HD 180902~c are given in \autoref{tbl:orbital_params}. 

We have identified this planet as a planet candidate due to its high FAP, 1.2\%. With the current observations, we are not fully convinced by the 15 day planet: the periodogram does not show a very strong signal, and the phase curve is sparsely sampled. Additionally, a 15 day period raises concerns of stellar rotation timescales and potentially activity-induced RV variations. Additionally, with a poorly-constrained stellar binary, it is possible that this 15 day signal would be resolved simply with a more accurate fit to the stellar companion. With more observations, we should be able to confirm the existence of this planet. We further note that it has a relatively high transit probability. A detected transit, either ground-based or from \emph{TESS}, would be able to confirm this planet, which is why we have included this candidate signal. 

\subsection{A possible sub-Jupiter orbiting HD 196645}
HD 196645 is a K0 subgiant with V$=7.80$, B-V$=0.91$ \citep{SIMBAD}. It has effective temperature $\Teff = 5041$~K, and surface gravity $\logg = 3.43$. A summary of its stellar parameters can be found in \autoref{tbl:stellar_params}.

\begin{figure*}
\centering
\includegraphics[width=\columnwidth]{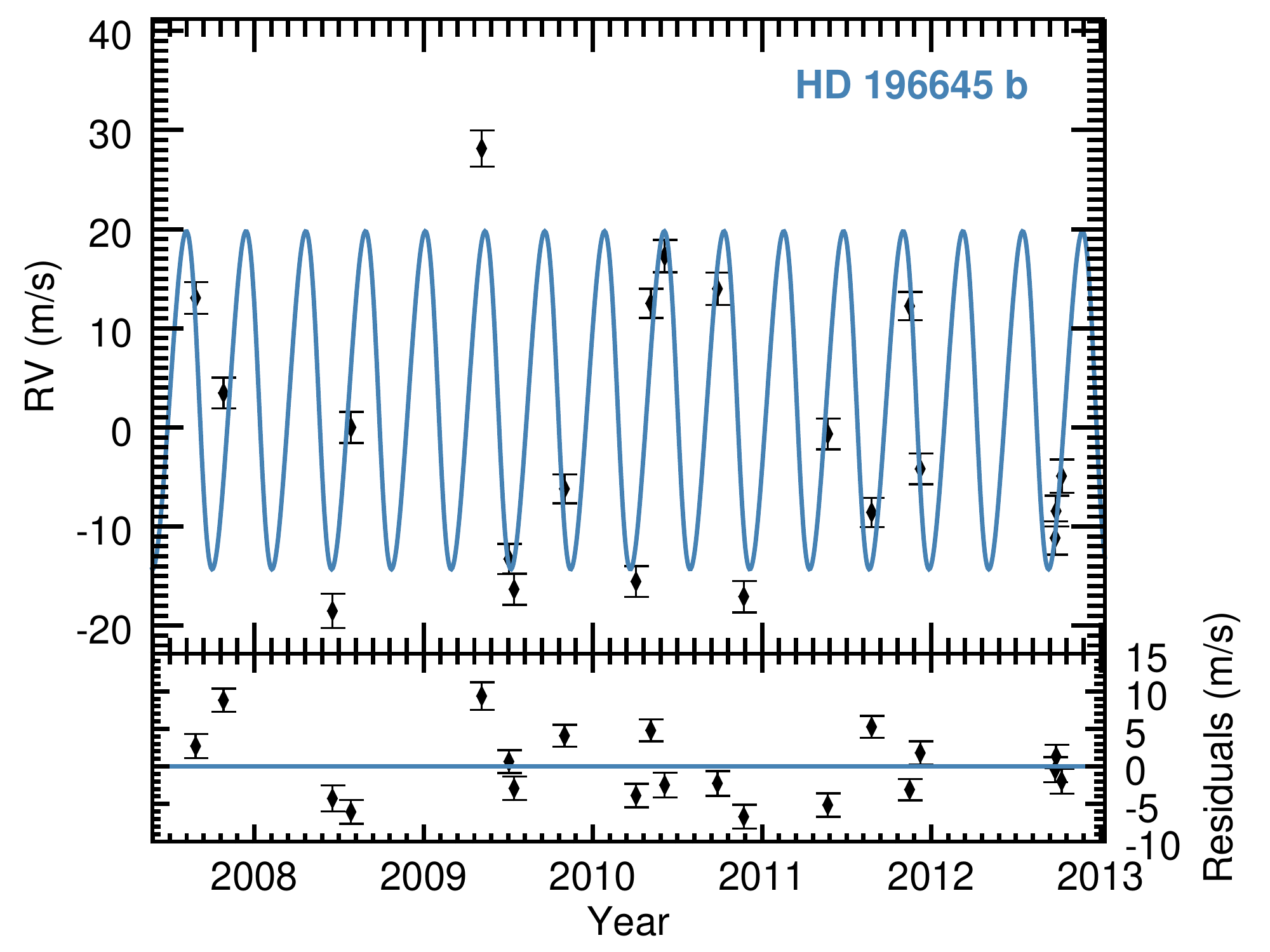}
\includegraphics[width=\columnwidth]{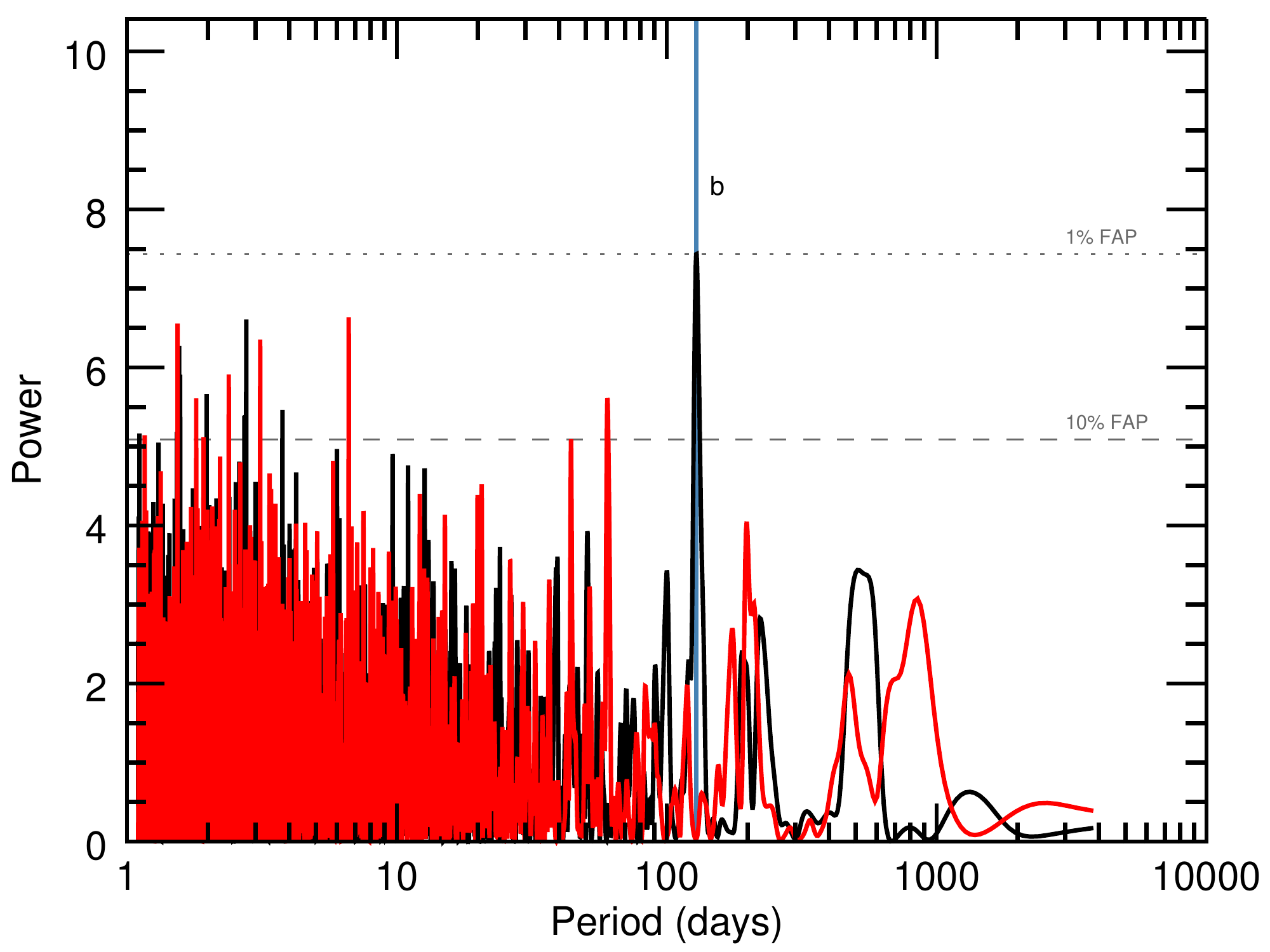}
\caption{(\emph{Left}) Time series and best-fit orbital solution for HD 196645~b, with a 128.94 day period, eccentricity of 0.106, and minimum mass $0.497~$M$_{\mathrm{Jup}}$. The residuals are shown in the bottom panel, which have an RV RMS $ = 4.7$~m$\,$s$^{-1}$. The remaining best-fit parameters can be found in \autoref{tbl:orbital_params}. The phase curve can be seen in \autoref{fig:196645_phase_curve}. (\emph{Right}) Periodogram of HD 196645 RV data before (black) and after (red) subtracting the best-fit planet parameters for HD 196645~b. The vertical line indicates the best-fit period of HD 196645~b.}
\label{fig:196645_time_series}
\end{figure*}

\begin{figure}
\includegraphics[width=\columnwidth]{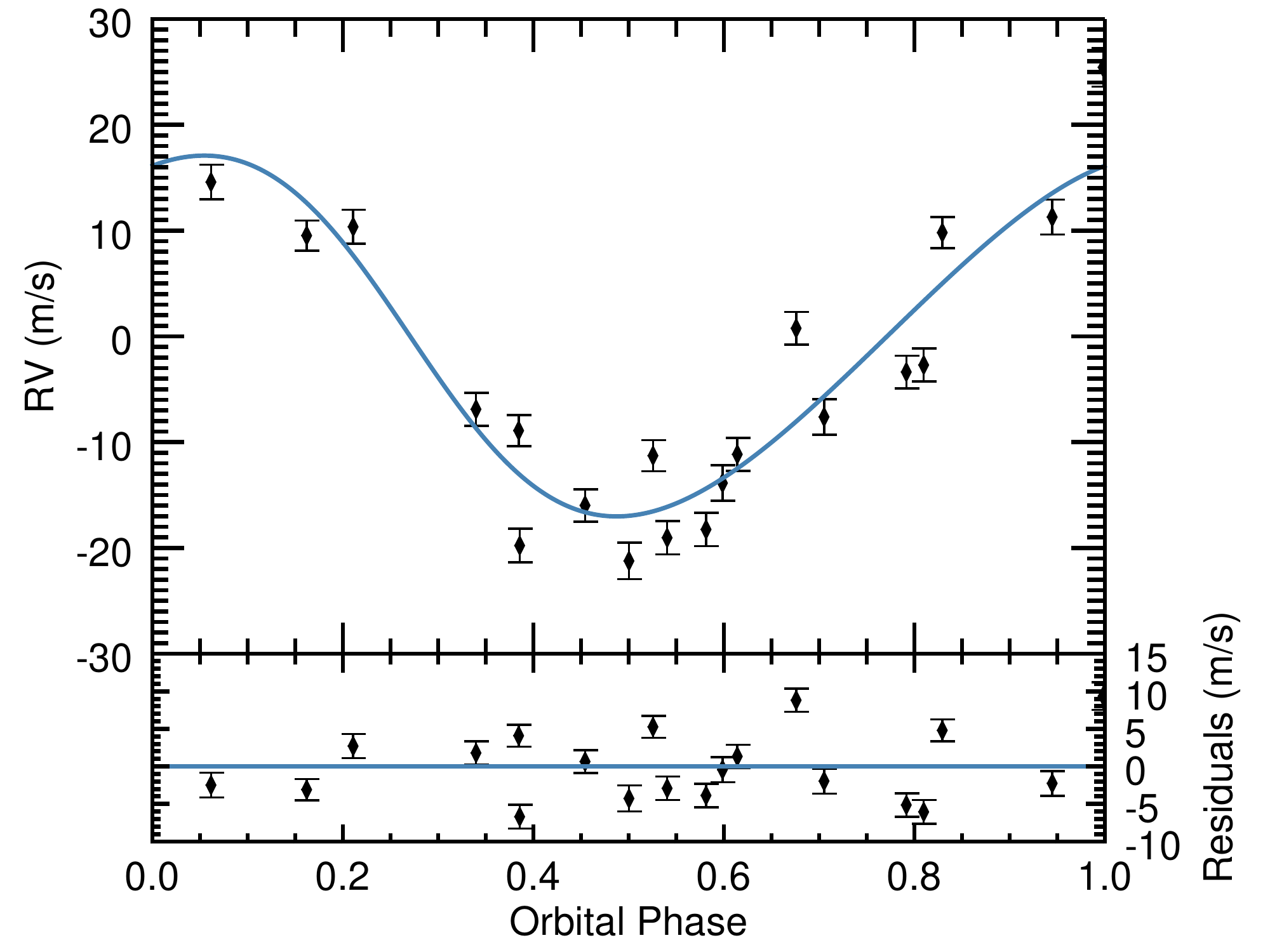}
\caption{Phase-folded velocities of the best-fit solution for HD 196645~b.}
\label{fig:196645_phase_curve}
\end{figure}

It has 20 observations that span nearly 6 years. The best-fit orbital solution yields a m$_{p}\sin{i} =0.497\pm44$~M$_{\mathrm{Jup}}$ planet on a $128.94\pm0.41$ day period, with relatively low eccentricity, $e=0.106\pm0.091$. The full set of orbital parameters are given in \autoref{tbl:orbital_params}. \autoref{fig:196645_time_series} shows the time series of HD 196645~b, with the initial and final periodogram for this system. For clarity, \autoref{fig:196645_phase_curve} also shows the phase-folded velocities for HD 196645~b. For this star, the low semi-amplitude and few observations keep us from definitively claiming this planet. It has FAP of 1.0\% and is close to our $D >10$ threshold at $D\sim 13.6$, which in combination lead to its candidate status. Continued observations will likely add significance to the periodogram peak and result in better phase coverage.

\subsection{A possible Jupiter orbiting HD 207077}
HD 207077 is a G8 subgiant with V $=8.24$, B-V $= 0.870$ \citep{SIMBAD}. It has effective temperature $\Teff = 5067$~K, and surface gravity $\logg = 3.27$. A summary of its stellar parameters can be found in \autoref{tbl:stellar_params}.

\begin{figure*}
\centering
\includegraphics[width=\columnwidth]{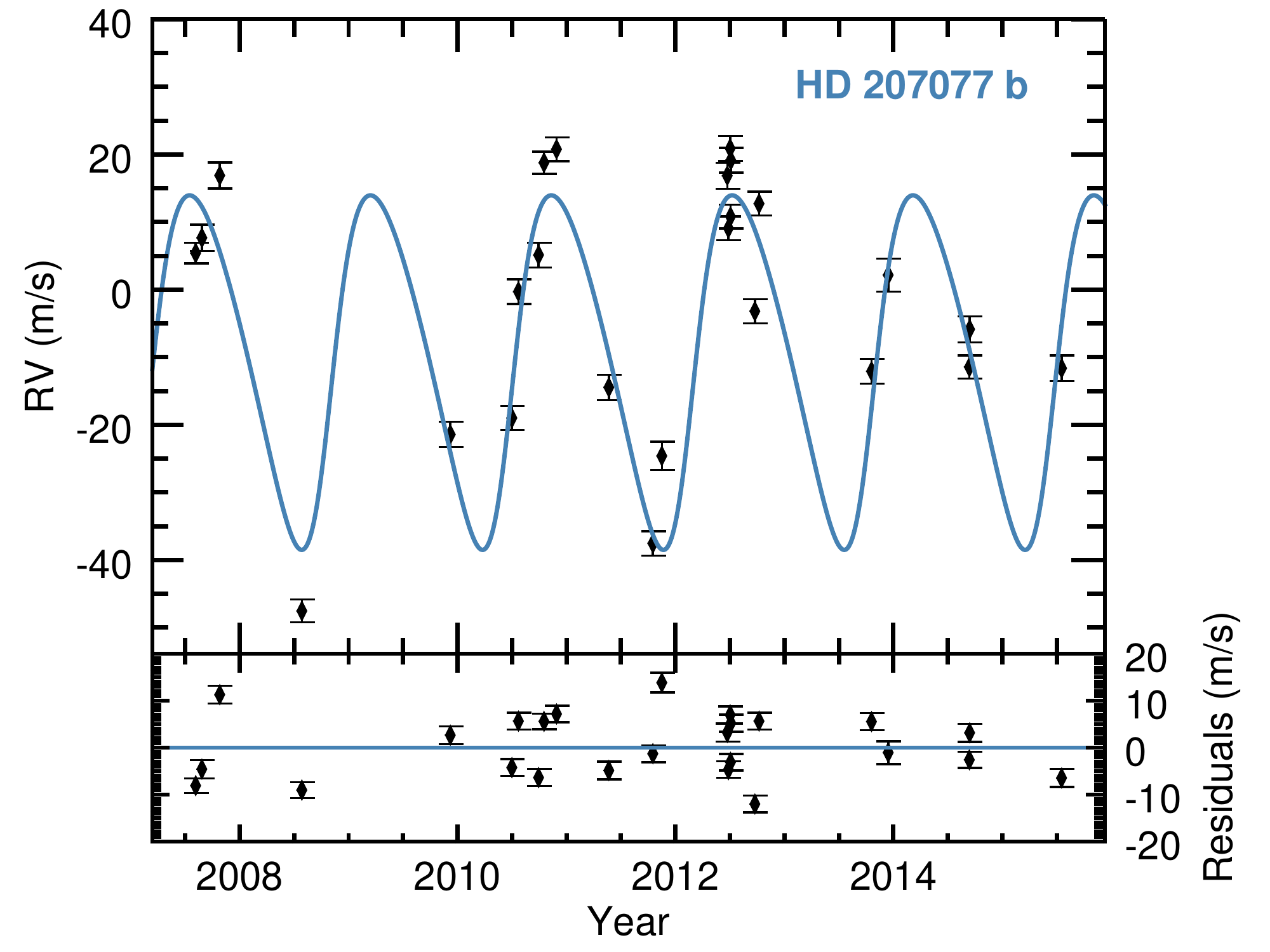}
\includegraphics[width=\columnwidth]{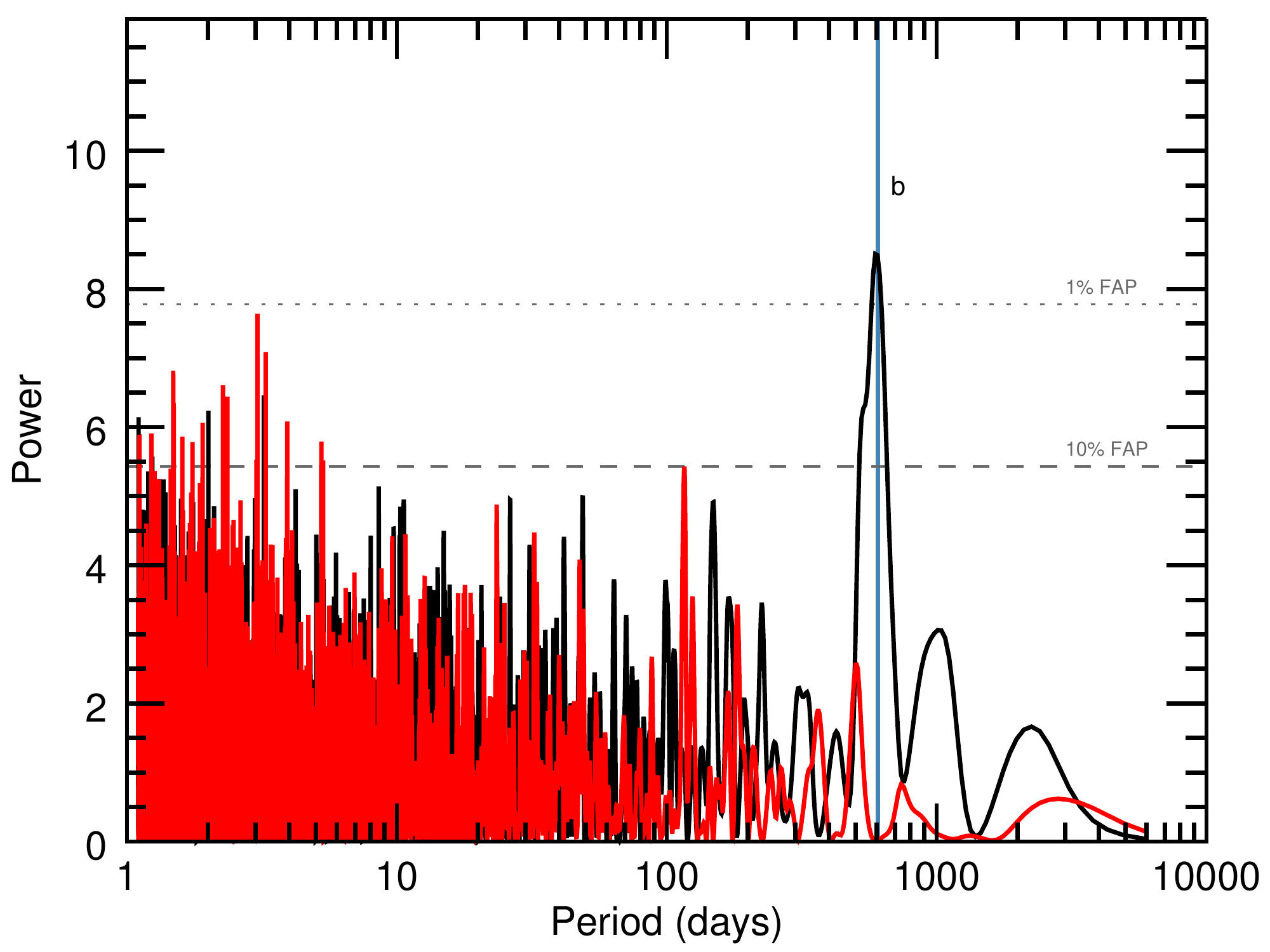}
\caption{(\emph{Left}) Time series and best-fit orbital solution for HD 207077~b, which has period P~$= 606.3$ days, eccentricity  $e=0.204$, and minimum mass m$_{p}\sin{i} =1.16~$M$_{\mathrm{Jup}}$. The residuals are shown in the bottom panel, which have an RV RMS $ = 6.7$~m$\,$s$^{-1}$. The remaining best-fit parameters can be found in \autoref{tbl:orbital_params}. (\emph{Right}) Periodogram of HD 207077 RV data before (black) and after (red) subtracting the best-fit planet parameters for HD 207077~b. The vertical line indicates the best-fit period of HD 207077~b.}
\label{fig:207077_time_series}
\end{figure*}

It has observations starting in 2007 that span roughly 8 years.  HD 207077 was previously identified in \citet{Butler2017} as having a planet candidate. The best-fit orbital solution yields a m$_{p}\sin{i} =1.16\pm0.10$~M$_{\mathrm{Jup}}$ planet on a $606.3\pm3.8$ day period, with eccentricity, $e=0.204\pm0.099$. The full set of orbital parameters are given in \autoref{tbl:orbital_params}. \autoref{fig:207077_time_series} shows the time series of HD 207077 b, with the initial and final periodogram for this system. Despite passing all of our planet detection thresholds, we are conservative for this star due to poor phase coverage and have classified it as a candidate. Continued observations will likely add to the periodogram peak and result in better phase coverage and confirming the planet status. We additionally note that this star has a relatively high RV RMS compared to similar stars (Luhn et al. 2018b, in prep.). Comparing the expected level of \emph{stellar} jitter from similar stars to the \emph{measured} RV RMS for this star indicates further evidence for an unsubtracted planet in the data that is responsible for the artificially high RV RMS.

\subsection{A possible additional planet orbiting HD 33142}
\begin{figure*}
\centering
\includegraphics[width=\columnwidth]{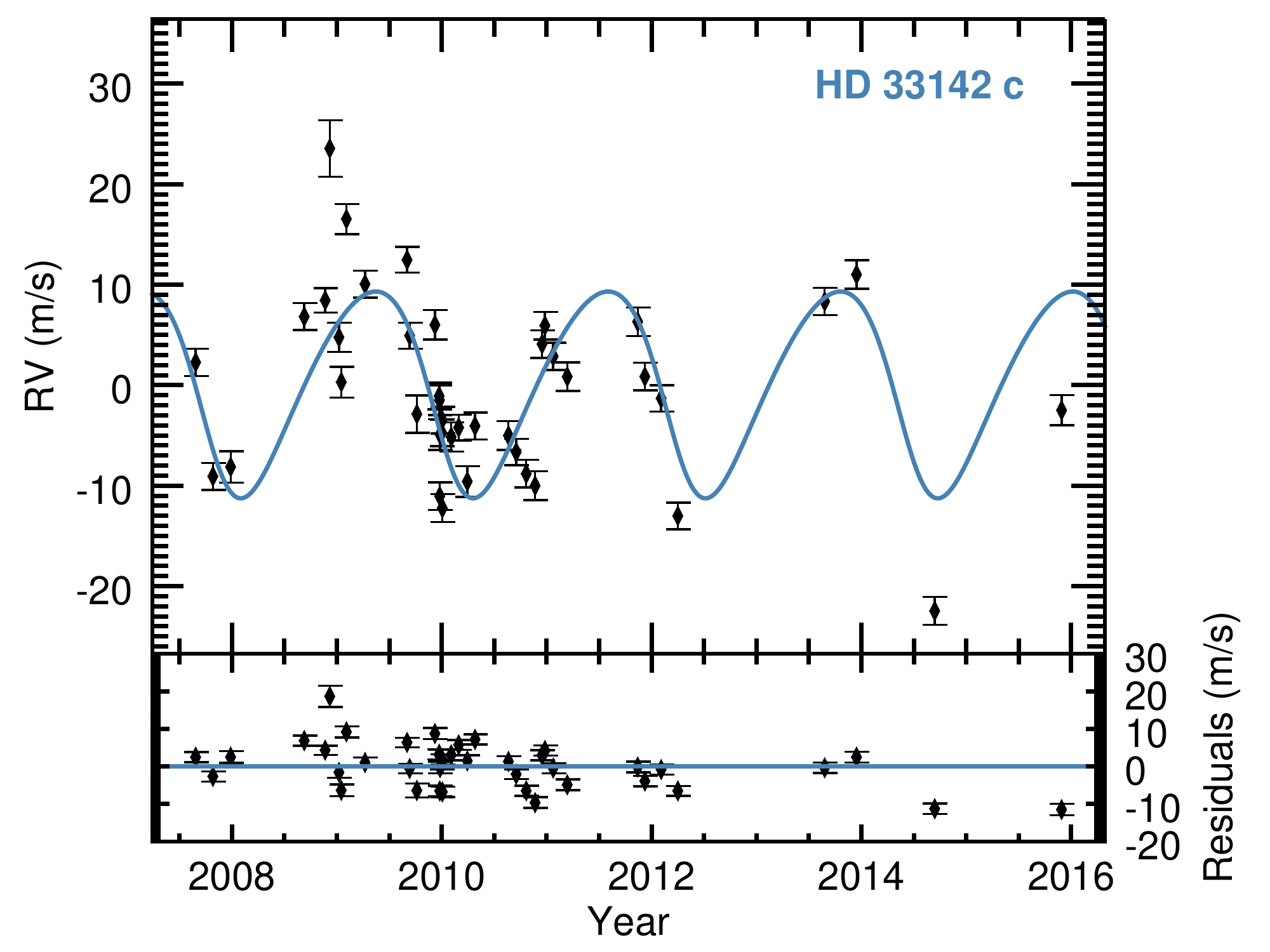}
\includegraphics[width=\columnwidth]{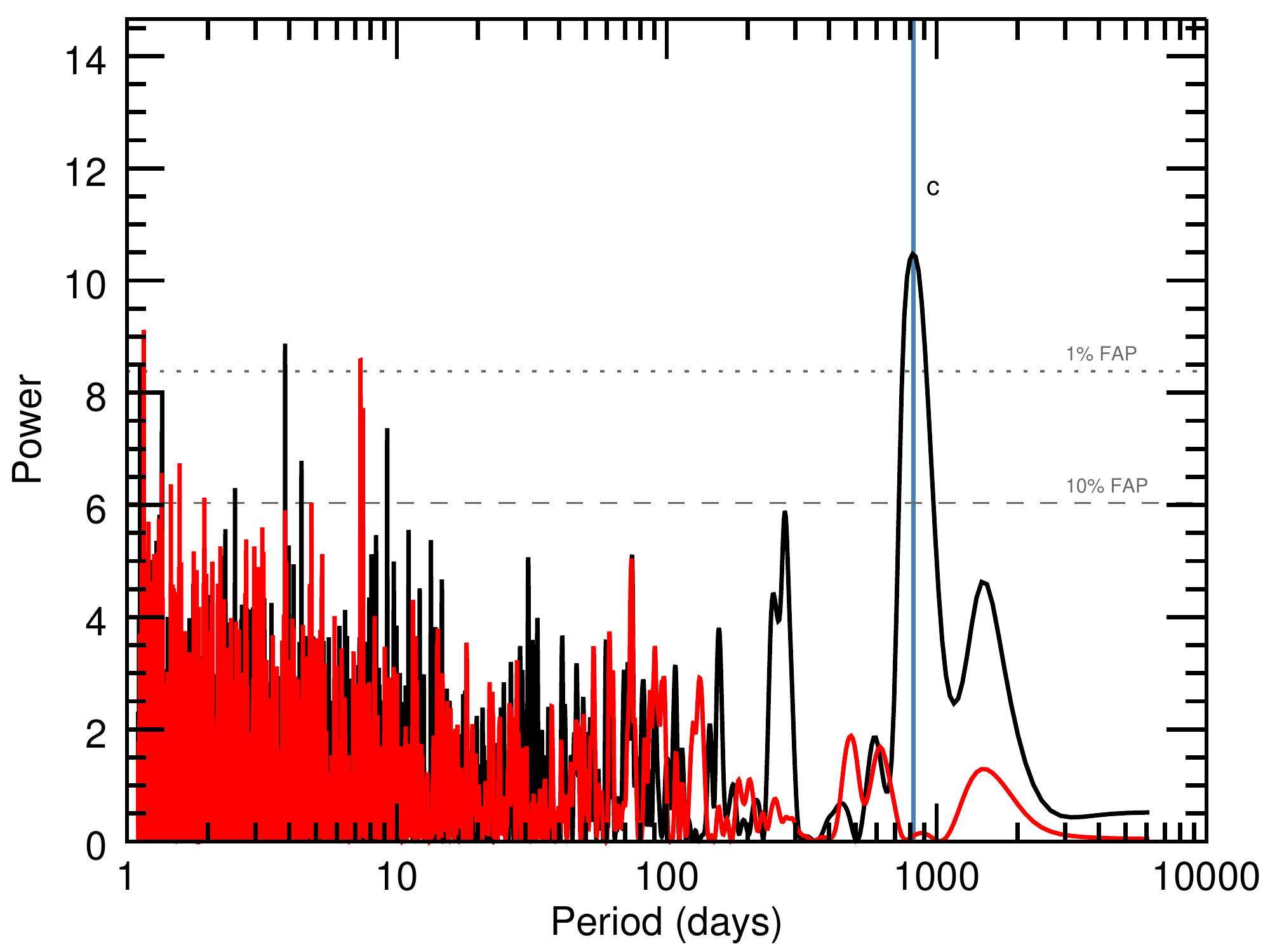}
\caption{Residuals after subtracting out the previously discovered planet HD 33142~b (period P~$=326.0$ days, eccentricity $e=0.066$, and minimum mass m$_{p}\sin{i} =1.306~$M$_{\mathrm{Jup}}$), showing evidence of a second planet. (\emph{Left}) Time series for HD 33142~c with period $809$ days, eccentricity  $e=0.16$, and minimum mass m$_{p}\sin{i} =0.59~$M$_{\mathrm{Jup}}$. (\emph{Right}) Periodogram after subtracting out the best fit orbital parameters for the previously discovered planet HD 33142~b (black) and the final periodogram after subtracting out the best two-planet fit (red). Note the black peak at 800 days, the starting guess used in the 2-planet fit.}
\label{fig:33142_time_series}
\end{figure*}

\begin{figure}
\includegraphics[width=\columnwidth]{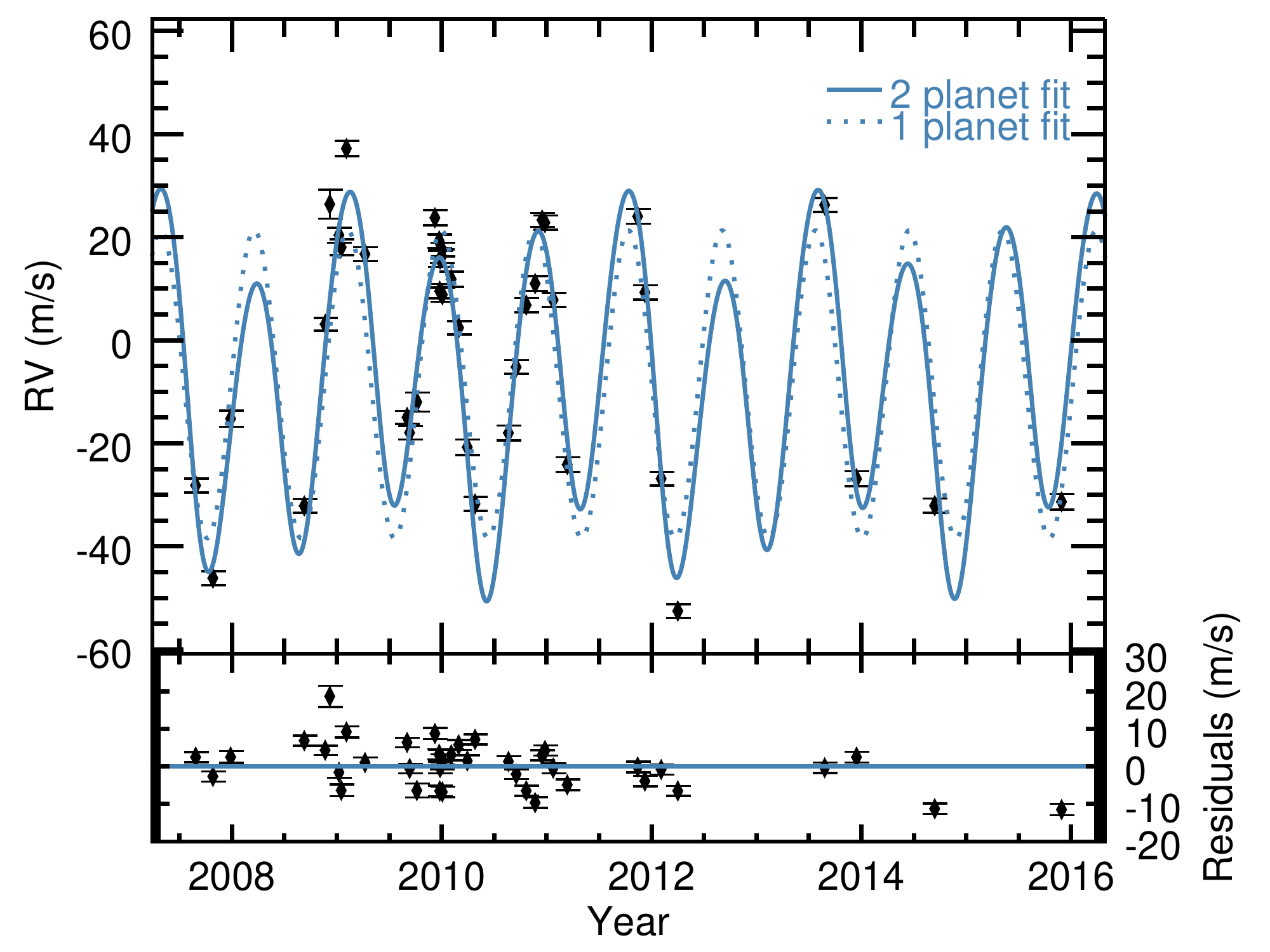}
\caption{Full time series for HD 33142 and best 2-planet fit. The dotted line shows for reference the best one-planet fit. The bottom panel shows the residuals to the best 2-planet fit.}
\label{fig:33142_full_time_series}
\end{figure}

HD 33142 is a K0 subgiant \citep{SIMBAD} with 1 previously known planet \citep{Johnson2011}. It has a V magnitude V $= 7.96$, B-V $= 0.945$, effective temperature $\Teff = 4978$~K, and $\logg=3.40$. It has a mass of 1.41~M$_{\odot}$. Additional stellar properties can be found in \autoref{tbl:stellar_params}. There are 40 observations spanning about 8 years, and we make use of the additional 7 points since the initial publication. The original fit for this planet in \citet{Johnson2011} noted unusually high jitter in this star and claimed evidence of a planet near 900 days, which we show in the periodogram with the residuals to our best fit to HD 33142~b in the right panel of \autoref{fig:33142_time_series} . As can be seen, we find evidence for a second planet near 800 days. Refitting the RVs with a two-planet fit starting with a second planet near 800 days resulted in an improvement in the reduced $\chi^2$ from 38 to 19. The minimum mass for HD 33142~c is $0.59\pm0.10$ M$_{\mathrm{Jup}}$ and it has a period of $809$ days and eccentricity  $e=0.16$. See \autoref{tbl:orbital_params} for more orbital and planet parameters and uncertainties. The time series for the second planet (HD 33142~c) is shown in \autoref{fig:33142_time_series} with the full time series and 2-planet fit shown in \autoref{fig:33142_full_time_series}. This planet candidate does not meet our detection threshold $D > 10$ with $D =9.4$. Additionally, it is not clear from the phase curve nor the full time series that the second planet is indeed present and with further observations we should know for certain. 

\section{Non-planetary Companions to Subgiant Stars}\label{sec:binaries}
\begin{figure*}
\centering
\includegraphics[width=\textwidth]{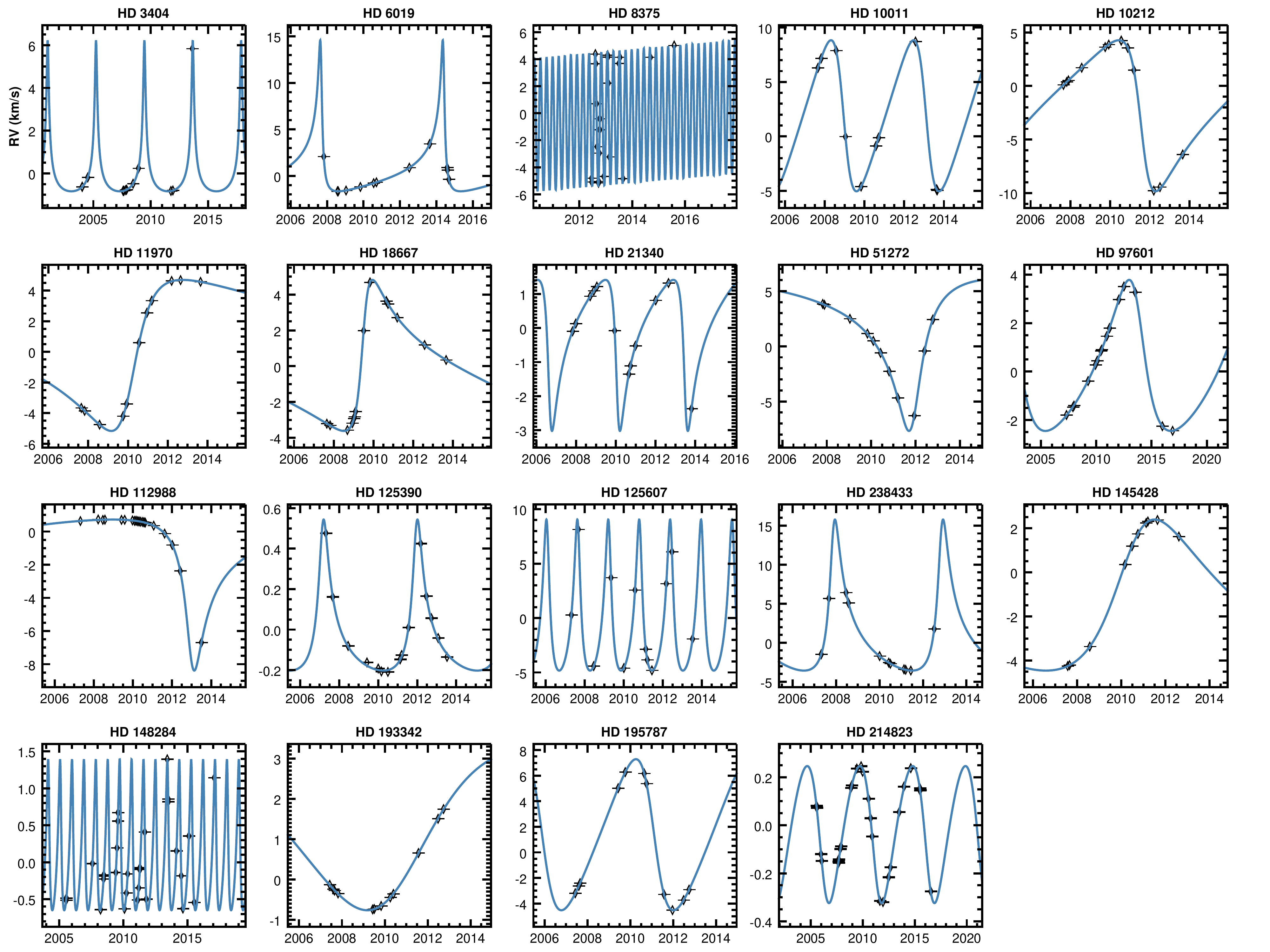}
\caption{Full time series for all stellar binaries in this work. Best fit orbital parameters for each system can be found in \autoref{tbl:orbital_params}. Note that the y-axes have units of km/s. Vertical error bars are added for clarity as in previous figures. The horizontal marks do not show error in time.}
\label{fig:subgiant_binaries}
\end{figure*}

A number of stars in our sample of subgiants showed evidence of stellar binary companions (m$\,\sin{i} > 13$~M$_{\mathrm{Jup}}$). All binaries are listed in \autoref{tbl:orbital_params} and their time series can be seen in \autoref{fig:subgiant_binaries}. In the following paragraphs, we describe those with minimum mass less than 100~M$_{\mathrm{Jup}}$, as these are likely brown dwarf candidates and are likely of special interest to the exoplanet community. 

\subsection{A 22 $M_{\mathrm{Jup}}$ brown dwarf orbiting HD 125390}
HD 125390 is a G7 star \citep{SIMBAD} with V-band magnitude 8.21, effective temperature $\Teff = 4850$~K, and surface gravity $\logg = 3.13$. It has a mass of 1.12~M$_{\odot}$ and has 15 observations over a  6 year span. We present the discovery of a $22.16\pm0.96$~M$_{\mathrm{Jup}}$ brown dwarf companion to this star with period $1756.2\pm3.9$ days. The rest of the orbital parameters can be found in \autoref{tbl:orbital_params}. The time series for this companion is shown in \autoref{fig:subgiant_binaries}.

\subsection{HD 148284}

HD 148284 is a K0 star \citep{SIMBAD} with V-band magnitude 9.01, effective temperature $\Teff = 5572$~K, and surface gravity $\logg = 3.97$. It has a mass of 1.02~M$_{\odot}$ and has 30 observations over an 11 year span. We present the discovery of a $34.5\pm0.96$~M$_{\mathrm{Jup}}$ brown dwarf companion to this star with period $339.302\pm0.026$ days. The rest of the orbital parameters can be found in \autoref{tbl:orbital_params}. The time series for this companion is shown in \autoref{fig:subgiant_binaries}.

\subsection{HD 214823}

HD 214823 is a G0 star \citep{SIMBAD} with V-band magnitude 8.07, effective temperature $\Teff = 5933$~K, and surface gravity $\logg = 3.92$. It has a mass of 1.31~M$_{\odot}$ and has 28 observations over an 11 year span. We present the discovery of a $20.56\pm0.32$~M$_{\mathrm{Jup}}$ brown dwarf companion to this star with period $1853.9\pm1.6$ days. The rest of the orbital parameters and uncertainties can be found in \autoref{tbl:orbital_params}. The time series for this companion is shown in \autoref{fig:subgiant_binaries}.

\subsection{Brown dwarf candidate orbiting HD 180902}
We have already remarked on this system in \autoref{sec:hd180902}, however we wish to make a final remark here. The minimum mass for the stellar companion HD 180902~B ($98.7\pm7.6$ M$_{\mathrm{Jup}}$) places it in as a candidate brown dwarf. However, the orbit for this companion is poorly-constrained and so it is likely that the mass could substantially change with continued observations.

\section{Transit Times, Probabilities, Depths, and Durations of CPS Subgiants with Known RV Planets}\label{sec:transit_params}
Here we describe the transit parameters for subgiant companions used in this work, all of which are are given in \autoref{tbl:transit_params}. At typical separations of planets in this paper (1-2 au), the average value for the transit probability is roughly 2.2\%\footnote{We have excluded binary companions and only included planet companions in this rough calculation. We have also done a literature search for detections of transits (null or positive) for those in our sample with transit probability $> 10\%$ and have removed those from this calculation.}. For the 60 planets in \autoref{tbl:transit_params} it is likely that 1 or 2 will in fact transit. Observing a planet in transit provides a wealth of additional information about the system, most notably the size and mass (without the $\sin{i}$ dependence) of the planet, which provides a bulk density for the planet. As an example, KELT-11~b (also referred to as HD 93396) is a highly inflated planet  --- which we know only because we have mass \emph{and} radius information from the combined RV and transit data --- on a short-period orbit around a subgiant star \citep{Pepper2017}. With transit observations of RV planets on similarly close orbits around subgiants we can understand how unique inflated planets like KELT-11~b are and whether the increased insolation from the evolved subgiant host plays a role in inflating the planet. Furthermore, transits provide a model-independent measure of the stellar density, which combined with the radius from parallax would produce a model-independent stellar mass \citep{Seager2003}. Using Gaia parallaxes, reasonable SED measurements, and precise transits, errors in the stellar radius would be ~1\%, and mass errors ~3\% \citep{Beatty2017}.

In order to determine the best transit parameters, we refit all radial velocity data for known subgiants. The reasons for this were two-fold: 1) several of these planets have additional observations since their published discovery, and 2) updated stellar parameters from \citet{Brewer2016} allow for more accurate planet masses and reduced uncertainties in predicted transit time durations. All best-fit orbital parameters in this work were obtained using the IDL RVLIN package \citep{Wright2009} in conjunction with the BOOTTRAN \citep{Wang2012} package, which works with RVLIN and uses the bootstrapping technique to find best fit parameters and uncertainties of radial velocity fits. The best-fit orbital solutions for each planet is given in \autoref{tbl:orbital_params} and we include all RV observations used in this work in \autoref{tbl:all_rvs}. Once we fit the RV data and obtained new orbital parameters, the final remaining step was to estimate radii in order to calculate estimated transit parameters. We used the Python code FORECAST \citep{Chen2017} to estimate radii from minimum masses provided by the RV fits. We note that our transit calculations are meant to be estimates and as such, we merely used the mean radius from FORECAST and do not include radius errors\footnote{Typical radius errors were 0.2 R$_{\mathrm{Jup}}$, which results in less than 0.01\% uncertainty in the transit probabilities.}.  The BOOTTRAN package also outputs the Barycentric Julian Date (BJD) of the next $n$ times of inferior conjunction and their uncertainties. For simplicity we use the terms ``conjunction time" and ``transit time" synonymously despite the fact that the planet may not actually transit. Using BOOTTRAN, we predict the next 3 transits for each planet starting after the launch of \emph{TESS} and report those with uncertainties as carried through by BOOTTRAN in Julian Date (Columns 3-8). They require space-based observations because the majority of these planets are at 1-2 au around inflated subgiants so their transit depths will likely be difficult to observe with ground-based telescopes, and transit durations and ingress/egress durations will be rather long, again posing a difficulty for ground-based transits. Despite the short photometric baseline from surveys like \emph{TESS} ($\sim$27 days for the shortest), they may manage to catch one of these planets in transit. Furthermore, the longer period RV planets have large uncertainties on their predicted transit times which would pose further problems for ground-based observing. We calculate the transit parameters as given in \citet{Seager2010}, including the \emph{a priori} transit probability 
\begin{equation}
\tau_{\mathrm{pr}} = \frac{\left(R_{\star}+R_{\mathrm{p}}\right)}{a} \frac{\left(1+e \sin{\omega}\right)}{1-e^2}
\label{eqn:tr_prob}
\end{equation}
(see also \citet{Seagroves2003,Barnes2007}) which is given in Column 9. True \emph{a posteriori} probabilities \citep{Stevens2013} can be computed with knowledge of the underlying mass function for planets orbiting subgiants with comparable periods and host star masses. Since the exoplanet mass function has negative slope, this means our transit probabilities as given in \autoref{eqn:tr_prob} are slightly underestimated. The transit depth is 
\begin{equation}
\tau_{\mathrm{depth}} = \left(\frac{R_{\mathrm{p}}}{R_{\star}}\right)^2
\label{eqn:tr_depth}
\end{equation}
and is given in Column 10. We also calculate transit and ingress/egress durations which are given by
\begin{equation}
\tau_{\mathrm{dur}} = \frac{P}{\pi} \arcsin\left(\frac{R_{\mathrm{\star}}}{a \sqrt{(1+\frac{R_{\mathrm{p}}}{R_{\star}})^2.}}\right)
\label{eqn:duration}
\end{equation}
\begin{equation}
\Delta \tau_{12} = \frac{P R_{\mathrm{p}}}{\pi a}
\label{eqn:ing_duration}
\end{equation}
and can be found in Columns 11 and 12. We note that the duration equations have been simplified to assume a circular orbit and excludes grazing transits, which are acceptable for first-order calculations. Finally, the three transit times given in Columns 3, 5, and 7 are listed again in YYYY/MM/DD format (UTC) for quick reference.

While the large majority of planets have probabilities 1-2\%, we highlight 3 planets with transit probabilities greater than 9\%: HD 102956~b (26.16\%), HD 180902~c (19.37\%), and HD 163607~b (9.55\%)\footnote{We have not included HD 93396 (KELT-11), which we list as having a 70\% transit probability, since it is known to transit and was in fact discovered via transits \citep{Pepper2017}. We have also excluded HD 88133~b, which has transit probability 22.89\% as it has been found to be non-transiting \citep{Piskorz2016}. Additional planets excluded from this list as they have been found to be non-transiting include HD 185269~b \citep{Johnson2006,Moutou2006} with 13\% transit probability, HD 38529~b \citep{Henry2013} with 13\% transit probability.}. These are all planets with periods less than about 75 days, relatively short for evolved stars of intermediate mass. With the exception of HD 180902~c, which is a large hot neptune candidate, they also have predicted depths of $\sim800$ ppm or more which makes these more amenable for ground-based observations than the other planets. Although 800 ppm is still a challenge for ground-based observing, it has been demonstrated on mid-class telescopes with diffuser-assisted photometry \citep{Stefansson2017}. As short period RV planets, their periods are more tightly constrained, and so transit time uncertainties are relatively small ($\ll 1$ day, with the exception of HD 180902~c). We emphasize that the majority of the planets around subgiants have semi-major axes of 1 au or more, which in combination with the large stellar radii leads to long transit durations, small transit depths, and lower transit probabilities\footnote{The increased stellar radii actually helps to inflate the transit probabilities, but the large separations ultimately account for the low probabilities.}. In addition, many of the uncertainties in the midtransit times are on the order of 10 days (anywhere from 13 days to 150 days), which makes hunting for them with ground-based telescopes difficult. These factors are what make these planets more amenable to a space-based survey like \emph{TESS}, which may incidentally catch a single transit for one of these planets.

Refitting each known planet as we did means that \autoref{tbl:orbital_params}, which contains the best-fit orbital parameters for all CPS subgiant stars with known planets, lists the most precise and up-to-date orbital solutions for these planets.

\section{Summary and Conclusions}\label{sec:conclusions}
In this work, we presented the discovery of 15 new planetary signals around subgiant stars, 8 of which are planet candidates requiring further observations, increasing the number of known RV planets around subgiants by $\sim$15\% (25\% including the candidates). Of special importance are possibly the first 3-planet systems around a subgiant star, HD 163607 and HD 4917. As observations on subgiants continue and more long-period planets are discovered, these systems will be useful in determining how multi-planet systems evolve as their host star leaves the main sequence. We lastly note several stellar companions of brown dwarf to nearly solar mass size and provide orbital parameters for these systems as well.

In this work, we have calculated transit parameters (transit times, probabilities, depths, and durations, see \autoref{tbl:transit_params}) for all known planets around California Planet Search subgiant stars ($3 < \logg < 4$). We find that 3 planets have relatively high transit probabilities ($ \gtrsim 10\%$). These planets have the best chance of having a transit observed from a ground-based telescope. The remaining 50 planets in general all have lower transit probabilities (1-2\%), longer transit durations ($\sim50$~hr), smaller transit depths (of order 500 ppm), and more uncertain transit times (tens of days). The combination of these factors indicates that these are challenging to observe from the ground, but instead will make good targets for future space-based missions like \emph{TESS}. In predicting transit parameters, we have made use of additional RV observations since the planets' initial discoveries and updated stellar parameters to refit all planets, resulting in updated orbital parameters for all planets (\autoref{tbl:orbital_params}).

\acknowledgments{We would like to thank the anonymous referee whose suggestions have greatly enhanced this paper. We would like to thank Thomas Beatty and Arpita Roy for their discussions and helpful comments. Many of the new planets and planet candidates in this work have come to fruition with the last several years of observations. These observations would not have taken place without Andrew Howard shepherding the observations and extending the time baseline for these systems. We also wish to thank Alan Reyes for his help observing. We also wish to thank Scott Gaudi for pointing out that our transit probabilities are \emph{a priori} transit probabilities and Dan Stevens for clarification on how the \emph{a posteriori} probabilities would differ. The authors wish to recognize and acknowledge the very significant cultural role and reverence that the summit of Mauna Kea has always had within the indigenous Hawaiian community. We are most fortunate to have the opportunity to conduct observations from this mountain.

 This research has made use of the SIMBAD database, operated at CDS, Strasbourg, France; the Exoplanet Orbit Database and the Exoplanet Data Explorer at exoplanets.org.; and of NASA's Astrophysics Data System Bibliographic Services. This material is based upon work supported by the National Science Foundation Graduate Research Fellowship Program under Grant No. DGE1255832.}

\software{RVLIN \citep{Wright2009}, BOOTTRAN \citep{Wang2012}, FORECAST \citep{Chen2017}}

\bibliography{\string~/Google_Drive/Research/library}

\startlongtable
\begin{deluxetable*}{c c c c c c c c c c c}
\tablecaption{Stellar Parameters for Subgiants with Known Companions \label{tbl:stellar_params}}
\tabletypesize{\scriptsize}
\tablehead{
	 \colhead{Star} 	&  \colhead{RA}   & \colhead{DEC}  & \colhead{V} & \colhead{Mass} & \colhead{Radius} & \colhead{$\Teff$} & \colhead{$\logg$}	& \colhead{$\logRHK$} & \colhead{S$_{\mathrm{HK}}$} & \colhead{$\sigma_{\mathrm{RV}}$} \\
	  \colhead{} 	&  \colhead{deg}    	& \colhead{deg}  	 & \colhead{mag} & \colhead{M$_{\odot}$} & \colhead{R$_{\odot}$} & \colhead{K} & \colhead{} & \colhead{} & \colhead{} &\colhead{(m/s)} }
	\colnumbers
	\startdata
	           HD 10697 &      26.23260 &      20.08315 &    6.27 &    1.13 &    1.79 &    5600 &    3.96 &     -5.04 &     0.16 &      7.856         \\
           HD 38529 &      86.64547 &       1.16819 &    5.95 &    1.41 &    2.56 &    5541 &    3.77 &     -5.03 &     0.16 &      9.531         \\
          HD 114613 &     198.01328 &     -37.80302 &    4.85 &    1.27 &    2.14 &    5641 &    3.87 &     -5.08 &     0.15 &      5.675         \\
          HD 117176 &     202.10753 &      13.77879 &    4.97 &    1.09 &    1.89 &    5495 &    3.95 &     -5.12 &     0.14 &      3.974         \\
          HD 159868 &     264.74802 &     -43.14551 &    7.24 &    1.19 &    2.13 &    5534 &    3.92 &     -5.02 &     0.16 &      5.485         \\
          HD 175541 &     283.92035 &       4.26533 &    8.02 &    1.39 &    4.19 &    5013 &    3.33 &     -5.18 &     0.14 &      6.516         \\
          HD 190228 &     300.75323 &      28.30686 &    7.30 &    1.14 &    2.51 &    5238 &    3.72 &     -4.98 &     0.17 &      4.314         \\
            HD 1502 &       4.82111 &      14.05475 &    8.36 &    1.46 &    4.67 &    4947 &    3.18 &     -5.15 &     0.14 &     11.328         \\
\enddata
	\tablecomments{Host star parameters for all planets listed in \autoref{tbl:transit_params}. Column 1 lists the star name. Columns 2 and 3 give the coordinates. Data in Columns 4-10 come from \citet{Brewer2016}. Column 4 gives the V-band magnitude. Columns 5, 6, and 7 list the mass, radius, and effective temperature of the star. Column 8 gives the spectroscopic surface gravity. Columns 9 and 10 give two measures of chromospheric activity. We note that for most subgiants in this sample, the $\logRHK$ value is not calibrated, which is why we have also included S$_{\mathrm{HK}}$. Finally, Column 11 gives the measured RV RMS after subtracting the planetary signal.}
\end{deluxetable*}

\begin{longrotatetable}
\begin{deluxetable*}{c c c c c c c c c c c c r}
\tablecaption{Orbital Parameters \label{tbl:orbital_params}}
\tabletypesize{\scriptsize}
\tablehead{
	 \colhead{Star} 	&  \colhead{Com}   & \colhead{m$\sin{i}$}  & \colhead{R} & \colhead{P} &	\colhead{a} & \colhead{T$_{\mathrm{p}}$}	& \colhead{e} & \colhead{$\omega$} & \colhead{K} & \colhead{$\gamma$} & \colhead{d$v$d$t$} & Orbit Ref \\
	  \colhead{} 	&  \colhead{}    	& \colhead{(M$_{\mathrm{Jup}}$)}  	 & \colhead{R$_{\mathrm{Jup}}$} & \colhead{(days)} & \colhead{} & \colhead{(JD$-2440000$)} & \colhead{} & \colhead{degrees} & \colhead{(m/s)} & \colhead{(m/s)} & \colhead{}}
	 \colnumbers
	 \startdata
	           HD 10697 &   b &         6.383(78) &             1.152 &       1075.69(82) &         2.140(88) &         11492(17) &        0.1043(83) &        114.9(6.0) &        116.9(1.3) &         -9.15(82) &                 0 &      \citet{Wittenmyer2009}         \\
           HD 38529 &   b &         0.797(15) &             1.268 &      14.30944(30) &        0.1294(58) &      14384.59(16) &         0.280(17) &         89.8(4.4) &         55.2(1.1) &          8.83(74) &                 0 &          \citet{Wright2009}         \\
                --- &   c &         12.99(15) &             1.123 &       2136.1(3.1) &          3.64(16) &      12255.9(8.1) &        0.3407(69) &         18.4(1.4) &        172.4(1.3) &          0.0(0.0) &                 0 &          \citet{Wright2009}         \\
          HD 114613 &   b &         0.357(32) &             1.182 &         4000(120) &          5.34(26) &         15150(90) &         0.458(89) &           196(11) &          4.38(40) &          2.62(65) &                 0 &      \citet{Wittenmyer2014}         \\
          HD 117176 &   b &         7.416(57) &             1.142 &      116.6880(54) &         0.481(22) &     15291.721(86) &        0.3988(19) &        359.83(29) &        316.20(78) &        -51.25(53) &                 0 &          \citet{Butler2006}         \\
          HD 159868 &   b &         2.218(59) &             1.202 &       1184.1(7.1) &          2.32(14) &        13670(190) &         0.024(19) &            63(76) &         37.92(97) &         -6.42(92) &                 0 &      \citet{Wittenmyer2012}         \\
                --- &   c &         0.768(44) &             1.269 &        351.0(1.1) &         1.032(64) &         13225(13) &         0.184(37) &           272(16) &         20.0(1.1) &          0.0(0.0) &                 0 &      \citet{Wittenmyer2012}         \\
\enddata
	\tablecomments{Orbital parameters and uncertainties for all planets listed in \autoref{tbl:transit_params}. Uncertainties are given in parenthetical format, whereby the value has been rounded to the appropriate significant digit, and the listed uncertainty is of the same order (e.g., 1.93(17) is the same as writing $1.93 \pm 0.17$)). Columns 1 and 2 list the star and planet designations, as in \autoref{tbl:transit_params}. Column 3 lists the minimum mass of the planet. Column 4 gives the estimated radius from FORECAST \citep{Chen2017} used for transit calculations. Column 5 gives the best-fit period. Column 6 gives the calculated semi-major axis of the planet's orbit. Columns 7-12 give the remaining best-fit orbital parameters: time of periastron crossing (Column 7), eccentricity (Column 8), argument of periastron (Column 9), semi-amplitude (Column 10), the systemic velocity (Column 11), and the linear trend parameter (Column 12). Finally, Column 13 lists the reference used as initial input orbital parameters for the known planets. New orbital companions are listed as ``This Work" if they are planetary companions (m$\sin{i} < 14$ M$_{\mathrm{Jup}}$) or "Binary" if their minimum mass indicates they are a brown dwarf or larger mass. We have not performed an extensive search on the previous literature of the binary companions. We have only included those that have well-constrained orbits. This is discussed more in \autoref{sec:binaries} and all binary time series are shown in \autoref{fig:subgiant_binaries}.}
\end{deluxetable*}
\end{longrotatetable}

\startlongtable
\begin{deluxetable}{c c c c c}
\tablecaption{Radial Velocities \label{tbl:all_rvs}}
\tabletypesize{\scriptsize}
\tablehead{
	 \colhead{Star} 	&  \colhead{Date}    & \colhead{RV}  	 & \colhead{$\sigma_{\mathrm{RV}}$} & \colhead{Tel} \\
	  \colhead{} 	&  \colhead{(JD-2440000)}    & \colhead{(m/s)}  	 & \colhead{(m/s)} & \colhead{}}
	  \colnumbers
	  \startdata
	  \input{all_latex_subgiant_planet_RVs.txt}
	  \enddata
	  \tablecomments{\autoref{tbl:all_rvs} is published in its entirety in the machine-readable format. A portion is shown here for guidance regarding its form and content.}
\end{deluxetable}

\begin{longrotatetable}
\startlongtable
\begin{deluxetable*}{c c c c c c c c c c c c c c c}
\tablecaption{Transit Parameters for Known Planets Around Subgiants \label{tbl:transit_params}}
\tabletypesize{\scriptsize}
\tablehead{
	 \colhead{Star} 	&  \colhead{Com}   & \colhead{$\tau_1$ }  & \colhead{$\sigma_1$} & \colhead{$\tau_2$} & \colhead{$\sigma_2$} & \colhead{$\tau_3$} & \colhead{$\sigma_3$}	& \colhead{$\tau_{\mathrm{pr}}$} & \colhead{$\tau_{\mathrm{depth}}$} & \colhead{$\tau_{\mathrm{dur}}$} & \colhead{$\Delta \tau_{12}$} & \colhead{$\tau_1$} & \colhead{$\tau_2$} & \colhead{$\tau_3$} \\
	  \colhead{} 	&  \colhead{}    	& \colhead{(JD)}  	 & \colhead{(JD)} & \colhead{(JD)} & \colhead{(JD)} & \colhead{(JD)} & \colhead{(JD)} & \colhead{(\%)} & \colhead{(ppm)} & \colhead{(hr)} & \colhead{(hr)} & \colhead{(yyyy/mm/dd)} & \colhead{(yyyy/mm/dd)} & \colhead{(yyyy/mm/dd)} }
	\colnumbers
	\startdata
	                    HD 10697 &   b &       2458961.13 &     6.46 &       2460036.82 &     7.17 &       2461112.51 &     7.91 &                        0.46 &      4360.578 &       30.9724 &        1.9186 &       2020/04/21 &       2023/04/02 &         2026/03/13         \\
  HD 38529 \tablenotemark{b} &   b &       2458405.55 &     0.12 &       2458419.86 &     0.12 &       2458434.17 &     0.12 &     13.44 \tablenotemark{b} &      2582.275 &        7.9461 &        0.3837 &       2018/10/14 &       2018/10/28 &         2018/11/11         \\
                         --- &   c &       2458889.11 &     6.20 &       2461025.19 &     9.31 &       2463161.28 &    12.42 &                        0.43 &      2023.902 &       47.3599 &        2.0389 &       2020/02/09 &       2025/12/15 &         2031/10/21         \\
                   HD 114613 &   b &       2458570.08 &   194.59 &       2462572.23 &   288.82 &       2466574.38 &   394.43 &                        0.22 &      3208.864 &       61.2665 &        3.2845 &       2019/03/27 &       2030/03/11 &         2041/02/23         \\
                   HD 117176 &   b &       2458457.10 &     0.18 &       2458573.79 &     0.18 &       2458690.48 &     0.19 &                        2.31 &      3843.355 &       15.8918 &        0.9276 &       2018/12/04 &       2019/03/31 &         2019/07/25         \\
                   HD 159868 &   b &       2458487.74 &    24.49 &       2459671.82 &    31.03 &       2460855.89 &    37.79 &                        0.46 &      3349.915 &       40.0036 &        2.1887 &       2019/01/04 &       2022/04/02 &         2025/06/29         \\
                         --- &   c &       2458662.80 &    12.74 &       2459013.76 &    13.50 &       2459364.73 &    14.29 &                        0.86 &      3738.315 &       32.9151 &        1.8965 &       2019/06/28 &       2020/06/13 &         2021/05/30         \\
\enddata
	\tablenotetext{a}{Planet is known to transit.}
	\tablenotetext{b}{Planet is known to not transit.}
	\tablecomments{Columns 1 and 2 list the star and companion for which transit parameters are computed. A * symbol indicates a stellar companion rather than a planetary companion. Columns 3-8 list the next three transit times and uncertainties after May 1, 2018 in BJD. Columns 9-12 list transit probabilities, depths, durations and ingress/egress durations, respectively. The transit probabilities are calculated purely from the best-fit radial velocity parameters. We have performed a literature search for those planets with transit probabilities greater than 10\%; those that are known to transit or not transit are marked accordingly. Finally, Columns 13-15 list the date of the next three transits in yyyy/mm/dd format (UTC) for easy reference.}
\end{deluxetable*}
\end{longrotatetable}

\end{document}